\documentclass[fleqn,12pt]{article}
\usepackage{amsmath,amssymb,amsfonts,amsthm, bbm, mathrsfs, natbib, cancel, sectsty, color, enumitem, array, lscape, hyperref, graphicx, enumitem, framed, booktabs, float}
\usepackage{ifthen, appendix}
\usepackage{setspace}
\usepackage[justification=centering]{caption}
\usepackage{subcaption}
\hypersetup{pdfborder = {0 0 0},colorlinks=true,linkcolor=blue,urlcolor=blue,citecolor=blue}

\usepackage[top=.95in,bottom=.95in,left=1in,right=1in]{geometry}

\usepackage{xr}
\newtheorem{theorem}{Theorem}

\newtheorem{corollary}{Corollary}

\newtheorem{assumption}{Assumption}

\theoremstyle{definition}
\newtheorem{remark_tmp}{Remark}
\newenvironment{remark}
	{ \begin{remark_tmp} 	}
	{ 
		
		\qed 
		\end{remark_tmp} 
	}

\allowdisplaybreaks[1]

\newcommand{\forloop}[5][1]{\setcounter{#2}{#3}\ifthenelse{#4}{#5\addtocounter{#2}{#1}\forloop[#1]{#2}{\value{#2}}{#4}{#5}}{}}

\newcommand{\plim}{\text{plim}}

\newcommand{\I}{\mathbbm{1}}
\newcommand{\Ihat}{\hat{\mathbbm{1}}}

\newcommand{\pto}{\to_\mathbb{P}}
\newcommand{\dto}{\to_{\text{d}}}

\newcommand{\E}{\mathbb{E}}
\newcommand{\V}{\mathbb{V}}
\renewcommand{\P}{\mathbb{P}}

\newcommand{\dz}{{d}}

\newcommand{\sumjd}{\sum_{j=1}^{J_t^{\dz}}}
\newcommand{\sumt}{\sum_{t=1}^T}

\newcommand{\bz}{\mathbf{z}}
\newcommand{\bx}{\mathbf{x}}
\newcommand{\ba}{\mathbf{a}}
\newcommand{\bbeta}{\boldsymbol{\beta}}

\newcommand{\bF}{\mathbf{f}}

\begin{document}
\setcounter{page}{0}

\newgeometry{left=.75in, right=.75in, top=.75in, bottom=.75in}

\captionsetup[subfigure]{labelformat=empty}
\captionsetup{justification=raggedright, singlelinecheck=false}



\title{\vspace{0in}{\Large \textbf{Characteristic-Sorted Portfolios: Estimation and Inference}\thanks{First draft: March 28, 2015. The authors would like to thank Tobias Adrian, Francisco Barillas, Tim Bollerslev, Nina Boyarchenko, Jules van Binsbergen, John Campbell, Kent Daniel, Fernando Duarte, Eric Ghysels, Stefano Giglio, Peter Hansen, Ralph Koijen, Gabriele La Spada, Jonathan Lewellen,  Jia Li, Erik Loualiche, Stefan Nagel, Andrew Patton, Karen Shen, George Tauchen, Stijn Van Nieuwerburgh, Peter Van Tassel, S. Viswanathan, Erik Vogt, Brian Weller, Jonathan Wright as well as seminar and conference participants at the 2015 Interactions Conference, the 2016 MFA Annual Meetings, University of Miami, and Duke University for helpful comments and discussions. Skanda Amarnath, Evan Friedman, and Rui Yu provided excellent research assistance. Last but not least, we thank the editor, Olivier Coibion, and three anonymous reviewers for their comments. The views expressed in this paper are those of the authors and do not necessarily represent those of the Federal Reserve Bank of New York, the Federal Reserve System or AQR Capital Management LLC. Cattaneo gratefully acknowledges financial support from the National Science Foundation through grant SES 1459931. Farrell gratefully acknowledges financial support from the Richard N. Rosett and John E. Jeuck Fellowships.}}}

\author{
Matias D. Cattaneo\thanks{Department of Operations Research and Financial Engineering, Princeton University; {\tt cattaneo@princeton.edu}}\and
Richard K. Crump\thanks{Federal Reserve Bank of New York; {\tt richard.crump@ny.frb.org}}\and
Max H. Farrell\thanks{Booth School of Business, University of Chicago; {\tt max.farrell@chicagobooth.edu}}\and
Ernst Schaumburg\thanks{AQR Capital Management; {\tt ernst.schaumburg@gmail.com}}}
\date{\today}
\maketitle

\vspace{-.3in}
\begin{abstract}
\doublespacing Portfolio sorting is ubiquitous in the empirical finance literature, where it has been widely used to identify pricing anomalies. Despite its popularity, little attention has been paid to the statistical properties of the procedure.  We develop a general framework for portfolio sorting by casting it as a nonparametric estimator. We present valid asymptotic inference methods and a valid mean square error expansion of the estimator leading to an optimal choice for the number of portfolios. In practical settings, the optimal choice may be much larger than the standard choices of 5 or 10. To illustrate the relevance of our results, we revisit the size and momentum anomalies.\end{abstract}

\begin{small}
\textbf{Keywords}: Portfolio sorts, nonparametric estimation, partitioning, tuning parameter selection.

\textbf{JEL Classification}: C12, C14

\end{small}

\thispagestyle{empty}

\restoregeometry
\newpage 
\doublespacing
\setcounter{page}{1}

\section{Introduction}

Portfolio sorting is an important tool of modern empirical finance.  It has been used to test fundamental theories in asset pricing, to establish a number of different pricing anomalies, and to identify profitable investment strategies. However, despite its ubiquity in the empirical finance literature, little attention has been paid to the statistical properties of the procedure. We endeavor to fill this gap by formalizing and investigating the properties of so-called characteristic-sorted portfolios---where portfolios of assets are constructed based on similar values for one or more idiosyncratic characteristics and the cross-section of portfolio returns is of primary interest. The empirical applications of characteristic-sorted portfolios are too numerous to list, but some of the seminal work applied to the cross-section of equity returns includes \citet*{Basu1977}, \citet*{Stattman1980}, \citet*{Banz1981}, \citet*{DeBondtThaler1985}, \citet*{Jegadeesh1990}, \citet*{FamaFrench1992}, and \citet*{JegadeeshTitman1993}. More recently, the procedure has been applied to other asset classes such as currencies and across different assets; furthermore, portfolio sorting remains a highly popular tool in empirical finance.

We develop a general, formal framework for portfolio sorting by casting the procedure as a nonparametric estimator. Sorting into portfolios has been informally recognized in the literature as a nonparametric alternative to imposing linearity on the relationship between returns and characteristics in recent years \citep[e.g.][]{FamaFrench2008,Cochrane2011}, but no formal framework is at present available in the literature. We impose sampling assumptions which are very general and can accommodate momentum and reversal effects, conditional heteroskedasticity in both the cross section and the time series, and idiosyncratic characteristics with a factor structure. Furthermore, our proposed framework allows for both estimated quantiles when forming the portfolios and additive linear-in-parameters conditioning variables entering the underlying model governing the relationship between returns and sorting characteristics. This latter feature of our proposed framework bridges the gap between portfolio sorts and cross-sectional regressions and will allow empirical researchers to investigate new candidate variables while controlling for existing anomalies already identified. More generally, our framework captures and formalizes the main aspects of common empirical work in finance employing portfolio sorts, and therefore gives the basis for a thorough analysis of the statistical properties of popular estimators and test statistics.

Employing our framework, we study the asymptotic properties of the portfolio-sorting estimator and related test statistics in settings with ``large'' cross-sectional and times-series sample sizes, as this is the most usual situation encountered in applied work. We first establish consistency and asymptotic normality of the estimator, explicitly allowing for estimated quantile-spaced portfolios, which reflects standard practice in empirical finance. In addition, we prove the validity of two distinct standard error estimators. The first is a ``plug-in'' variance estimator which is new to the literature. The second is the omnipresent \citet*{FamaMacBeth1973}-style variance estimator which treats the average portfolio returns as if they were draws from a single, uncorrelated time series.  Despite its widespread use, we are unaware of an existing proof of its validity for inference in this setting, although this finding is presaged by the results in \citet*{IM2010, IM2016}. All together, our first-order asymptotic results provide theory-based guidance to empirical researchers.

Once the portfolio sorting estimator is viewed through the lens of nonparametric estimation, it is clear that the choice of number of portfolios acts as the tuning parameter for the procedure and that an appropriate choice is paramount for drawing valid empirical conclusions. To address this issue, we obtain higher-order asymptotic mean square error expansions for the estimator which we employ to develop several optimal choices of the total number of portfolios for applications. These optimal choices balance bias and variance and will change depending on the prevalence of many common features of panel data in finance such as unbalanced panels, the relative number of cross-sectional observations versus time-series observations and the presence of conditional heteroskedasticity.  In practice, the common approach in the empirical finance literature is  to treat the choice of the number of portfolios as invariant to the data at hand---often following historical norms, such as $10$ portfolios when sorting on a single characteristic. This is summarized succinctly in \citet*[][p. 1061]{Cochrane2011}: ``Following Fama and French, a standard methodology has developed: Sort assets into portfolios based on a characteristic, look at the portfolio means (especially the \textit{1--10} portfolio alpha, information ratio, and t-statistic)...'' (emphasis added).  Thus, another contribution of our paper is to provide a simple, data-driven procedure which is optimal in an objective sense to choose the appropriate number of portfolios. Employing this data-driven procedure provides more power to discern a significant return differential in the data. The optimal choice will vary across time with the cross-sectional sample size and, all else equal, be larger for longer time series. Our results thus directly impact empirical practice by providing a transparent, objective, data-driven way to choose the number of portfolios which nonetheless captures intuitive, real-world concerns in data analysis.

We demonstrate the empirical relevance of our theoretical results by revisiting the size anomaly, where smaller firms earn higher returns than larger firms on average, and the momentum anomaly, where firms which have had better relative returns in the recent past also have higher future relative returns on average.  We find that in the universe of US stocks the size anomaly is significant using our methods and is robust to different sub-periods including the period from 1980--2015. Moreover, this conclusion would not be reached with the ad-hoc, yet standard, choice of ten portfolios; our results are thus crucial for data analysis. Our results suggest that the relationship is monotonically decreasing and convex; this is borne out graphically. As pointed out in the existing literature, the size anomaly is not robust in sub-samples which exclude ``smaller'' small firms (i.e., considering only firms listed on the NYSE).  We also find that in the universe of US stocks the momentum anomaly is significant, with the ``short'' side of the trade becoming more profitable in later sub-periods. Graphically, the relationship appears monotonically increasing and concave. We also show that the momentum anomaly is distinct from industry momentum by including the latter measure (along with its square and cube) as linear control variables in a portfolio sorting exercise.  In both empirical applications we find that the optimal number of portfolios varies substantially over time and is much larger than the standard choice of ten routinely used in the empirical finance literature, and more importantly, that substantive conclusions change with the number of portfolios chosen for analysis. In the case of the size anomaly, the optimal number of portfolios can be as small as about 50 in the 1920s and can rise to above 200 in the late 1990s. However, for the momentum anomaly, the optimal number of portfolios is about 10 in the 1920s and about 50 in the late 1990s.

The financial econometrics literature has primarily focused on the study of estimation and inference in (restricted) factor models featuring common risk factors and idiosyncratic loadings. For recent examples, see \citet*{ShankenZhou2007}, \citet*{Kleibergen2010}, \citet*{NagelSingleton2011}, \citet*{ConnorHagmannLinton2012}, \citet*{ACM2013}, \citet*{ALS2017}, \citet*{GKR2016} among others. In contrast, to our knowledge, we are the first to provide a formal framework and to analyze the standard empirical approach of (characteristic-based) portfolio sorting.  A few authors have investigated specific aspects of sorted portfolios. \citet*{LoMacKinlay1990} and \citet*{ConradCooperKaul2003} have studied the effects of data-snooping bias on empirical conclusions drawn from sorted portfolios and argue that they can be quite large. \citet*{Berk2000} investigates the power of testing asset pricing models using only the assets within a particular portfolio and argues that this approach biases results in favor of rejecting the model being studied. More recently, \citet*{PattonTimmermann2010} and \citet*{RomanoWolf2013} have proposed tests of monotonicity in the average cross-section of returns taking the sorted portfolios themselves as \textit{given}.  Finally, there is a large literature attempting to discriminate between factor-based and characteristic-based explanations for return anomalies.  The empirical implementations in this literature often use characteristic-sorted portfolios as test assets although this approach is not universally advocated (see, for example, \citet*{LewellenNagelShanken2010} and \citet*{Kleibergen2010}).

The paper is organized as follows.  Section \ref{sec:overview} describes our framework and provides a brief overview of our new results. The more general framework is presented in Section \ref{sec:framework}. Then Sections \ref{sec:first order} and \ref{sec:mse} treat first-order asymptotic theory and mean square error expansions, respectively; the latter provides guidance on implementation. Section \ref{sec:Applications} provides our empirical results and Section \ref{sec:Conclusion} concludes and discusses further work.

\section{Motivation and Overview of Results}
	\label{sec:overview}

This section provides motivation for our study of portfolio sorting and a simplified overview of our results. The premise behind portfolio sorting is to discover whether expected returns of an asset are related to a certain characteristic. A natural, and popular, way to investigate this is to sort observed returns by the characteristic value, divide the assets into portfolios according to the characteristic, and then compare differences in average returns across the portfolios. This methodological approach has found wide popularity in the empirical finance literature not least because it utilizes a basic building block of modern finance, a portfolio of assets, which produces an intuitive estimator of the relationship between asset returns and characteristics. The main goal of this paper is to provide a formal framework and develop rigorous inference results for this procedure. All assumptions and technical results are discussed in detail in the following sections, but omitted here for ease of exposition.

To begin, suppose we observe both the return, $R$, and value of a single continuous characteristic, $z$, for $n$ assets over $T$ time periods, that are related through a regression-type model of the form
\begin{equation}
	\label{eqn:ModelNP0}
	R_{it} = \mu(z_{it}) + \varepsilon_{it}, \qquad i=1,\ldots,n, \quad t=1,\ldots,T.
\end{equation}
Here $\mu(\cdot)$ is the unknown object of interest that dictates how expected returns vary with the characteristic, and is assumed to be twice continuously differentiable. The general results given in the next section cover a wide range of inference targets and extend the model of equation \eqref{eqn:ModelNP0} to include multiple sorting characteristics, conditioning variables, and unbalanced panels, among other features commonly encountered in empirical finance.

To understand the relationship between expected returns and the characteristic at hand, characterized by the unknown function $\mu(z)$, we first form portfolios by partitioning the support of $z$ into quantile-spaced bins. While it is possible to form portfolios in other ways, quantile spacing is the standard technique in empirical finance; our goal is to develop theory that mimics empirical practice as closely as possible. For each period $t$, it is common practice to form $J$ disjoint portfolios, denoted by $P_{jt}$, as follows: $P_{jt} = \left[z_{\left(\left\lfloor{n(j-1)/J}\right\rfloor\right)t} , z_{\left(\left\lfloor{nj/J}\right\rfloor\right)t}\right)$ if $j=1, \ldots,J-1$, and $P_{Jt} = \left[z_{\left(\left\lfloor{n(J-1)/J}\right\rfloor\right)t} , z_{(n)t}\right]$, where $z_{(\ell)t}$ denotes the $\ell$-th order statistic of the sample of characteristics $\{z_{it}:1\leq i\leq n\}$ at each time period $t=1,2,\cdots,T$, and $\left\lfloor\cdot\right\rfloor$ denotes the floor operator. In other words, each portfolio is a random interval containing roughly $(100/J)$-percent of the observations at each moment in time. This means that the position and length of the portfolios vary over time, but is set automatically, while the number of such portfolios ($J$) must be chosen by the researcher. A careful (asymptotic) analysis of portfolio-sorting estimators requires accounting for the randomness introduced in the construction of the portfolios, as we do in more detail below.

With the portfolios thus formed, we estimate $\mu(z_*)$ at some fixed point $z_*$ with the average returns within the portfolio containing $z_*$. Here $z_*$ represents the evaluation point that is of interest to the empirical researcher.  For example, one might be interested in expected returns for those individual assets with a very high value of a characteristic.  Over time, exactly which portfolio includes assets with characteristic $z_*$ may change. If we let $P_{jt}^*$ represent the appropriate portfolio at each time $t$ then the basic portfolio-sorted estimate is
\begin{equation}
	\label{eqn:overview estimator}
	\hat{\mu}(z_*) = \frac{1}{T} \sumt \hat{\mu}_t(z_*),   		\qquad\qquad  		  \hat{\mu}_t(z_*) = \frac{1}{N_{jt}^*} \sum_{i: z_{it} \in P_{jt}^*} R_{it},
\end{equation}
where $N^{*}_{jt}$ is the number of assets in $P^{*}_{jt}$ at time $t$. If $J\leq n$, this estimator is well-defined, as there are (roughly) $n/J$ assets in all portfolios. The main motivation for using a sample average of each individual estimator is so that the procedure more closely mimics the actual practice of portfolio choice (where future returns are unknown) and because of the highly unbalanced nature of financial panel data. That said, this estimator (as well as the more general version below) can be simply implemented using ordinary least squares (or weighted least squares in the case of value-weighted portfolios).

The starting point of our formalization is the realization that each $\hat{\mu}_t(z_*)$, $t=1,\ldots,T$, is a nonparametric estimate of the regression function $\mu(z_*)$, using a technique known as \emph{partitioned regression}. Studied recently by \citet*{Cattaneo-Farrell_2013_JoE}, the partition regression estimator estimates $\mu(z_*)$ using observations that are ``close'' to $z_*$, which at present means that they are in the same portfolio. A key lesson is that $J$ is the tuning parameter of this nonparametric procedure, akin to the bandwidth in kernel-based estimators or the number of terms in a sieve estimator (such as knots for spline regression). It is well documented that nonparametric inference is sensitive to tuning parameter choices, and empirical finance is no exception. For smaller $J$, the variance of $\hat{\mu}_t(z_*)$ will be low, as a relatively large portion of the sample is in each portfolio, but this also implies that the portfolio includes assets with characteristics quite far from $z_*$, implying an increased bias; on the other hand, a larger $J$ will decrease bias, but inflate variance. For each cross section, $\hat{\mu}_t(\cdot)$ is a step function with $J$ ``rungs'', each an average return within a portfolio. While estimation of $\mu(\cdot)$ could be performed with a variety of nonparametric estimators (such as kernel or series regression), our goal is to explicitly analyze portfolio sorting. Such methods are not immune from tuning parameter choice sensitivity, and may require stronger assumptions than portfolio sorting. From a practitioner's perspective, the estimator has the advantage that it has a direct interpretation as a return on a portfolio which is an economically meaningful object.

\begin{figure}[h!]
\vspace{-.75cm}
	\caption{\doublespacing\textbf{Introductory Example}\\[0.005in] \normalsize{This figure shows the true (dashed line) and estimated function (solid line). The left panels show the $n=500$ data points (gray dots) and the middle panels display the estimated function for each time period (light gray line).  Breakpoints are chosen as estimated quantiles of $z$ where $z\sim \mathcal{B}eta\left(1,1\right)/$$z\sim \mathcal{B}eta\left(1.2,1.2\right)$ for odd/even time periods.}}
	\label{fig:overview}
	\begin{subfigure}[t]{0.33\columnwidth}
		\centering	\caption{$J=4,$ $T=1$}
		\includegraphics[trim={6cm 7cm 6cm 7cm}, scale=.25]{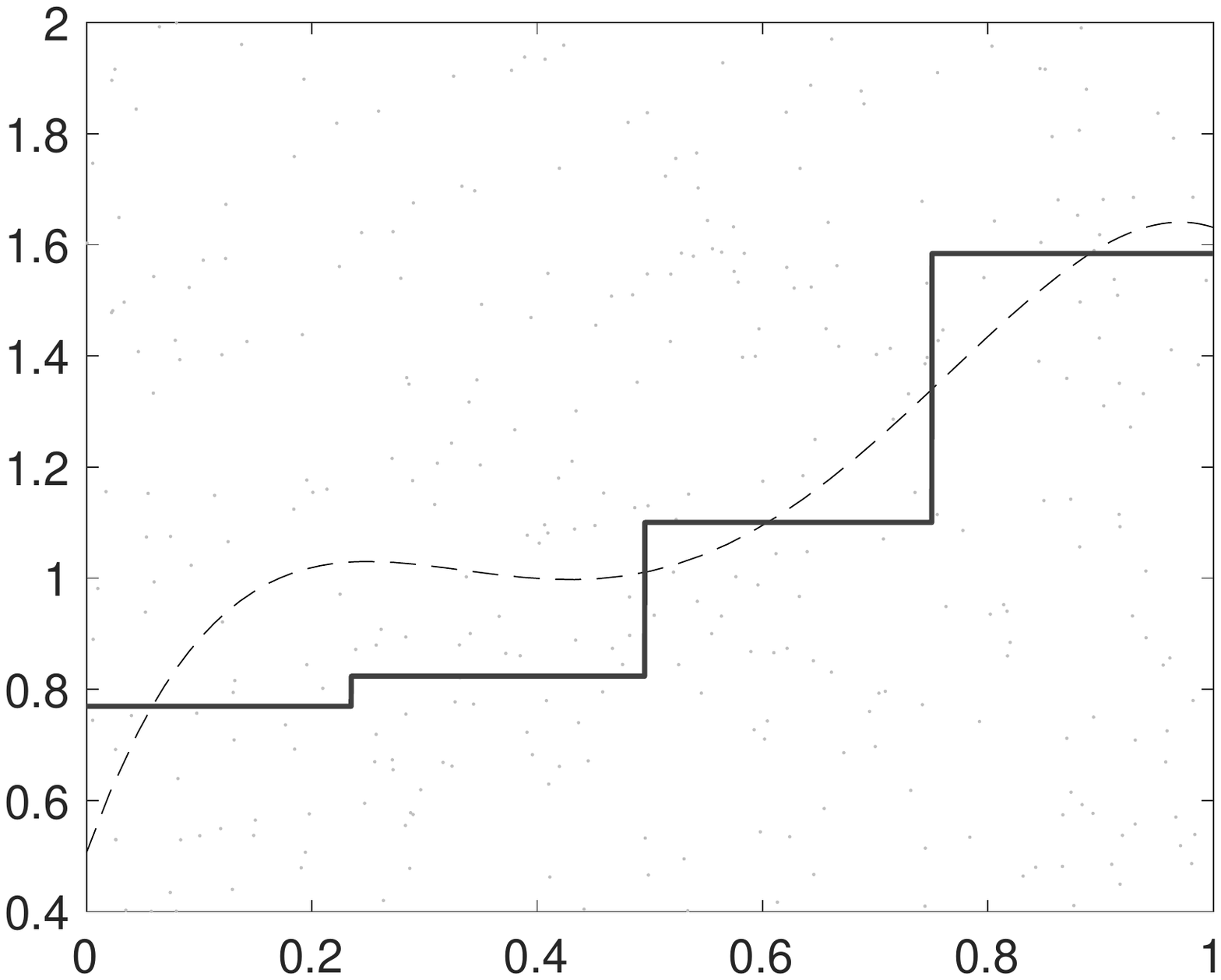}
	\end{subfigure}\hfill
	\begin{subfigure}[t]{0.33\columnwidth}
		\centering	\caption{$J=4,$ $T=2$}
		\includegraphics[trim={6cm 7cm 6cm 7cm}, scale=.25]{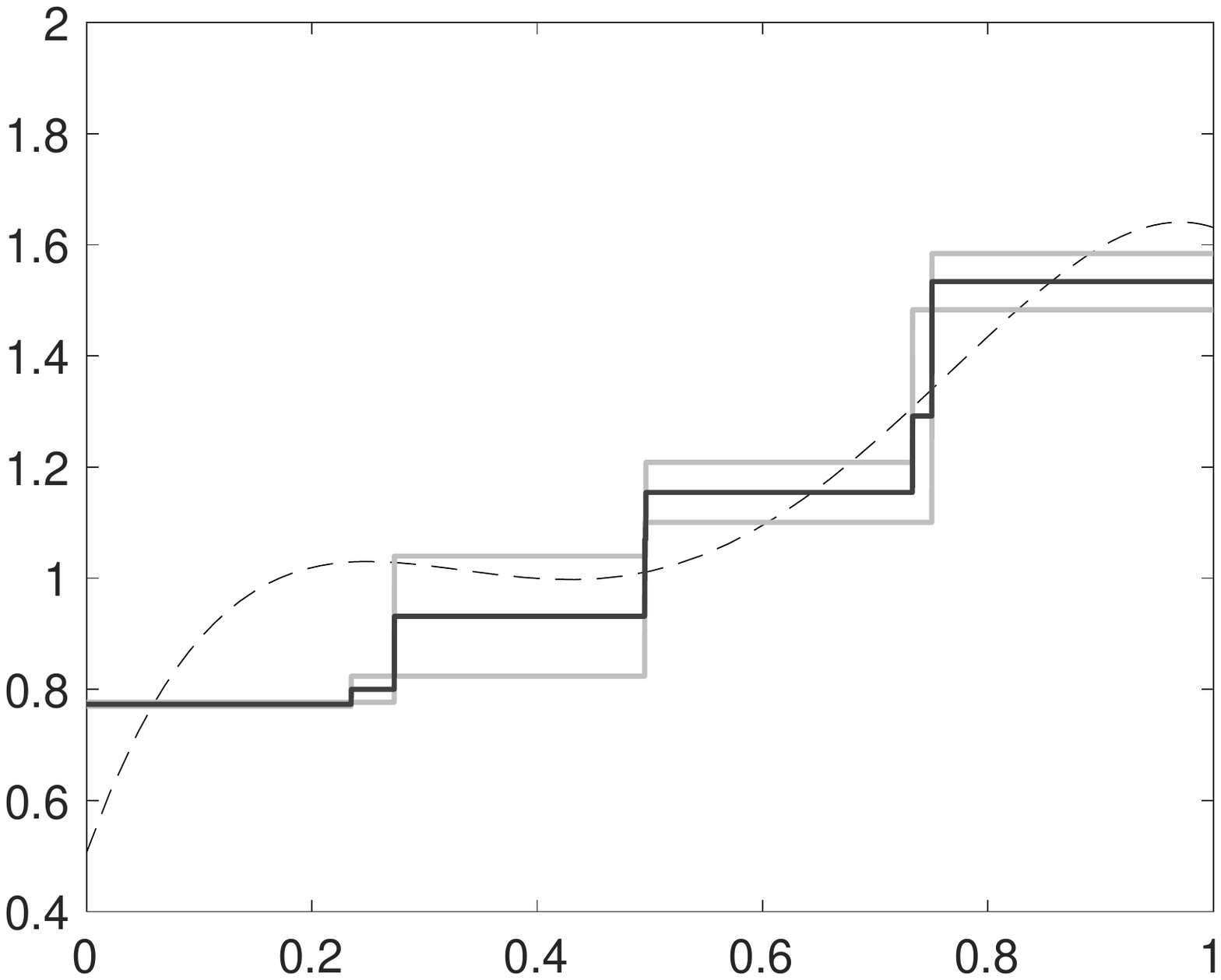}
	\end{subfigure}\hfill
	\begin{subfigure}[t]{0.33\columnwidth}
		\centering	\caption{$J=4,$ $T=50$}
		\includegraphics[trim={6cm 7cm 6cm 7cm}, scale=.25]{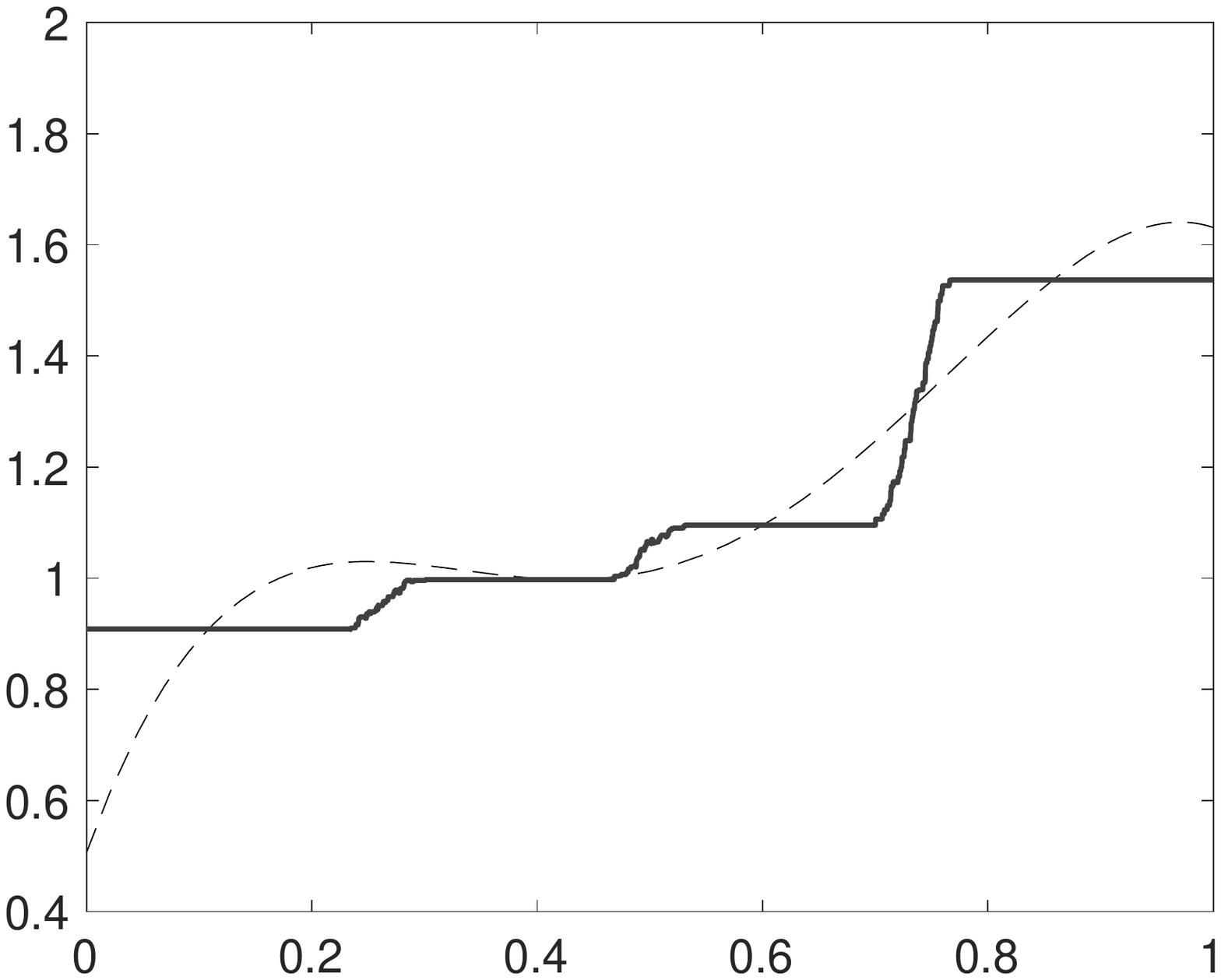}
	\end{subfigure}\\[0.15in]
	\begin{subfigure}[t]{0.33\columnwidth}
		\centering	\caption{$J=10,$ $T=1$}
		\includegraphics[trim={6cm 7cm 6cm 7cm}, scale=.25]{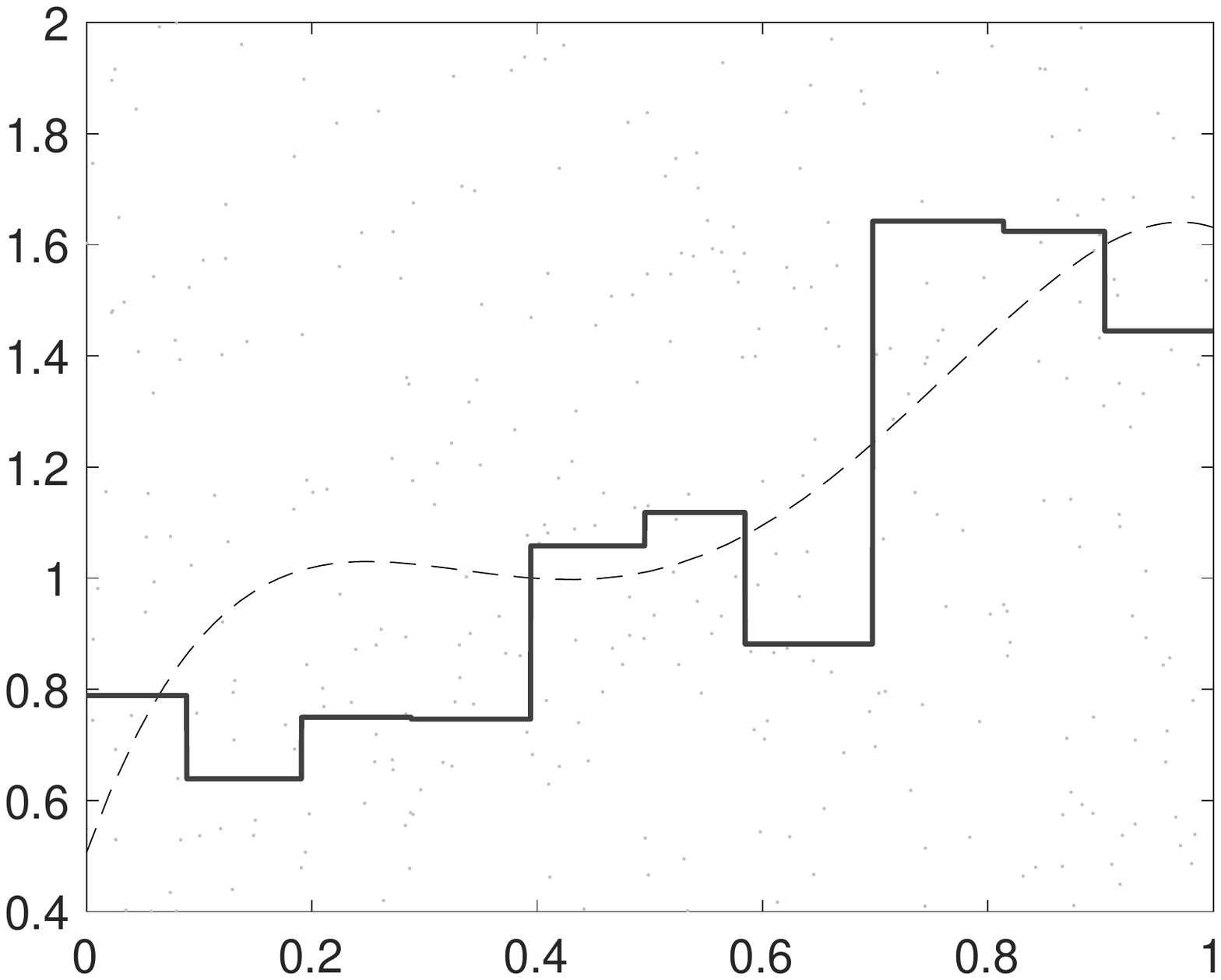}
	\end{subfigure}\hfill
	\begin{subfigure}[t]{0.33\columnwidth}
		\centering	\caption{$J=10,$ $T=2$}
		\includegraphics[trim={6cm 7cm 6cm 7cm}, scale=.25]{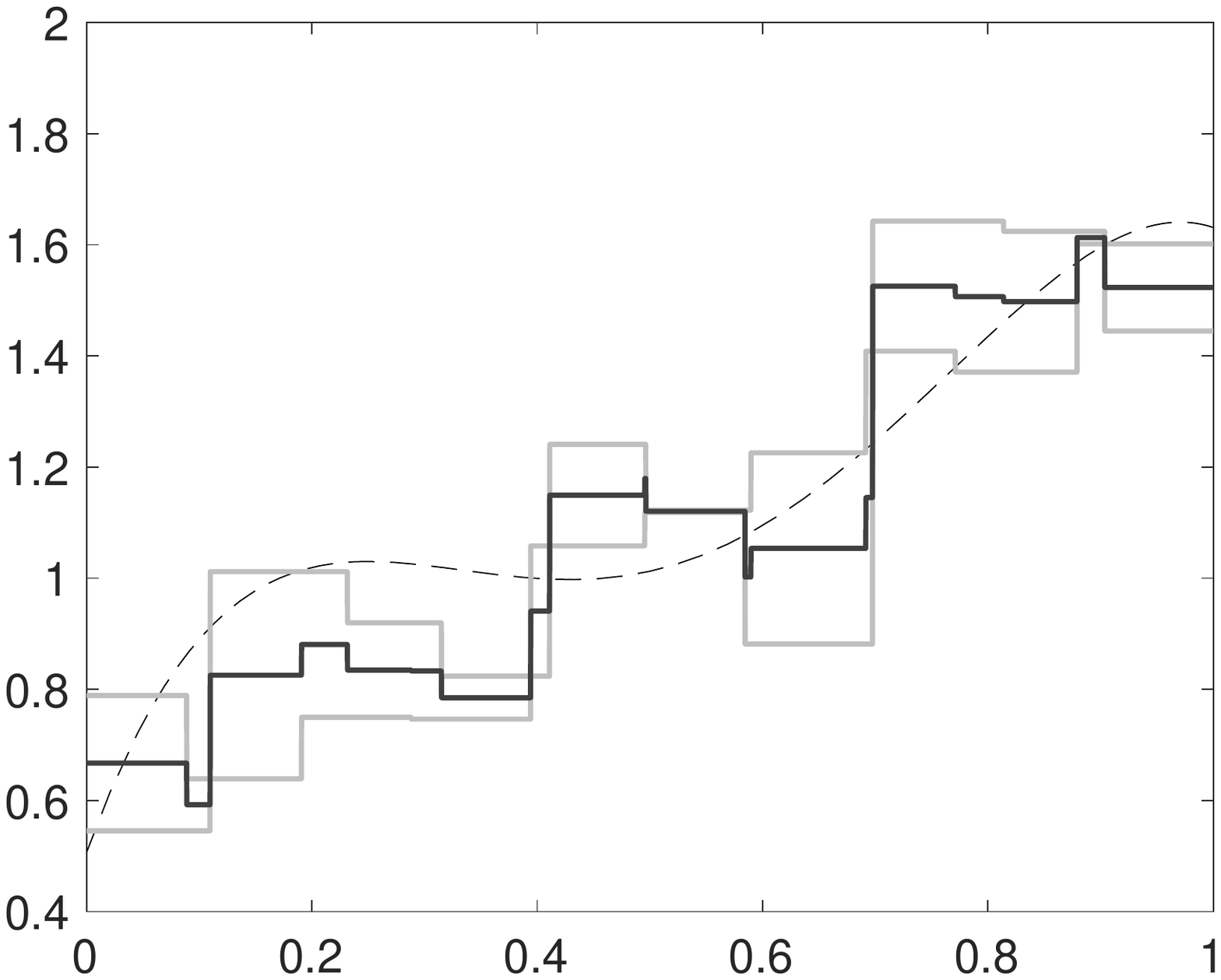}
	\end{subfigure}\hfill
	\begin{subfigure}[t]{0.33\columnwidth}
		\centering	\caption{$J=10,$ $T=50$}
		\includegraphics[trim={6cm 7cm 6cm 7cm}, scale=.25]{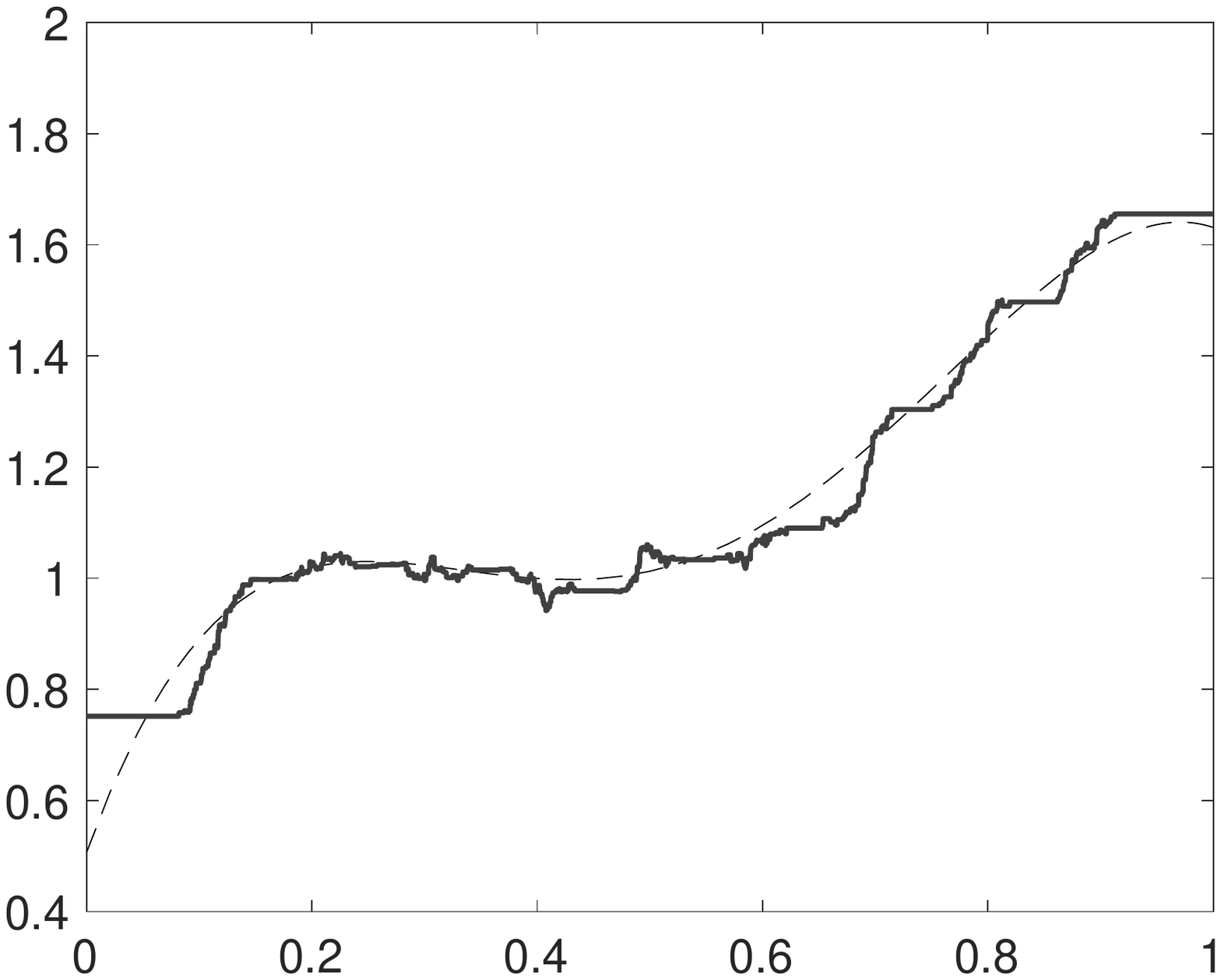}
	\end{subfigure}\hfill
\end{figure}

Moving beyond the cross section, the same structure and lessons holds for the full $\hat{\mu}(z_*)$ of equation \eqref{eqn:overview estimator}, but with dramatically different results. Consider Figure \ref{fig:overview}. Panel (a) shows a single realization of $\hat{\mu}_t(\cdot)$, with $J=4$, \emph{for a single cross section}. Moving to panel (b), we see that averaging over only two time periods results in a more complex estimator, as the portfolios are formed separately for each cross section. Finally, panel (c) shows the result with $T=50$ (though a typical application may have $T$ in the hundreds). Throughout, $J$ is fixed, but the increase in $T$ acts to smooth the fit; this point appears to be poorly recognized in practice, and makes clear that the choice of $J$ must depend on $T$. Next, for the same choices of $n$ and $T$, Panels (d)--(f) repeat the exercise but with $J=10$. Comparing panels in the top row to the bottom of Figure  \ref{fig:overview} shows the bias-variance tradeoff discussed above. Figure \ref{fig:overview} makes clear that $J$ must depend on the features of the data at hand. We show that consistency of $\hat{\mu}(\cdot)$ requires that $J$ diverge with $n$ and $T$ fast enough to remove bias but not so quickly that the variance explodes. We detail practicable choices of $J$ later in the paper.
\medskip

With the portfolios and estimator defined, by far the most common object of interest in the empirical finance literature is the expected returns in the highest portfolio less those in the lowest, which is then either (informally) interpreted as a test of monotonicity of the function $\mu(z)$ or used to construct factors based on the characteristic $z$. These are different goals (inference and point estimation, respectively), and thus require different choices of $J$. 

First, consider the test of monotonicity, which is also interpreted as the return from a strategy of buying the spread portfolio: long one dollar of the higher expected return portfolio and short one dollar of the lower expected return portfolio. Formally, we wish to conduct the hypothesis test:
\begin{equation}
	\label{eqn:hypothesis}
	{\sf H}_0: \mu(z_H) - \mu(z_L) = 0  		\qquad \text{vs.\ } \qquad {\sf H}_1:   \mu(z_H) - \mu(z_L) \neq 0,
\end{equation}
where $z_L<z_H$ denote ``low'' and ``high'' evaluation points. (In practice, $z_L$ and $z_H$ are usually far apart and never within the same portfolio.) Statistical significance in this context is intimately related to the economic significance of the trading strategy, as measured by the Sharpe ratio.
Our general framework allows for a richer class of estimands (see Remark \ref{rem:partial means}) but this estimand will remain our focus throughout the paper because it is the most relevant to empirical researchers.

Our main result establishes asymptotic validity for testing \eqref{eqn:hypothesis} using portfolio sorting with estimated quantiles. Namely, it follows from (the more general) Theorem \ref{thm:clt} that
\[
	\mathcal{T} = \frac{ \bigl[ \hat{\mu}(z_H) - \hat{\mu}(z_L) \bigr]  -  \bigl[\mu(z_H) - \mu(z_L) \bigr]}{\sqrt{\hat{V}(z_H) + \hat{V}(z_L)} }  \dto \mathcal{N}(0,1), 
\]
provided that $J \log(\max(J,T))/n \to 0$ and $n T / J^3 \to 0$, and other regularity conditions hold. The growth restrictions on $J$ formalize the bias-variance trade off in this problem. 

Consistent variance estimation can be done in several ways. The structure of the estimator implies that the variance of $\hat{\mu}(z_H) - \hat{\mu}(z_L)$ is the sum of each pointwise variance, and that $\hat{V}(z) \asymp J / (nT)$. We show that the commonly-used \citet*{FamaMacBeth1973} variance estimator, given by
	\[\hat{V}_\mathtt{FM}(z) = \frac{1}{T^2} \sumt (\hat{\mu}_t(z)-\hat{\mu}(z))^2,\] 
is indeed valid for Studentization, as is a novel plug-in approach. Both are given in equation \eqref{eqn:standard errors} below. See Theorem \ref{thm:se} for complete discussion. To the best of our knowledge, these results are all new to the literature.

\begin{figure}[t]
\caption{\doublespacing\textbf{Optimal Portfolios Counts}\\[0.05in] \normalsize{This figure shows the optimal number of portfolios for the estimated relationsh between the cross section of equity returns and lagged market equity (equation \eqref{eqn:size}, left column), and 12-2 momentum (equation \eqref{eqn:momentum}, right column). $J_t$ has been chosen based on equation \eqref{eqn:HOMSE}, $z_H = \Phi^{-1}(.975)$, $z_L = \Phi^{-1}(.025)$.}}
\label{fig:JtStar}
\begin{subfigure}[t]{0.5\columnwidth}
\centering	\caption{\textbf{Size Anomaly}\\\textit{1926--2015}}
\includegraphics[trim={6cm 7cm 6cm 7cm}, scale=.40]{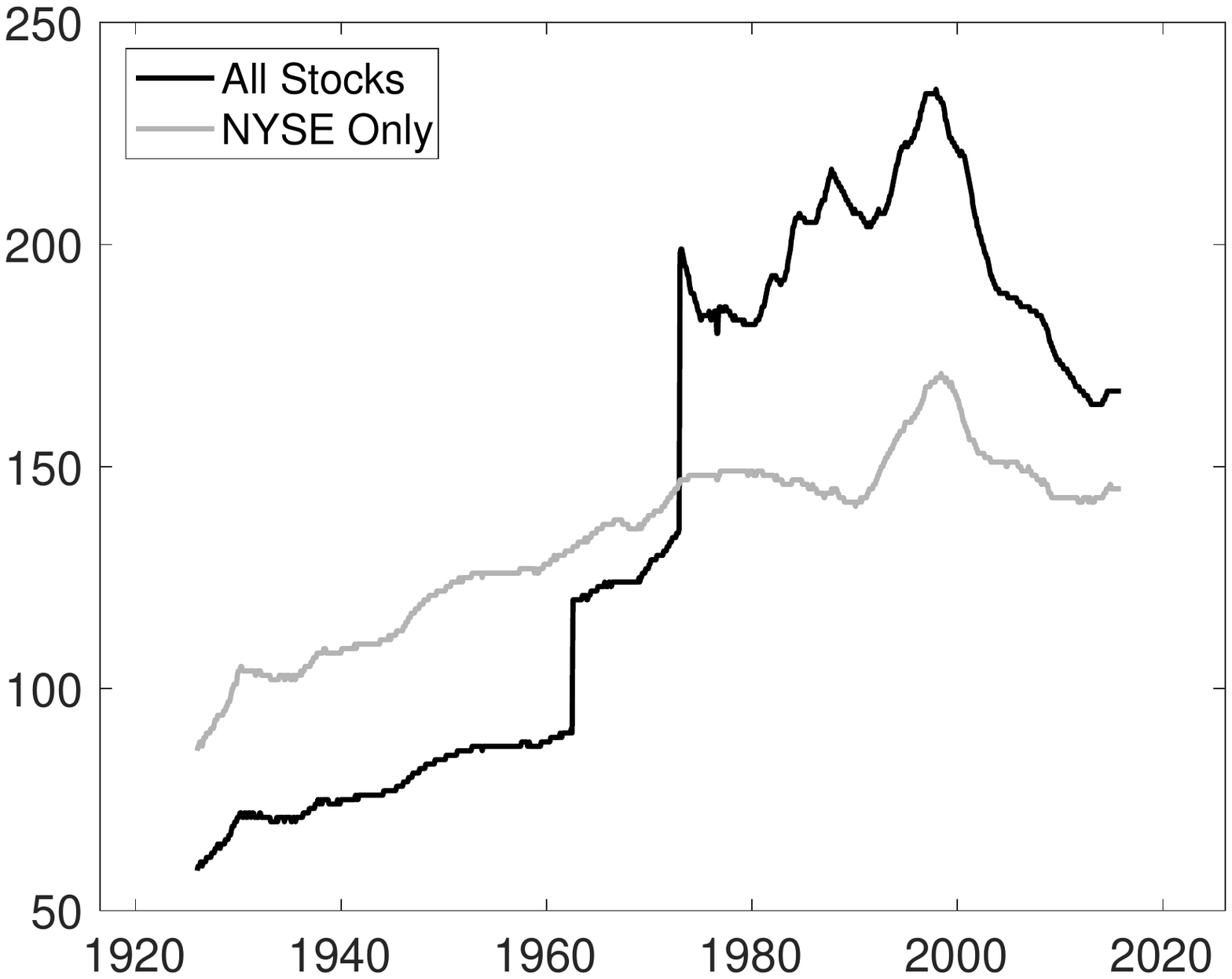}
\end{subfigure}\hfill
\begin{subfigure}[t]{0.5\columnwidth}
\centering	\caption{\textbf{Momentum Anomaly}\\\textit{1926--2015}}
\includegraphics[trim={6cm 7cm 6cm 7cm}, scale=.40]{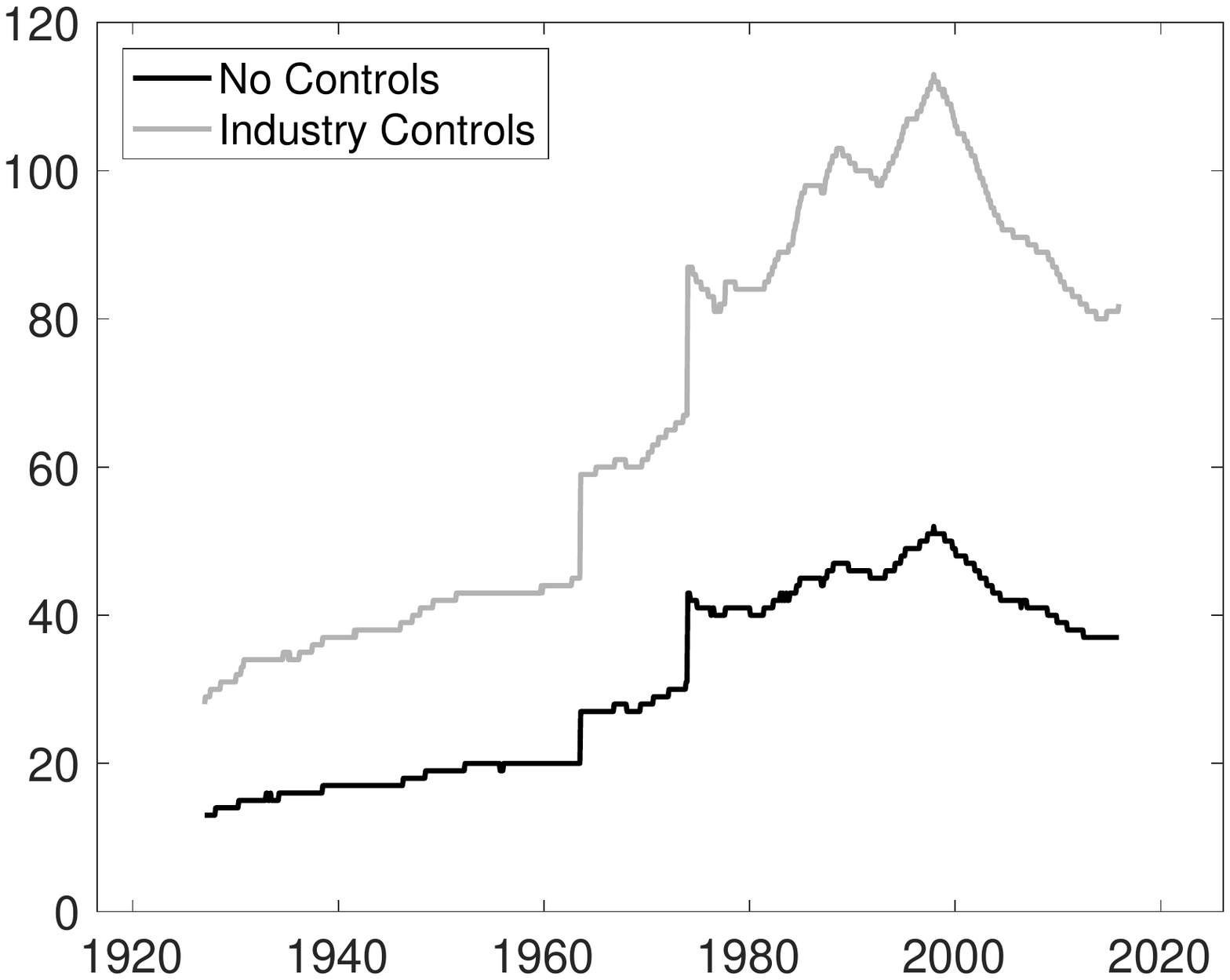}
\end{subfigure}\\[0.25in]
\begin{subfigure}[t]{0.5\columnwidth}
\centering	\caption{\textit{1967--2015}}
\includegraphics[trim={6cm 7cm 6cm 7cm}, scale=.40]{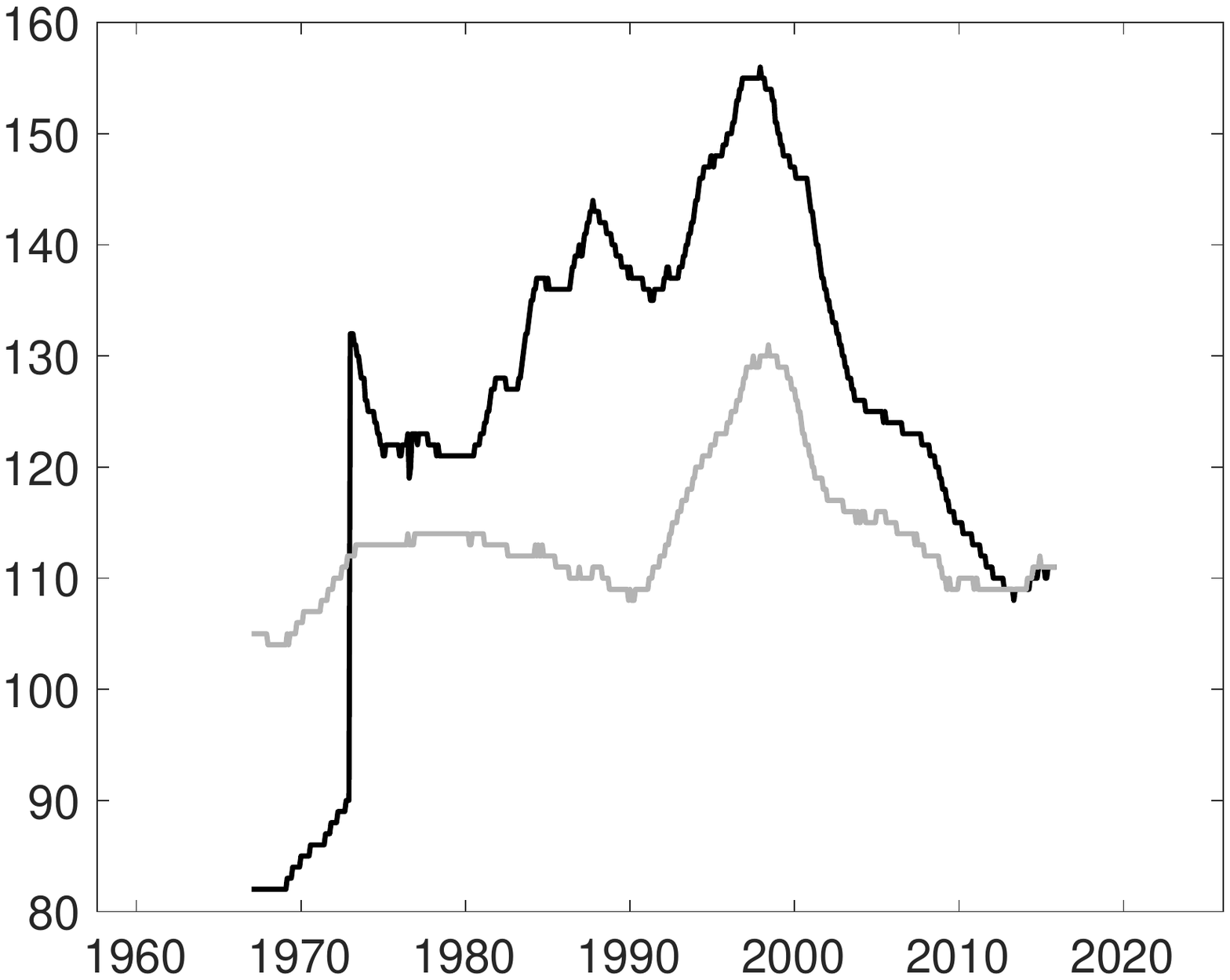}
\end{subfigure}\hfill
\begin{subfigure}[t]{0.5\columnwidth}
\centering	\caption{\textit{1967--2015}}
\includegraphics[trim={6cm 7cm 6cm 7cm}, scale=.40]{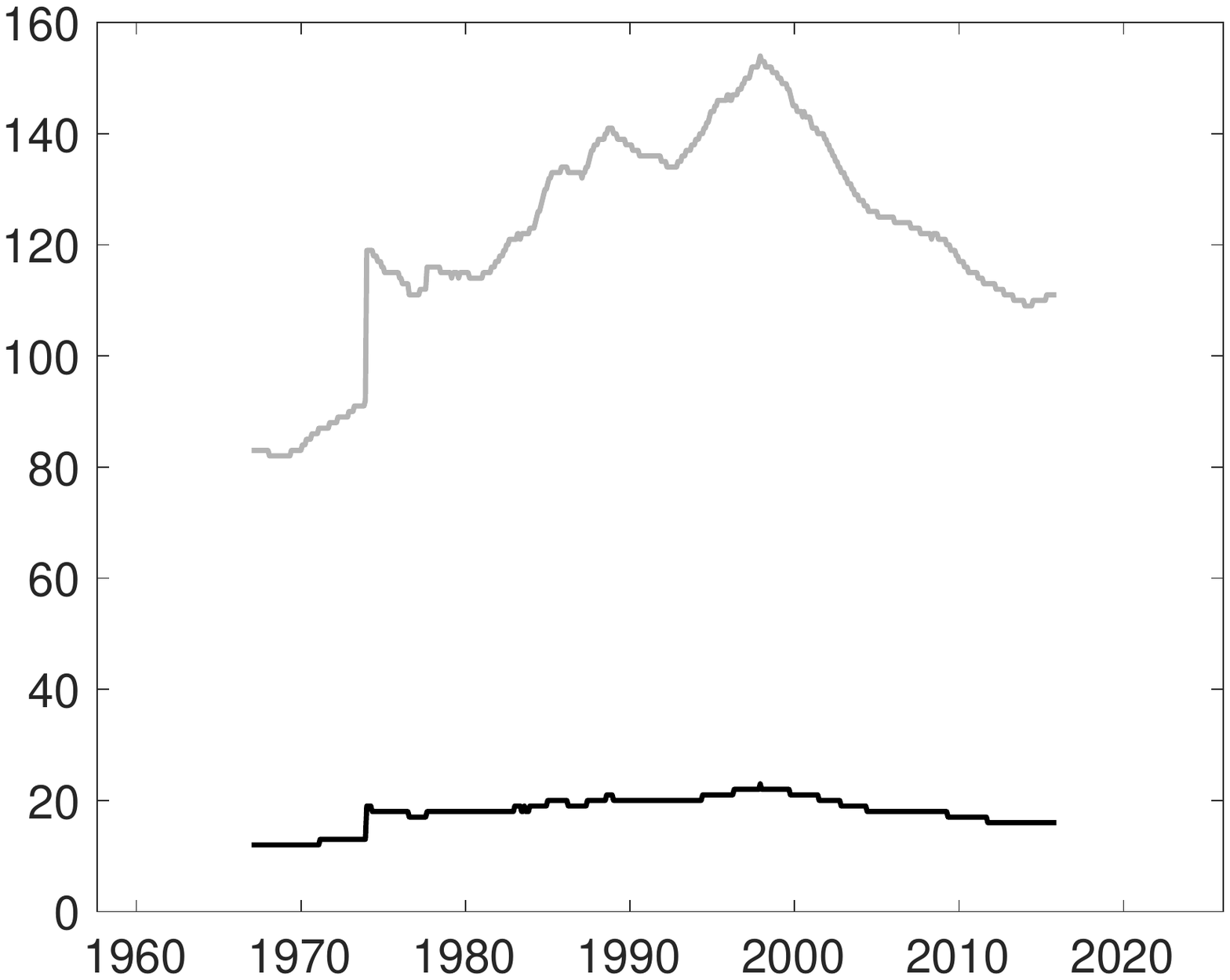}
\end{subfigure}\\[0.25in]
\begin{subfigure}[t]{0.5\columnwidth}
\centering	\caption{\textit{1980--2015}}
\includegraphics[trim={6cm 7cm 6cm 7cm}, scale=.40]{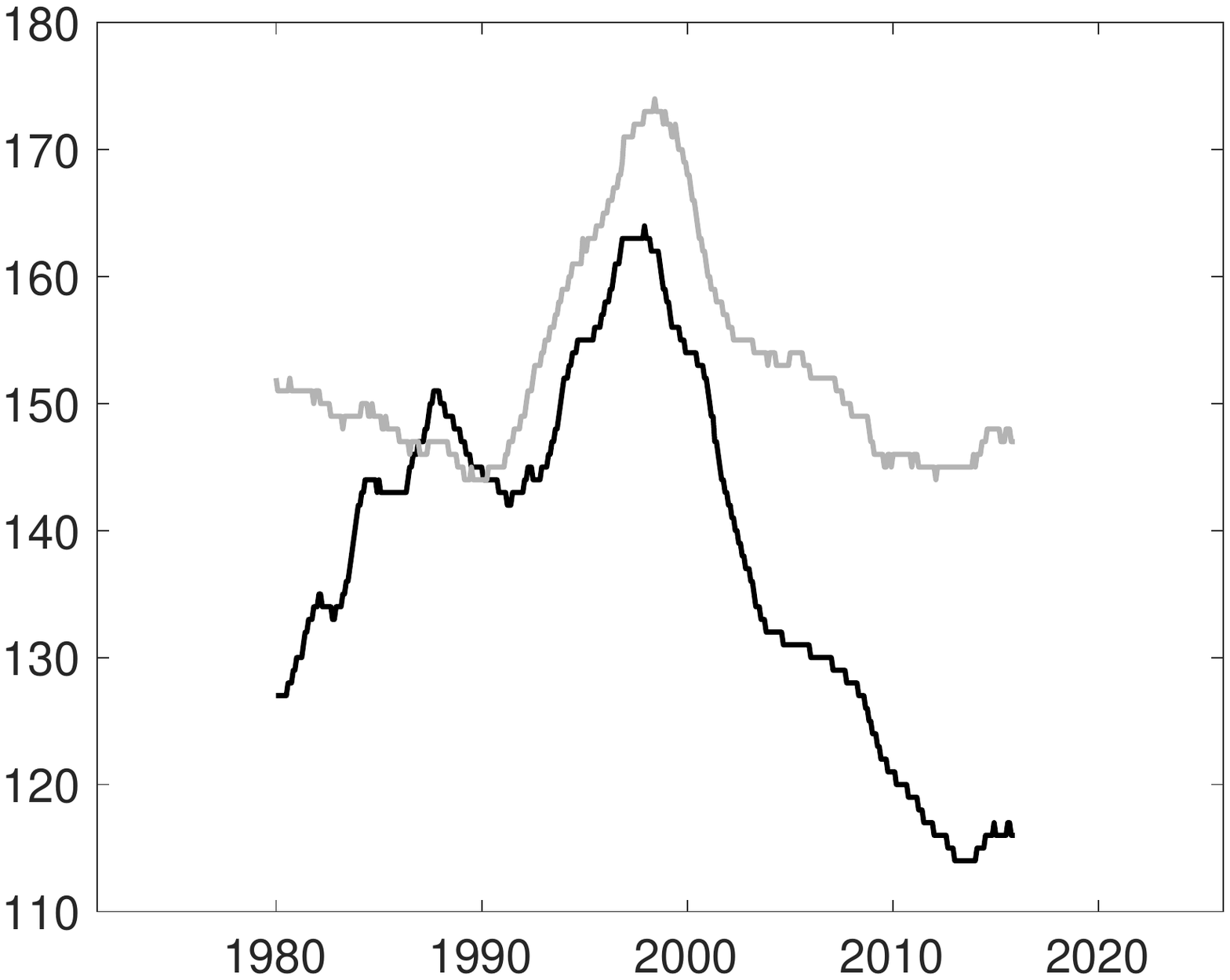}
\end{subfigure}\hfill
\begin{subfigure}[t]{0.5\columnwidth}
\centering	\caption{\textit{1980--2015}}
\includegraphics[trim={6cm 7cm 6cm 7cm}, scale=.40]{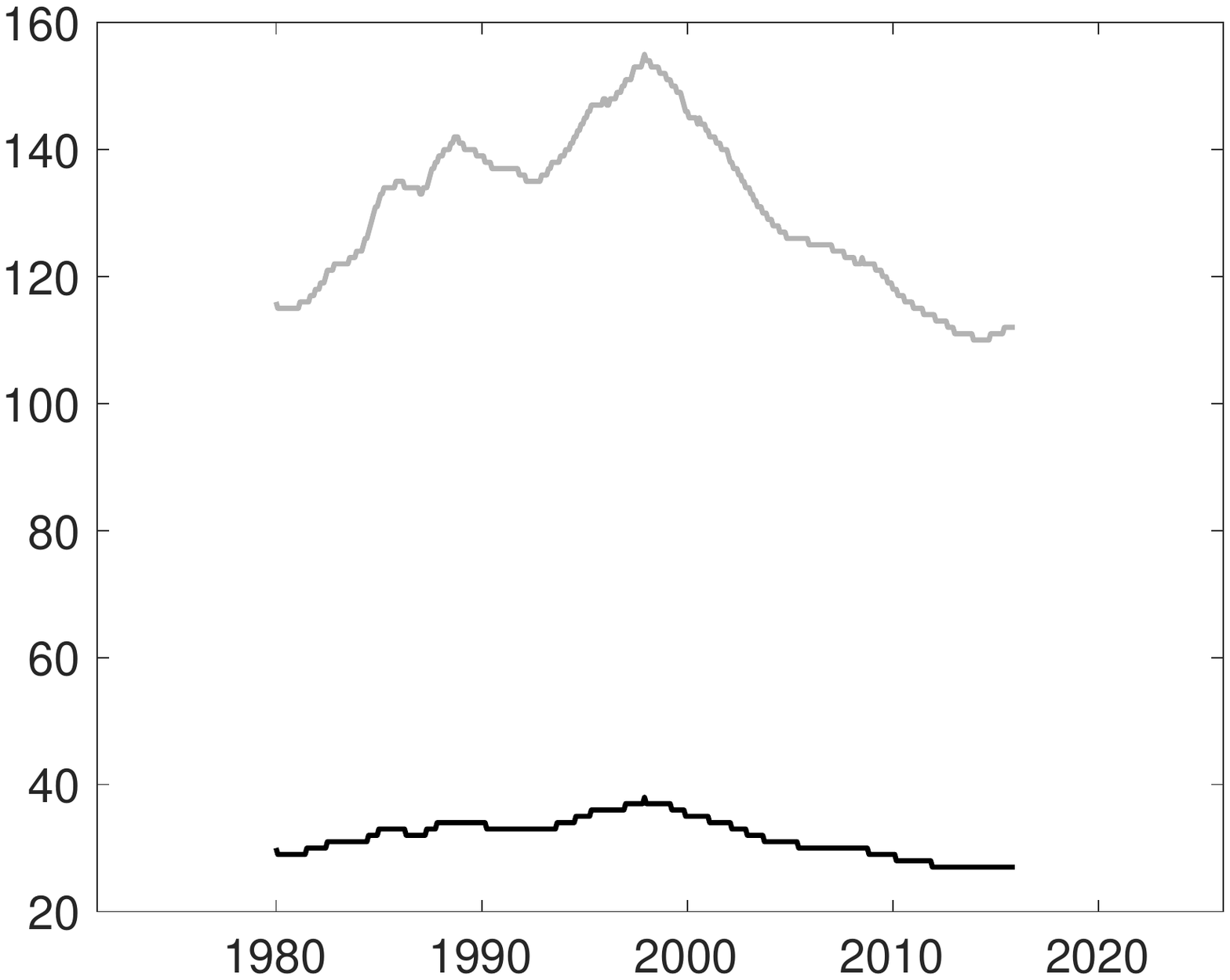}
\end{subfigure}\hfill
\end{figure}

Beyond first-order validity, we also provide explicit, practicable guidance for choice of $J$ via higher-order mean square error (MSE) expansions. To our knowledge, this represents the first theory-founded choice of $J$ for implementing portfolio sorting based inference. The literature typically employs \emph{ad hoc} choices, and often $J=10$ (see quotation from \citet*{Cochrane2011} above). However, given the nonparametric nature of the problem, $J$ should depend on the features of the data, and moreover, should change over time because cross-sectional sample sizes vary substantially. To make this clear notationally, we will write $J_t$ for the number of portfolios in period $t$. Even if these facts are recognized by empirical researchers, and the need for $J\neq10$ is clear, a lack of principled tools may be holding back practice. Our results fill this gap by providing a transparent, data-driven method of portfolio choice, so that practitioners who wish to use something other than ten may do so in a replicable, objective way. For example, in our data, $n$ ranges from 500 to nearly 8,000 (see Section \ref{sec:Applications} and Figure \ref{fig:nt} 
in the Supplemental Appendix) and the optimal choice of $J_t$, for example, varies from 13 to 52 for the momentum anomaly (Figure \ref{fig:JtStar}).

In the context of hypothesis testing, as in equation \eqref{eqn:hypothesis}, we find that the optimal number of portfolios obeys
	  \[J_t^\star = K^\star n_t^{1/2} T^{1/4} , \qquad t=1,2,\cdots,T,\]
where the constant $K^\star$ depends on the data generating process. It is easy to check that $J_t^\star$ satisfies the conditions above (i.e., those for Theorem \ref{thm:clt}). In Section \ref{sec:mse} we detail the constant terms and discuss implementation in applications. Turning to factor construction, we find a different choice of $J$ will be optimal, namely
	  \[J_t^{\star\star} = K^{\star\star} n_t^{1/3} T^{1/3}, \qquad t=1,2,\cdots,T,\]
where, again, portfolios are chosen separately at each time, $K^{\star}$ depends on the data generating process, and implementation is discussed in Section \ref{sec:mse}. The major difference here is that for point estimation, the optimal number of portfolios, $J_t^{\star\star}$, diverges more slowly than for hypothesis testing, $J_t^{\star}$, in typical applications where the cross-sectional sample size is much larger than the number of time-series observations. The bias-variance trade off, though still present of course, manifests differently because this is a point estimation problem, rather than one of inference. In particular, the divergence rate will often be slower. This formal choice is a further contribution of our paper, and is new to the literature. However, it does seem that, at least informally, the status quo is to use fewer portfolios for factor construction than for testing. See Remark \ref{rem:factors} for further discussion.

\begin{figure}[H]
\caption{\doublespacing\textbf{Momentum Anomaly Example}\\[0.005in] \normalsize{This figure shows the estimated relation between equity returns and 12-2 momentum. The left column shows $\hat{\mu}(z)$ using $J^{\star}_t$; the right column shows the estimated relation using the standard  implementation with $J=10$.  All returns are in monthly changes and all portfolios are value weighted based on lagged market equity. The sample period is 1927--2015.}}
\label{fig:momentum_example}
\begin{subfigure}[t]{0.5\columnwidth}
\centering	\caption{\emph{$\hat{\mu}(z)$ with Optimal $J$}}
\includegraphics[trim={6cm 7cm 6cm 7cm}, scale=.35]{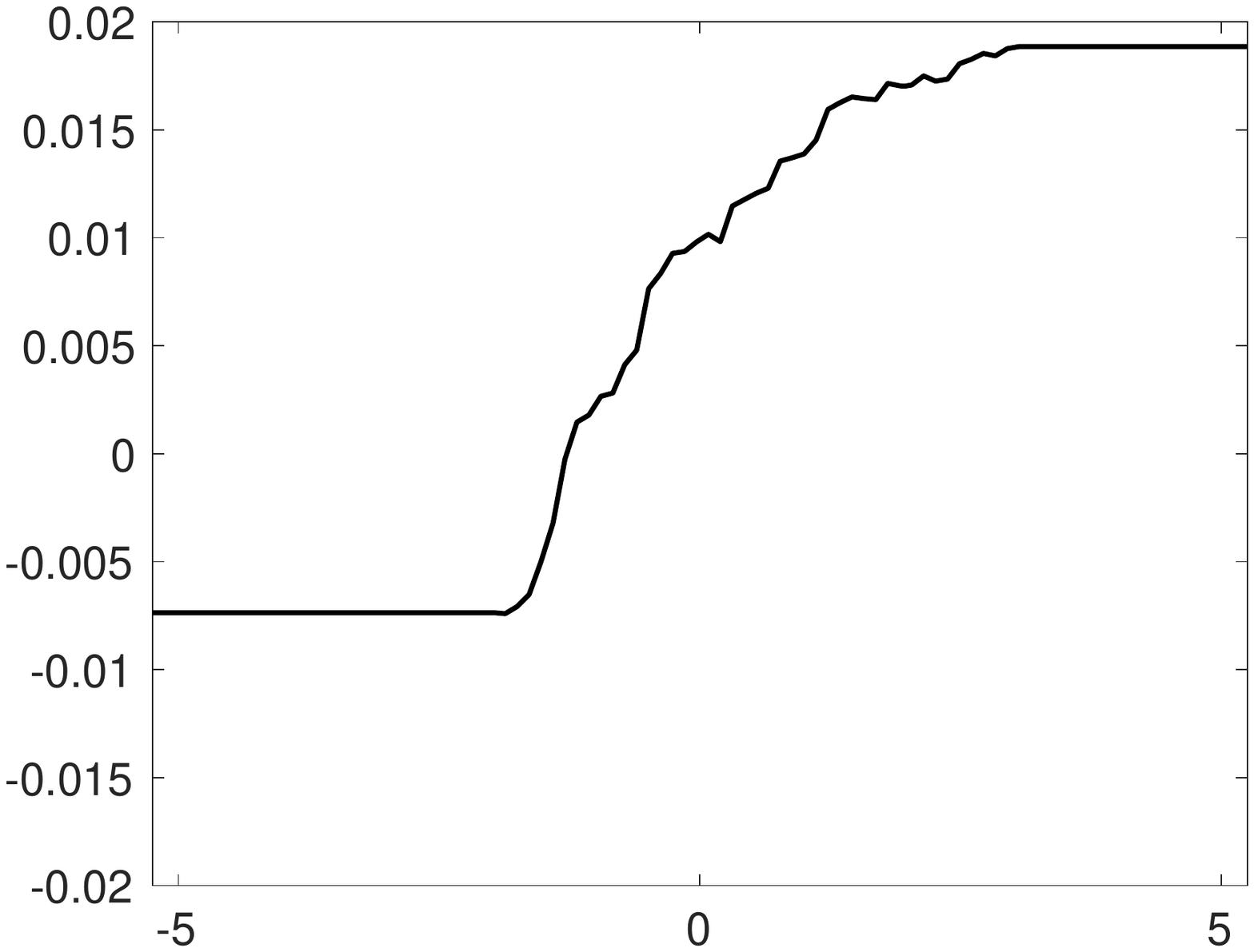}
\end{subfigure}\hfill
\begin{subfigure}[t]{0.5\columnwidth}
\centering	\caption{\emph{Standard Implementation with $J=10$}}
\includegraphics[trim={6cm 7cm 6cm 7cm}, scale=.35]{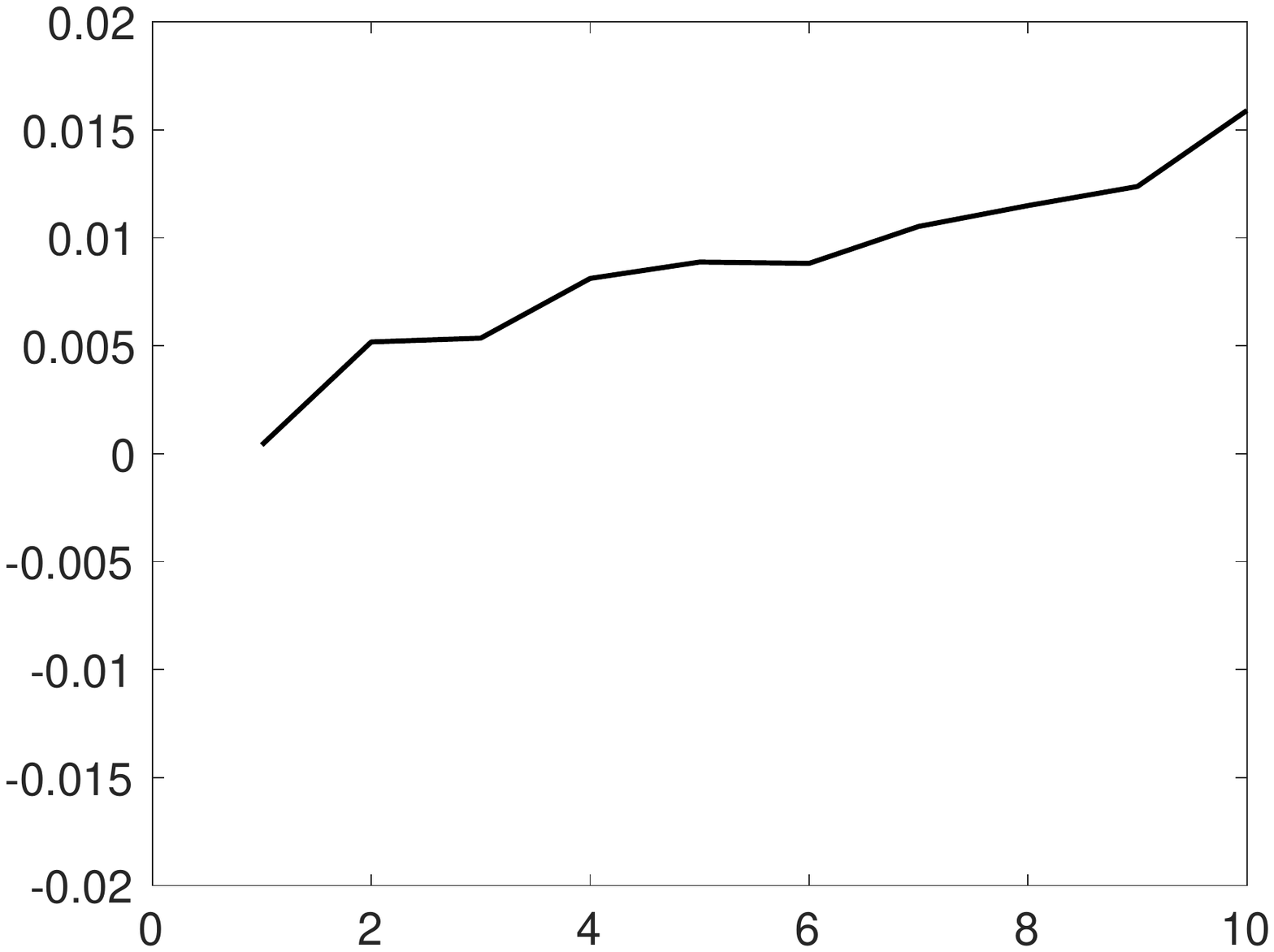}
\end{subfigure}\hfill
\end{figure}

We illustrate the use and importance of our results in our empirical applications (Section \ref{sec:Applications}). As a preview, consider the momentum anomaly. We find that in the universe of US stocks the momentum anomaly results in statistically significant average returns, both overall, and also individually for the long side and short side of the trade (see Table  \ref{tab:empirical}).  Graphically, the relationship between past relative returns and current returns appears monotonically increasing and concave, shown in Figure \ref{fig:momentum_example}. Alongside we show the results using the standard approach based on 10 portfolios.  This makes clear that these same conclusions would not be reached using the conventional estimator.
\medskip

Finally, we note that when $z_H$ and $z_L$ are always in the extreme portfolios, the estimator $\hat{\mu}(z_H) - \hat{\mu}(z_L)$, based on \eqref{eqn:overview estimator}, is exactly the standard portfolio sorting estimator that enjoys widespread use in empirical finance. We exploit the assumed structure that $\mu(z)$ is constant over time as a function of the characteristic value itself, which allows for intuitive and interpretable estimation and inference about $\mu(z)$ at $z \neq \{z_L, z_H\}$. The analogous assumption implicitly required in standard portfolio sorting is that $\mu(\cdot)$ is constant over time as a function of the (random) cross-sectional order statistic of the characteristics, i.e., the ranks. These two overlap in the special case when $z_H$ and $z_L$ are always in the extreme portfolios. We could accommodate this case but with substantial notational complexity. Moreover, the key insights obtained in this paper by formalizing and analyzing the portfolio sorting estimator would not be affected. In these broad terms then, the main contribution of our paper is a formal asymptotic treatment of the standard portfolio-sorts test on $\hat{\mu}(z_H) - \hat{\mu}(z_L)$, but a further contribution is to show how portfolio sorting can be used for a much wider range of inference targets and correspondingly to allow for inference on additional testable hypotheses generated by theory (e.g., shape restrictions).

An alternative interpretation that unifies the two approaches, which researchers may hold implicitly, is as inference on the grand mean at that point, even if $\mu(\cdot)$ is not constant in $z$ itself or in its rank. That is, recalling \eqref{eqn:overview estimator}, we interpret the estimand as (the limit of) $\bar{\mu}(z_*) = \sumt \mu_t(z_*) / T$. When $z_H$ and $z_L$ are always in the extreme portfolios this interpretation may be natural and the quantity $\hat{\mu}(z_H) - \hat{\mu}(z_L)$ directly interpretable. Our method accommodates this interpretation without substantive change.

\begin{remark}[Analogy to Cross-Sectional Regressions]
	\label{rem:csregressions}	
	The assumption that $\mu(z)$ is constant over time as a function of the characteristic value is perfectly aligned with the practice of cross-sectional (or Fama-MacBeth) regressions (\citet*{FamaMacBeth1973}). This approach is motivated by a model of the form: $R_{it} = \zeta z_{it} + \varepsilon_{it}$,  $i = 1,\ldots, n_t$, $t = 1, \ldots, T$, where $z_{it}$ is the value of the characteristic (or a vector of characteristics, more generally). Thus, cross-sectional regressions are then nested in equation \eqref{eqn:ModelNP0} under the assumption that $\mu(\cdot)$ is linear in the characteristics (see also Remark \ref{rem:csregressions2} below).
\end{remark}

\section{General Asset Returns Model and Sorting Estimator}
	\label{sec:framework}

In this section we study a more general model and develop a correspondingly general characteristic-sorted portfolio estimator. We extend beyond the simple case of the previous section in two directions. First, we allow for multiple sorting characteristics, such that $z_{it}$ is replaced by $\bz_{it} \in \mathcal{Z} \subset \mathbb{R}^\dz$. This extension is important because sorting on two variables is quite common in empirical work, and further, we can capture and quantify the empirical reality that sorting is very rarely done on more than two characteristics because this leads to empty portfolios. Intuitively, the nonparametric partitioning estimator, like all others, suffers from the curse of dimensionality, and performance deteriorates as $\dz$ increases, as we can make precise (see also Section \ref{sec:conditional sorts} and Remark \ref{rem:csregressions2}). To address this issue, our second generalization is to allow for other conditioning variables, denoted by $\bx_{it} \in \mathbb{R}^{d_x}$, to enter the model in a flexible parametric fashion. 

Formally, our model for asset returns is
\begin{equation}
	\label{eqn:model}
	R_{it} = \mu(\bz_{it}) + \bx_{it}'\bbeta_t + \varepsilon_{it},   \qquad  i=1,2,\cdots,n_t,       \qquad t=1,2,\cdots,T.
\end{equation}
This model retains the nonparametric structure on $\mu(\bz)$ as in equation \eqref{eqn:ModelNP0}, with the same interpretation (though now conditional on $\bx_{it}$). Notice that the vector $\bx_{it}$ may contain both basic conditioning variables as well as transformations thereof (e.g., interactions and/or power expansions), thus providing a flexible parametric approach to modeling these variables and providing a bridge to cross-sectional regressions from portfolio sorting. Cross-sectional regressions are popular because their linear structure means a larger number of variables can be incorporated compared to the nonparametric nature of portfolio sorting (i.e. cross-sectional regressions do not suffer the curse of dimensionality). Model \eqref{eqn:model} keeps this property while retaining the nonparametric flexibility and spirit of portfolio sorting. Indeed, the parameters $\bbeta_t$ are estimable at the parametric rate, in contrast to the nonparametric rate for $\mu(\bz)$. The additive separability of the conditioning variables, common to both approaches, is the crucial restriction that enables this. Furthermore, due to the linear structure, the sorting estimator can be easily implemented via ordinary least squares, as discussed below.

As in the prior section, the main hypothesis of interest in the empirical finance literature is the presence of a large discrepancy in expected returns between a lower and a higher portfolio. To put \eqref{eqn:hypothesis} into the present, formalized notation, let $\bz_L<\bz_H$ be two values at or near the lower and upper (observed) boundary points. We are then interested in testing ${\sf H}_0: \mu(\bz_H) - \mu(\bz_L)=0$ against the two-sided alternative. Of course, our results also cover other linear transformations such as the ``diff-in-diff'' approach: e.g., for $\dz=2$, the estimand $\mu(\bz_{1H}, \bz_{2H}) - \mu(\bz_{1H}, \bz_{2L}) - ( \mu(\bz_{1L}, \bz_{2H}) - \mu(\bz_{1L}, \bz_{2L}))$. See \citet*{Nagel2005} for an example of the latter, and Remark \ref{rem:partial means} below for further discussion on other potential hypotheses of interest. We will frame much of our discussion around the main hypothesis ${\sf H}_0$ for concreteness, while still providing generic results that may be used for other inference targets.

The framework is completed with the following assumption governing the data-generating process, which also includes regularity conditions for our asymptotic results.
\begin{assumption}[Data Generating Process]
	\label{ass:dgp} 
	Let the sigma fields $\mathcal{F}_t=\sigma(\bF_t)$ be generated from a sequence of unobserved (possibly dependent) random vectors $\{\bF_t:t=0,1,\cdots,T\}$. For $t = 1, 2, \ldots, T$, the following conditions hold.
	\begin{enumerate}[label=(\alph{*}), ref=\ref{ass:dgp}(\alph{*})]
	
		\item  Conditional on $\mathcal{F}_t$, $\{(R_{it},\bz_{it}',\bx_{it}')':i=1,2,\cdots,n_t\}$ are i.i.d. satisfying Model \eqref{eqn:model}.
		
		\item $\E[\varepsilon_{it} | \bz_{it}, \bx_{it}, \mathcal{F}_t] = 0$; uniformly in $t$, $\Omega_{\mathrm{uu},t}=\E[\V(\bx_{it}|\bz_{it},\mathcal{F}_t)]$ is bounded and its minimum eigenvalue is bounded away from zero,  $\sigma_{it}^2=\E[|\varepsilon_{it}|^2|\bz_{it},\bx_{it},\mathcal{F}_t]$ is bounded and bounded away from zero, $\E[|\varepsilon_{it}|^{2+\phi}|\bz_{it},\bx_{it},\mathcal{F}_t]$ is bounded for some $\phi>0$, and $\E[\ba'\bx_{it} | \bz_{it}, \mathcal{F}_t]$ is sub-Gaussian for all $\ba\in\mathbb{R}^{d_x}$.
		
		\item Conditional on $\mathcal{F}_t$, $\bz_{it}$ has time-invariant support, denoted $\mathcal{Z}$, and continuous Lebesgue density bounded away from zero.

		\item $\mu(\bz)$ is twice continuously differentiable; $|\E[x_{it,\ell} | \bz_{it}=\bz,\mathcal{F}_t] - \E[x_{it,\ell} | \bz_{it}=\bz',\mathcal{F}_t]| \leq C \| \bz - \bz' \|$ for all $\bz, \bz' \in \mathcal{Z}$ where $x_{it,\ell}$ is the $\ell$th element of $\bx_{it}$ and the constant $C$ is not a function of $t$ or $\mathcal{F}_t$. 
		
	\end{enumerate}
\end{assumption}

These conditions allow for considerable flexibility in the behavior of the time series of returns and the cross-sectional dependence. Indeed, \citet*[][p. 1552]{Andrews2005}, using the same condition in a single cross-section, called Assumption \ref{ass:dgp}(a) ``surprisingly general''. The set up allows for dependence and conditional heteroskedasticity across assets and time. For example, if $\bF_t$ were to include a business cycle variable then we could allow for a common business-cycle component in the idiosyncratic variance of returns. As another example, the sampling assumptions allow for a factor structure in the $\bz_{it}$ variables. Perhaps most importantly, we do not impose that returns are independent or even uncorrelated over time. Our assumptions accommodate momentum or reversal effects whereby an asset's past \textit{relative} return predicts its future \textit{relative} return, which corresponds to lagged returns entering $\bz_{it}$ (see, for example, \citet*{DeBondtThaler1985}, \citet*{Jegadeesh1990}, \citet*{Lehmann1990}, \citet*{JegadeeshTitman1993, JegadeeshTitman2001}).

Assumption \ref{ass:dgp} requires that the density of $\bz_{it}$ be bounded away from zero, for each $t=1,2,\dots,T$, which is useful to form (asymptotically) not empty portfolios. The assumption that the support of the characteristics is the same across time-series observations is common when studying panel data. The other restrictions are mostly regularity conditions standard in the (cross-sectional) semi-/non-parametric literature, related to boundedness of moments and smoothness conditions of unknown functions. These conditions are not materially stronger than typically imposed, despite the complex nature of the estimation and the use of an estimated set of basis functions in the nonparametric step (due to the estimated quantiles).

In the context of model \eqref{eqn:model}, the portfolio sorting estimator of $\mu(\bz)$ retains the structure given above in \eqref{eqn:overview estimator}, but first the conditioning variables must be projected out. Thus, the cross-sectional estimator $\hat{\mu}_t(\bz)$ can be constructed by simple ordinary least squares: regressing $R_{it}$ on $J_t^{\dz}$ dummies indicating whether $\bz_{it}$ is in portfolio $j$, along with the $d_x$ control variables $\bx_{it}$. Note that, in contrast to Section \ref{sec:overview}, we allow $J = J_t$ to vary over time, in line with having an unbalanced panel. This is particularly important for applications to equities as these data tend to be very unbalanced with cross sections much larger later in the sample as they are at the beginning of the sample.  For example, in our empirical applications the largest cross-sectional sample size is approximately fifteen times the smallest.

The multiple-characteristic portfolios are formed as the Cartesian products of marginal intervals. That is, we first partition each characteristic into $J_t$ intervals, using its \emph{marginal} quantiles, and then form $J_t^{\dz}$ portfolios by taking the Cartesian products of all such intervals. We retain the notation $P_{jt} \subset \mathbb{R}^{\dz}$ for a typical portfolio, where here $j = 1, 2, \ldots, J_t^{\dz}$. For $\dz > 1$, even if $J^{\dz} < n$, these portfolios are not uniformly guaranteed to contain any assets, and this concern for ``empty'' portfolios can be found in the empirical literature \citep*[see, for example,][p. 31]{Goyal2012}. Our construction mimics empirical practice, and we formalize the constraints on $J$ that ensure nonempty portfolios (a variance condition), while simultaneously controlling bias. While the problem of a large $J$ implying empty portfolios has been recognized (though never studied), the idea of controlling bias appears to be poorly understood. However, in our framework the nonparametric bias arises naturally and is amenable to study. Conditional sorts have been used to ``overcome'' the empty portfolio issue, but these are different conceptually, as discussed below.

With the portfolios thus formed, we can define the final portfolio sorting estimator of $\mu(\bz)$, for a point of interest $\bz \in \mathcal{Z}$. First, with an eye to reinforcing the estimated portfolio breakpoints, for a given portfolio $P_{jt}$, $j = 1, 2, \ldots, J_t^{\dz}, t = 1, \ldots, T$, let $\Ihat_{jt}(\bz) = \I\{ \bz \in P_{jt} \}$ indicate that the point $\bz$ is in $P_{jt}$, and let $N_{jt} =  \sum_{i=1}^{n_t}  \Ihat_{jt}(\bz_{it})$ denote its (random) sample size. The portfolio sorting estimator is then defined as
\begin{equation}
	\label{eqn:muhat}
	\hat{\mu}(\bz)  =  \frac{1}{T} \sumt  \hat{\mu}_t(\bz),  		\quad\qquad   		\hat{\mu}_t(\bz) = \sumjd  \frac{1}{N_{jt} } \sum_{i=1}^{n_t} \hat{\mathbf{1}}_{jt} \Ihat_{jt}(\bz) \Ihat_{jt}(\bz_{it}) (R_{it} - \bx_{it}'\hat{\bbeta}_t),
\end{equation}
where
\begin{align}
	\begin{split}
		\label{eqn:matrixes}
		& \hat{\bbeta}_t = (\mathbf{X}_t' \mathbf{M}_{t} \mathbf{X}_t)^{-1} \mathbf{X}_t' \mathbf{M}_{t} \mathbf{R}_t,  		  \quad\qquad  		\mathbf{R}_t = [R_{1t}, \ldots, R_{n_t t}]',   		\\
		&\mathbf{X}_t=[\bx_{1t},\bx_{2t},\cdots,\bx_{n_t t}]',  		\quad\qquad  		   \mathbf{M}_t = \mathbf{I}_{n_t} - \hat{\mathbf{B}}_t (\hat{\mathbf{B}}_t'\hat{\mathbf{B}}_t)^{-1}\hat{\mathbf{B}}_t' ,
	\end{split}
\end{align}
and $\hat{\mathbf{B}}_t=\hat{\mathbf{B}}_t\left(\bz_t\right)$ with $\bz_t=[\bz_{1t},\bz_{2t},\cdots,\bz_{n_t t}]'$ is the $n_t\times J_t^d$ matrix with $(i,j)$ element equal to $\Ihat_{j t}(z_{i t})$, characterizing the portfolios for the characteristics $z_{it}$. The indicator function $ \hat{\mathbf{1}}_{jt}$ ensures that all necessary inverses exist, and thus takes the value one if $P_{jt}$ is nonempty and $(\mathbf{X}_t' \mathbf{M}_{t} \mathbf{X}_t / n_t)$ is invertible. Both events occur with probability approaching one (see the Supplemental Appendix). It is established there that $N_{jt} \asymp n_t/J_t^{\dz} $ with probability approaching one, for all $j$ and $t$.

\begin{remark}[Implementation and Weighted Portfolios]
Despite the notational complexity, the estimator $\hat{\mu}_t(\bz)$ is implemented as a standard linear regression of the outcome $R_{it}$ on the $J_t^{\dz} + d_x$ covariates $\hat{\mathbf{B}}_t$ and $\mathbf{X}_t$. It is the product of the indicator functions $\Ihat_{jt}(\bz) \Ihat_{jt}(\bz_{it})$ that enforces the nonparametric nature of the estimator: only $\bz_{it}$ in the same portfolio as $\bz$, and hence ``close'', are used. The estimator can easily accommodate weighting schemes, such as weighting assets by market capitalization or inversely by their estimated (conditional) heteroskedasticity.  For notational ease we present our theory without portfolio weights, but all empirical results in Section \ref{sec:Applications} are based on the value-weighted portfolio estimator.
\end{remark}

It worth emphasizing that the nonparametric estimator $\hat{\mu}_t(\bz)$ of \eqref{eqn:muhat} is nonstandard. At first glance, it appears to be the nonparametric portion of the usual partially linear model, using the partitioning regression estimator as the first stage ($\hat{\bbeta}_t$ would be the parametric part). However, the partitioning estimator here is formed using estimated quantiles, which makes the ``basis'' functions of our nonparametric estimator nonstandard, and renders prior results from the literature inapplicable.


\begin{remark}[Connection to Other Anomalies Adjustments]
A number of authors have attempted to control for existing anomalies by first regressing their proposed anomaly variable on existing variables, and sorting on the residuals.
This is fundamentally (and analytically) different from what we study in this paper and this approach does not, in general, enjoy the usual interpretation of estimating the effect of $\bz_{it}$ on $R_{it}$ controlling for additional variables.  In contrast, our framework retains the standard interpretation through the additive separability assumption as described by model \eqref{eqn:model}.
\end{remark}

\subsection{Conditional Sorts}
	\label{sec:conditional sorts}
	
A common practice in empirical finance is to perform what are called ``conditional'' portfolio sorts. These are done by first sorting on one characteristic, and then within each portfolio separately, sorting on a second characteristic, and so forth (usually only two characteristics are considered). In each successive sort, quantile-spaced portfolios are used. In this subsection we discuss how our framework relates to conditional sorts, based on two distinct interpretations of conditional sorting: first as conditional testing and second as a mechanical solution to empty portfolios.

To fix ideas, consider firm size and credit rating. Small firms are less likely to have high credit ratings, and so in the ``high'' credit rating portfolio there may be no truly small firms. Directly applying \eqref{eqn:muhat} would thus yield empty portfolios. Conditional sorts ``solve'' the empty portfolios problem by construction: first sorting by rating and then within each rating-based portfolio, by size, but have the feature that the issue that the ``small firm'' portfolio within the highest rating portfolio will typically have larger firms than conditional on lower ratings.

But this may not present a problem if we seek to study whether smaller firms still earn higher average returns if we keep credit rating fixed (Section \ref{sec:Applications} finds evidence for the size anomaly marginally). To answer this question we could test the ``high minus low'' hypothesis within each credit-based portfolio. Our framework directly applies here, that is, the results and discussion in the following subsections, provided one is careful to interpret the results conditionally on the first sort. Further, if $\mu(\cdot)$ is truly monotonic in size, then these conditional results can be extrapolated to ``fill'' the empty bins, but our theory does not justify this.

A second interpretation of conditional sorts is that they are designed solely to solve the problem of empty portfolios. This is distinct from the above, and our framework does not apply here because in this formulation of portfolio sorting, it is implicitly assumed that the function $\mu(\bz)$ is constant over time as a function of the conditional order statistics, within each portfolio (or interest is in a specific grand mean, as above, though here mixing qualitatively different firms). This is difficult to treat theoretically, as the (population) assumption on $\mu(\bz)$ must hold for \emph{each} conditional sort for the (estimated) portfolios already constructed. Moreover, it is not clear that this approach can be extended to other interesting estimands. Finally, it would likely be challenging for an economic theory to generate such a constrained (conditional) return generating process.

However, an alternative, and arguably more transparent approach to empty portfolios would be to assume additive separability of the function $\mu(\cdot)$ so that, if we denote the $\dz$ components of $\bz_{it}$ by $z_{it,1}, \ldots, z_{it,\dz}$, we suppose
\begin{equation}
	\label{eqn:addsep}
	R_{it} = \mu_1(z_{it,1}) + \cdots + \mu_d(z_{it,\dz}) + \varepsilon_{it} \qquad i = 1,\ldots, n_t, \quad t = 1, \ldots, T.
\end{equation}
and so each characteristic affects returns via their own unknown function, $\mu_\ell(\cdot)$, for $\ell = 1,\ldots,d$.  The resulting estimator is always defined for any value $\bz$ in the support and so too avoids the problem of empty portfolios (see also Remark \ref{rem:csregressions2}).

\section{First-order Asymptotic Theory}
	\label{sec:first order}

With the estimator fully described we now present consistency and asymptotic normality results, and two valid standard error estimators. To our knowledge, these results are all new to the literature. As discussed in Section \ref{sec:overview} the empirical literature contains numerous studies that implement exactly the tests validated by the results below, but such validation has heretofore been absent.

Beyond the definition of the model \eqref{eqn:model} and the conditions placed upon it by Assumption \ref{ass:dgp}, we will require certain rate restrictions for our asymptotic results. We now make these precise, grouped into the following two assumptions.

\begin{assumption}[Panel Structure]
	\label{ass:panel}
	The cross-sectional sample sizes diverge proportionally: for a sequence $n\to\infty$, $n_t = \kappa_t n$, with $\kappa_t \leq 1$ and uniformly bounded away from zero.
\end{assumption}

Assumption \ref{ass:panel} requires that the cross-sectional sample sizes grow proportionally. This ensures that each $\hat{\mu}_t(\cdot)$ contributes to the final estimate, and at the same rate. We will also restrict attention to $J_t = J_t(n_t, n, T)$, which implies there is a sequence $J \to \infty$ such at $J_t \propto J$ for all $t$. Neither of these are likely to be limiting in practice: our optimal choices depend on $n_t$ by design, and  there is little conceptual point in letting $J_t$ vary over time beyond accounting for panel imbalance. The notation $n$ and $J$ for common growth rates enables us to present compact and simplified regularity conditions, such as the following assumption, which formalizes the bias-variance requirements on the nonparametric estimator. All limits are taken as $n,T\to\infty$, unless otherwise noted. 

\begin{assumption}[Rate Restrictions]
	\label{ass:rates} 
	The sequences $n$, $T$, and $J$ obey: {\it (a)} $n^{-1}J^\dz\log(\max(J^\dz,T))\log(n)\rightarrow 0$, {\it (b)} $\sqrt{nT}J^{-(\dz/2+1)} \rightarrow 0$, and, if $d_x \geq 1$, {\it (c)} $T/n \rightarrow 0$.
\end{assumption}

Assumption \ref{ass:rates}(a) ensures that all $J_t$ grow slowly enough that the variance of the nonparametric estimator is well-controlled and all portfolios are nonempty, while \ref{ass:rates}(b) ensures the nonparametric smoothing bias is negligible. Finally, Assumption \ref{ass:rates}(c) restricts the rate at which $T$ can grow. This additional assumption is necessary for standard inference when linear conditioning variables are included in the model and $\dz=1$. When $\dz>1$ then it is implied by Assumptions \ref{ass:rates}(a) and \ref{ass:rates}(b).

In general, the performance of the portfolio sorting estimator may be severely compromised if the number of time series observations is large relative to the cross section and/or $\dz$ is large. To illustrate, suppose for the moment that $J \asymp n^A$ and $T \asymp n^B$. Assumptions \ref{ass:rates}(a) and (b) require that $A \in \left( (1 + B)/(2 + \dz) \; , \; 1/\dz \right)$, which amounts to requiring $B \dz < 2$. If the time series dimension is large, then the number of allowable sorting characteristics is limited. For example, if $B$ is near one, at most two sorting characteristics are allowed, and even then just barely, and may lead to a very poor distributional approximation. Thus, some caution should be taken when applying the estimator to applications with relatively few underlying assets.

Before stating the asymptotic normality result, it is useful to first give an explicit (conditional) variance formula:
\begin{equation}
	\label{eqn:variance}
	V(\bz) =  \frac{1}{T} \sumt \sumjd \frac{1}{N_{jt}} \sum_{i=1}^{n_t}   \hat{\mathbf{1}}_{jt}  \Ihat_{jt}(\bz) \Ihat_{jt}(\bz_{it})\sigma^2_{it}.
\end{equation}
This formula, and the distributional result below, are stated for a single point $\bz$. It is rare that a single $\mu(\bz)$ would be of interest, but these results will serve as building blocks for more general parameters of interest, such as the leading case of testing \eqref{eqn:hypothesis} treated explicitly below. An important consideration in any such analysis is the covariance between point estimators. The special structure of the portfolio sorting estimator (or partition regression estimator) is useful here: as long as $\bz$ and $\bz'$ are in different portfolios (which is the only interesting case), $\hat{\mu}(\bz)$ and $\hat{\mu}(\bz')$ are uncorrelated because $  \Ihat_{jt}(\bz)  \Ihat_{jt}(\bz') \equiv 0$. The partitioning estimator is, in this sense, a local nonparametric estimator as opposed to a global smoother. 

We can now state our first main result.
\begin{theorem}[Asymptotic Distribution]\label{thm:clt}
Suppose Assumptions \ref{ass:dgp}, \ref{ass:panel}, and \ref{ass:rates} hold. Then,
\[V(\bz)^{-1/2}(\hat{\mu}(\bz)-\mu(\bz))
  = \sumt  \sum_{i=1}^{n_t}\hat{w}_{it}(\bz)\varepsilon_{it} + o_\P(1) \dto \mathcal{N}(0,1), 
\]
where
\[ V(\bz)\asymp \frac{J^\dz}{nT}    \quad\qquad  \text{ and } \qquad\quad
   \hat{w}_{it}(\bz) = V^{-1/2}(\bz)\sumjd  \frac{1}{TN_{jt}} \hat{\mathbf{1}}_{jt} \Ihat_{jt}(\bz) \Ihat_{jt}(\bz_{it}).\]
\end{theorem}

Theorem \ref{thm:clt} shows that the properly normalized and centered estimator $\hat{\mu}(z)$ has a limiting normal distribution. The cost of the flexibility of the nonparametric specification between returns and (some) characteristics comes at the expense at slower convergence --- the factor $J^{-\dz/2}$.  Theorem \ref{thm:clt} also makes clear why Assumption \ref{ass:rates}(b) is necessary:  the bias of the estimator is of the order $J^{-1}$ and thus, once the rate $J^{-\dz/2}\sqrt{nT}$ is applied, Assumption \ref{ass:rates}(b) must hold to ensure that the bias can be ignored for the limiting normal distribution. This undersmoothing approach is typical for bias removal. The statement of the theorem includes a weighted average asymptotic representation for the estimator, which is useful for treatment of estimands beyond point-by-point $\mu(\bz)$, including linear functionals such as partial means, as discussed in Remark \ref{rem:partial means}.


The final missing piece of the pointwise first-order asymptotic theory is a valid standard error estimator. To this end, we consider two options. The first, due in this context to \citet*{FamaMacBeth1973}, makes use of the fact that $\hat{\mu}(\bz)$ is an average over $T$ ``observations'', while the second is a plug-in estimator based on an asymptotic approximation to the large sample variability of the portfolio estimator. Define
\begin{equation}
		\label{eqn:standard errors}
		\hat{V}_\mathtt{FM}(\bz)  = \frac{1}{T^2} \sumt (\hat{\mu}_t(\bz)-\hat{\mu}(\bz))^2
		\quad\text{and}\quad
		\hat{V}_\mathtt{PI}(\bz)  = \frac{1}{T^2} \sumt \sumjd \sum_{i=1}^{n_t} \hat{\mathbf{1}}_{jt}\frac{1}{N_{jt}^2} \Ihat_{jt}(\bz) \Ihat_{jt}(\bz_{it})\hat{\varepsilon}_{it}^2
\end{equation}
with $\hat{\varepsilon}_{it}=R_{it} - \hat{\mu}(\bz) - \bx_{it}'\hat{\bbeta}_t$. The following result establishes the validity of both options. 

\begin{theorem}[Standard Errors]\label{thm:se}
Suppose the assumptions of Theorem \ref{thm:clt} hold with $\phi=2 + \varrho$ for some $\varrho>0$. Then,
\[\frac{nT}{J^\dz}(\hat{V}_\mathtt{FM}(\bz) - V(\bz))\pto 0, \qquad\text{and}\quad
  \frac{nT}{J^\dz}(\hat{V}_\mathtt{PI}(\bz) - V(\bz))\pto 0.\]
\end{theorem}

The \citet*{FamaMacBeth1973} variance estimator is commonly used in empirical work, but this is the first proof of its validity. In contrast, $\hat{V}_\mathtt{PI}$ is the ``plug-in'' variance estimator based on the results in Theorem \ref{thm:clt}.  Theorem \ref{thm:se} shows that these variance estimators are asymptotically equivalent.  In a fixed sample, it is unclear which of the two estimators is preferred. $\hat{V}_\mathtt{FM}$ is simple to implement and very popular, while $\hat{V}_\mathtt{PI}$ is based on estimated residuals and may need a large cross-section. On the other hand, while we assume $T$ diverges, in line with common applications of sorting, it may be established that $\hat{V}_\mathtt{PI}$ is valid for fixed $T$, whereas $\hat{V}_\mathtt{FM}$ is only valid for large-$T$ panels. However, a related result is due to \citet*{IM2010}, who provided conditions under which the \citet*{FamaMacBeth1973} approach applied to cross-sectional regressions produces inference on a scalar parameter that is valid or conservative, depending on the assumptions imposed. Specifically, \citet*{IM2010}, in the context of cross-sectional regressions, show that for fixed $T$ and a specific range of size-$\alpha$ tests, the \citet*{FamaMacBeth1973} approach is valid, but potentially conservative. Our empirical results in Section \ref{sec:Applications} use $\hat{V}_\mathtt{FM}$ to form test statistics so as to be comparable to existing results in the literature.  In general, a consistent message of our results is that caution is warranted in cases applying portfolio sorting to applications with a very modest number of time periods or, as discussed above, when the number of time periods is ``large'' relative to the cross-sectional sample sizes.

Theorems \ref{thm:clt} and \ref{thm:se} lead directly to the following result, which treats the main case of interest under simple and easy-to-interpret conditions. 
\begin{corollary}
	\label{cor:top bottom}
	Let the conditions of Theorem \ref{thm:se} hold. Then,
	\[ \frac{ \bigl[ \hat{\mu}(\bz_H) - \hat{\mu}(\bz_L) \bigr]  -  \bigl[\mu(\bz_H) - \mu(\bz_L) \bigr]}{\sqrt{\hat{V}(\bz_H) + \hat{V}(\bz_L)} }  \dto \mathcal{N}(0,1), \]
	where $\hat{V}(\bz)$ may be $\hat{V}_\mathtt{FM}$ or $\hat{V}_\mathtt{PI}$ as defined in equation \eqref{eqn:standard errors}.
\end{corollary}
Section \ref{sec:overview} states this same result, simplified to the model \eqref{eqn:ModelNP0}. This result shows that testing ${\sf H}_0: \mu(\bz_H) - \mu(\bz_L) = 0$ against the two-sided alternative can proceed as standard: by rejecting ${\sf H_0}$ if $\left\vert \hat{\mu}(\bz_H) - \hat{\mu}(\bz_L) \right\vert$ greater than $1.96\times\sqrt{\hat{V}(\bz_H) + \hat{V}(\bz_L)}$. In this way, our work shows under precisely what conditions the standard portfolio sorting approach is valid, and perhaps more importantly, under what conditions it may fail.

\begin{remark}[Other Estimands]
	\label{rem:partial means}
	As we have discussed above, our general framework allows for other estimands aside from the ``high minus low'' return. For example, a popular estimand in the literature that may be easily treated by our results is the case of partial means, which arises when $\dz > 1$. If we denote the $\dz$ components of $\bz$ by $z^{(1)}, z^{(2)}, \ldots, z^{(\dz)}$, then for some subset of these of size $\delta < \dz$, the object of interest is $\int_{ \times_{\ell = 1}^\delta} \mu(\bz) w \bigl( z^{(1)}, z^{(2)}, \ldots, z^{(\delta)} \bigr) dz^{(1)} dz^{(2)} \cdots dz^{(\delta)}$, where the components of $\bz$ that are not integrated over are held fixed at some value, or linear combinations for different initial $\bz$ points. Prominent examples are the SMB and HML factors of the Fama/French 3 Factors. 
The weighting function $w(\cdots)$ is often taken to be the uniform density (based on value-weighted portfolios), but this need not be the case. For example, if $\dz = 2$, one component may be integrated over before testing the analogous hypothesis to \eqref{eqn:hypothesis}:
\[ {\sf H}_0 :  \int_{ z^{(1)} } \mu\bigl( z^{(1)}, z^{(2)}_H \bigr) w \bigl( z^{(1)} \bigr) dz^{(1)}  -   \int_{ z^{(1)} } \mu\bigl( z^{(1)}, z^{(2)}_L \bigr) w \bigl( z^{(1)} \bigr) dz^{(1)}  =  0.\]
In the case of factor construction this corresponds to a test of whether a factor is priced unconditionally. Theorems \ref{thm:clt} and \ref{thm:se} can be applied to provide valid inference.
\end{remark}

\begin{remark}[Strong Approximations]\label{rem:coupling}
Our asymptotic results apply to hypothesis tests that can be written as pointwise transformations of $\mu(\bz)$, with the leading case being \eqref{eqn:hypothesis}: ${\sf H}_0: \mu(\bz_H) - \mu(\bz_L)=0$. However, there are other hypotheses of interest in this context of portfolio sorting that require moving beyond pointwise results. Chief among these is directly testing monotonicity of $\mu(\cdot)$, rather than using $\mu(\bz_H) - \mu(\bz_L)$ as a proxy (see discussion in Section \ref{sec:overview}). Building on \citet*{Cattaneo-Farrell-Feng_2018_wp} and \citet*{Cattaneo-Crump-Farrell-Feng_2019_wp}, it may be possible to establish a valid strong approximation to the suitable centered and scaled stochastic process $\{\hat{\mu}(\bz):\bz\in\mathcal{Z}\}$. Such a result would require non-trivial additional technical work, but would allow us to test monotonicity, concavity, and many other hypotheses of interest, such as testing for a ``U-shaped'' relationship \citep{HongLimStein2000}, or for the existence of any profitable trading strategy via ${\sf H}_0: \left| \max_z  {\mu}(\bz) - \min_z  {\mu}(\bz) \right|=0$.
\end{remark}

\begin{remark}[Analogy to Cross-Sectional Regressions]
	\label{rem:csregressions2}
	As we have discussed in Remark \ref{rem:csregressions}, cross-sectional regressions are the ``parametric alternative'' to portfolio sorting. In practice, however, the more natural parametric alternative to portfolio sorts with more than one sorting variable---interaction effects in the linear specification---are rarely utilized. Thus the more exact ``nonparametric counterpart'' to the common implementation of cross-sectional regressions is the additively separable model introduced in equation \eqref{eqn:addsep} of Section \ref{sec:conditional sorts}.  The assumption of additive separability would have the effect of ameliorating the ``curse of dimensionality''; in fact, it can be shown that in this model the rate restrictions $J \log(\max(J,T))/n \to 0$ and $n T / J^3 \to 0$ (i.e., Assumption \ref{ass:rates}  when $d=1$) are sufficient to ensure consistency and asymptotic normality of the estimators, $\hat{\mu}_{\ell}(z)$, based on the additively separable model with $d\geq1$ characteristics.
\end{remark}

\section{Mean Square Expansions and Practical Guidance}
	\label{sec:mse}

With the first-order theoretical properties of the portfolio sorting estimator established, we now turn to issues of implementation. Chief among these is choice of the number of portfolios: with the estimator defined as in equation \eqref{eqn:muhat}, all that remains for the practitioner is to choose $J_t$. The results in the previous two sections have emphasized the key role played by choice of $J_t$ in obtaining valid inference. In contrast, the choice of $J_t$ in empirical studies has been \emph{ad hoc}, and almost always set to either 5 or 10 portfolios. Here we will provide simple, data-driven rules to guide the choice of the number of parameters. To aid in this, we will consider a mean square error expansion for the portfolio estimator, with a particular eye toward testing the central hypothesis of interest: ${\sf H}_0: \mu(\bz_H) - \mu(\bz_L) = 0$, as the starting point for constructing a plug-in optimal choice.

Our main result for this section is the following characterization of the mean square error of the portfolio sorting estimator. To simplify the calculations this section assumes that the quantiles are known (as opposed to being estimated in each cross-section). This simplification only affects the constants of the higher-order terms in the MSE-expansion but not the corresponding rates (see \citet*{Calonico-Cattaneo-Titiunik_2015_JASA} for a related example and more discussion). Recall that $n$ and $J$ represent the common growth rates of the $\{n_t\}$ and $\{J_t\}$, respectively.
\begin{theorem}
	\label{thm:mse}
	Suppose Assumptions \ref{ass:dgp}, \ref{ass:panel}, and \ref{ass:rates} hold, and that the marginal quantiles of $\bz$ are known. Then,
	\begin{align*}
		&\E\left[ \left. \left( \bigl[ \hat{\mu}(\bz_H) - \hat{\mu}(\bz_L) \bigr]  -  \bigl[\mu(\bz_H) - \mu(\bz_L) \bigr] \right)^2 \right| \mathfrak{Z}, \mathfrak{X}, \mathcal{F}_1, \ldots, \mathcal{F}_T  \right]\\
		& \qquad = \mathcal{V}^{(1)} \frac{J^{\dz}}{n T}  +  \mathcal{V}^{(2)}\frac{J^{2\dz}}{n^2 T}   +   \mathcal{B}^2 \frac{1}{J^2}    
		    +  \mathcal{C}\frac{J^{3\dz/2}}{n^{3/2} T^{3/2}} 		    
		    +  O_\P\left( \frac{1}{nT}\right)  +   o_\P\left( J^{-2} +  \frac{J^{2\dz}}{n^2 T} \right),
	\end{align*}
where $\mathfrak{Z} = (z_{11}, \ldots, z_{n_T T})$, $\mathfrak{X} = (x_{11}, \ldots, x_{n_T T})$ and
$\mathcal{B} = \sumt \mathcal{B}_t(\bz_H) -  \sumt \mathcal{B}_t(\bz_L)$ and $\mathcal{V}^{(\ell)} = \sumt \mathcal{V}_t^{(\ell)}(\bz_L) + \sumt \mathcal{V}_t^{(\ell)}(\bz_H)$, $\ell \in \{1,2\}$, and $\mathcal{B}_t(\bz)$, $\mathcal{V}_t^{(1)}(\bz)$, $\mathcal{V}_t^{(2)}(\bz)$ and $\mathcal{C}$ are defined in the Supplementary Appendix. The term $\mathcal{C}$ is (conditionally) mean zero, and the term of order $1/(nT)$ captures the limiting variability of $\sqrt{n/T} \sumt (\hat{\bbeta}_t - \bbeta_t)$, and does not depend on $J$.
\end{theorem}

Under the conditions in Theorem \ref{thm:mse}, and imposing appropriate regularity conditions on the time series structure (e.g., mixing conditions), it can be shown that
$\bar{\mathcal{B}} = \plim_{n,T\to\infty} \mathcal{B}$, $\bar{\mathcal{V}}^{(1)} = \plim_{n,T\to\infty} \mathcal{V}^{(1)}$, $\bar{\mathcal{V}}^{(2)} = \plim_{n,T\to\infty} \mathcal{V}^{(2)}$, 
where $\bar{\mathcal{B}}$, $\bar{\mathcal{V}}^{(1)}$ and $\bar{\mathcal{V}}^{(2)}$ are non-random and non-zero quantities. In this paper, however, we remain agnostic about the specific regularity conditions for convergence in probability to occur because our methods do not rely on them.

To obtain an optimal choice for the number of portfolios, note that the first variance term of the expansion will match the first-order asymptotic variance of Theorem \ref{thm:clt}, which suggests choosing $J$ to jointly minimize the next two terms of the expansion: the bias and higher order variance (see \citet*{CCJ2010_JASA} for another application of this logic). This approach is optimal in an inference-targeted sense because it minimizes the two leading terms not accounted for by the approximation in Theorem \ref{thm:clt}. For testing ${\sf H}_0: \mu(\bz_H) - \mu(\bz_L) = 0$ we find the optimal number of portfolios to be
\begin{equation}
	\label{eqn:optimal J}
	J_t^\star = \Biggl\lfloor  \left(  \frac{ \bar{\mathcal{B}}^2}{ d \bar{\mathcal{V}}^{(2)}}  \left( n_t^2 T \right)  \right)^{\tfrac{1}{2\dz+2}} \Biggr\rfloor,
\end{equation}
where $\left\lfloor \cdot \right\rfloor$ is the integer part of the expression. A simple choice for enforcing the same number of portfolios in all periods is to simply replace $n_t$ with $n$ in this expression. It is straightforward to verify that this choice of $J_t^\star$ satisfies Assumption \ref{ass:rates}: the condition required remains that $B \dz < 2$, for $T \asymp n^B$, which limits the number of sorting characteristics and/or the length of time series allowed (see discussion of Assumption \ref{ass:rates}). To gain intuition for $J^{\star}_t$, consider the simple case of a univariate, homoskedastic linear model: $\mu(\bz) = b z$, $\sigma^2_{it} = \sigma^2$. Then $\mathcal{B}^2 \propto |b|^2$ and $\mathcal{V} \propto \sigma^2$, and so a steeper line (larger $|b|$) calls for more portfolios whereas more idiosyncratic noise (larger $\sigma^2$) calls for fewer.

To make this choice practicable we can select $J$ to minimize a sample version of the MSE expansion underlying equation \eqref{eqn:optimal J},
\begin{equation}
	\label{eqn:HOMSE}
\widehat{\mathbb{MSE}}\bigl( \hat{\mu}(\bz_H) - \hat{\mu}(\bz_L); J \bigr) = 
\hat{\mathcal{V}}^{(2)}\frac{J^{2\dz}}{n^2 T}   +   \hat{\mathcal{B}}^2 \frac{1}{J^2}
\end{equation}
where the estimators, $\hat{\mathcal{V}}^{(2)}$ and $\hat{\mathcal{B}}$, will themselves be a function of $J$.  Thus, it is straightforward to search over a grid of values of $J$ and choose based on the minimum value of the expression in equation \eqref{eqn:HOMSE} (see the Supplementary Appendix for further details). Alternatively, if we had pilot estimates of ${\mathcal{V}}^{(2)}$ and ${\mathcal{B}}$, then we could directly utilize the formula in equation \eqref{eqn:optimal J} to obtain a choice for each $J_t$.

\begin{remark}[Undersmoothing]
	\label{rem:undersmoothing}
	A common practice throughout semi- and non-parametric analyses is to select a tuning parameter by undersmoothing a mean square error optimal choice. In theory, this is feasible, but it is necessarily \emph{ad hoc}; see \citet*{Calonico-Cattaneo-Farrell_2018_JASA,Calonico-Cattaneo-Farrell_2019_CER} for more discussion. In contrast, the choice of $J_t^\star$ of equation \eqref{eqn:optimal J} has the advantage of being optimal in an objective sense and appropriate for conducting inference. A possible alternative to $J_t^\star$ would be to choose $J$ by balancing $\left| \bar{\mathcal{B}} \right|$ against $\bar{\mathcal{V}}^{(1)}$; however, this would lead to a choice of $J_t \propto \left( n_t T \right)^{\frac{1}{\dz+1}}$ which would tend to result in a larger number of portfolios chosen as compared to $J_t^\star$.
\end{remark}

\begin{remark}[Parametric Component]
	An additional advantage of $J_t^\star$ is that for $\dz\leq2$ (the most common case in empirical applications) inference on the parametric component is also valid for this choice of $J$. It can be shown that for any real, nonzero vector $\ba \in \mathbb{R}^{d_x}$,
	\begin{equation}
		\frac{\frac{1}{T} \sumt \ba'(\hat{\bbeta}_t - \bbeta_t)}{\sqrt{\frac{1}{T^2}\sumt (\ba'(\hat{\bbeta}_t - \bbeta_t))^2}}
		\dto \mathcal{N}(0,1)
	\end{equation}
	An advantage of the \citet*{FamaMacBeth1973} variance estimator over a ``plug-in'' alternative in this context is that inference on $\frac{1}{T} \sumt  \bbeta_t$ may be conducted without having to estimate the conditional expectation of $\bx$ given $\bz$ nonparametrically.
\end{remark}

\begin{remark}[Constructing factors]
	\label{rem:factors}
	Theorem \ref{thm:mse} can be also be used when the goal is point estimation rather than inference. Using the leading variance term and the bias, we obtain 
		\[ J_t^{\star\star} = \Biggl\lfloor  \left(  \frac{ 2 \bar{\mathcal{B}}^2}{ d \bar{\mathcal{V}}^{(1)} }  \left( n_t T \right)  \right)^{\tfrac{1}{\dz+2}} \Biggr\rfloor,  \]
	which is different in the constants but more importantly also the rate of divergence: for example, when $\dz=1$ then $J_t^{\star\star} \propto  n_t^{1/3} T^{1/3}$ whereas $J_t^{\star} \propto  n_t^{1/2} T^{1/4}$. In applications such as equities where the cross sectional sample size is much larger than the number of time periods then it will be the case that $J_t^{\star\star} = o(J_t^{\star})$, i.e., that the optimal number of portfolios is smaller when constructing factors than when conducting inference on whether expected returns vary significantly with characteristics. Informally, this has been recognized in the empirical literature as the number of portfolios used to construct factors has been relatively small (e.g., \citet*{FamaFrench1993}). As discussed in the supplement, a feasible version of $J_t^{\star\star}$ can be constructed following the steps as in \eqref{eqn:HOMSE}, replacing $\hat{\mathcal{V}}^{(2)} J^{2\dz} / (n^2 T)$ with $\hat{\mathcal{V}}^{(1)} J^{\dz} / (n T)$.
\end{remark}

\section{Empirical Applications}
\label{sec:Applications}

In this section we revisit some notable equity anomaly variables that have been considered in the literature and demonstrate the empirical relevance of the theoretical discussion of the previous sections. We focus on the size anomaly \citep*[e.g.,][]{Banz1981,Reinganum1981} and the momentum anomaly \citep*[e.g.,][]{JegadeeshTitman1993}.

\subsection{Data and Variable Construction}

We use monthly data from the Center for Research in Security Prices (CRSP) over the sample period January 1926 to December 2015.  We restrict these data to those firms listed on the New York Stock Exchange (NYSE), American Stock Exchange (AMEX), or Nasdaq and use only returns on common shares (i.e., CRSP share code 10 or 11).  To deal with delisting returns we follow the procedure described in \citet*{BEM2016}. When forming market equity we use quotes when closing prices are not available and set to missing all observations with $0$ shares outstanding. When forming the momentum variable we follow the popular convention of defining momentum by the cumulative return from 12 months ago (i.e, $t-12$) until one month prior to the current month (i.e., $t-2$). The one-month gap is to avoid confounding the momentum anomaly variable with the short-term reversal anomaly \citep*{Jegadeesh1990,Lehmann1990}. We set to missing this variable if any monthly returns are missing over the period.  We also construct an industry momentum variable. To do so we use the definitions of the $38$ industry portfolios used in Ken French's data library which are based on four digit SIC codes. To construct the industry momentum variable we form a value weighted average of each individual firm's momentum variable within the industry.  We use 13-month lagged market capitalization to form weights so they are unaffected by any subsequent changes in price.

We implement the estimator introduced in Section \ref{sec:framework} as follows. Since the underlying data are monthly, then portfolios are always formed and then rebalanced at the end of each month. All portfolios, including those based on the standard implementation approach, are value weighted using lagged market equity. We implement the estimators based on the number of portfolios which minimizes our higher-order MSE criterion, described in equation \eqref{eqn:HOMSE} since our objective in this section is inference. 

Finally, it is important to fully characterize the nature of these data.  In particular, the equity return data represent a highly unbalanced panel over our sample period.  At the beginning of the sample the CRSP universe includes approximately $500$ firms, increases to nearly 8,000 firms in the late 1990s, and is currently at approximately 4,000 firms.  Moreover, there are sharp jumps in cross-sectional sample sizes that occur in 1962 and 1972 which reflect the addition of firms listed on the AMEX and Nasdaq to the sample (see Figure \ref{fig:nt} 
in the Supplemental Appendix). Even for the subset of firms listed on the NYSE, the panel is still highly unbalanced. At the beginning of the sample, there are about $500$ firms before rising to a high of approximately 2,000 firms, and is currently slightly below 1,500 firms.  



\subsection{Size Anomaly}

We first consider the size anomaly---where smaller firms earn higher returns than larger firms on average. To investigate the size anomaly we use market capitalization as our measure of size of the firm. Thus, following the notation of Section \ref{sec:framework} we have,
\begin{equation}
R_{it} = \mu (  \textsc{me}_{i(t-1)}  ) + \varepsilon_{it}, \qquad i = 1,\ldots,n_t, \quad t = 1, \ldots, T.
\label{eqn:size}
\end{equation}
Here, $\textsc{me}_{it}$, represents the market equity of firm $i$ at time $t$ transformed in the following way: $(i)$ the natural logarithm of market equity of firm $i$ at time $t$ is taken; $(ii)$ at each cross section $t=1,\ldots,T$, the natural logarithm of market equity is demeaned and normalized by the inverse of the cross-sectional standard deviation (i.e., a zscore is applied). This latter transformation is necessary in light of Assumption \ref{ass:dgp}(c) and ensures that the measure of the size of a firm is comparable over time.

\begin{figure}[t]
\caption{\doublespacing\textbf{Size Anomaly: All Stocks}\\[0.05in] \normalsize{This figure shows the estimated relation between the cross section of equity returns and lagged market equity (equation \eqref{eqn:size}).  The left column displays $\hat{\mu}(\cdot)$ where $J_t$ has been chosen based on equation \eqref{eqn:HOMSE}, $z_H = \Phi^{-1}(.975)$, $z_L = \Phi^{-1}(.025)$. The right column displays the estimated relation using the standard portfolio sorting implementation with $J=10$.  All returns are in monthly changes and all portfolios are value weighted based on lagged market equity.}}
\label{fig:sizeAll}
\begin{subfigure}[t]{0.5\columnwidth}
\centering	\caption{\textit{1926--2015}}
\includegraphics[trim={6cm 7cm 6cm 7cm}, scale=.35]{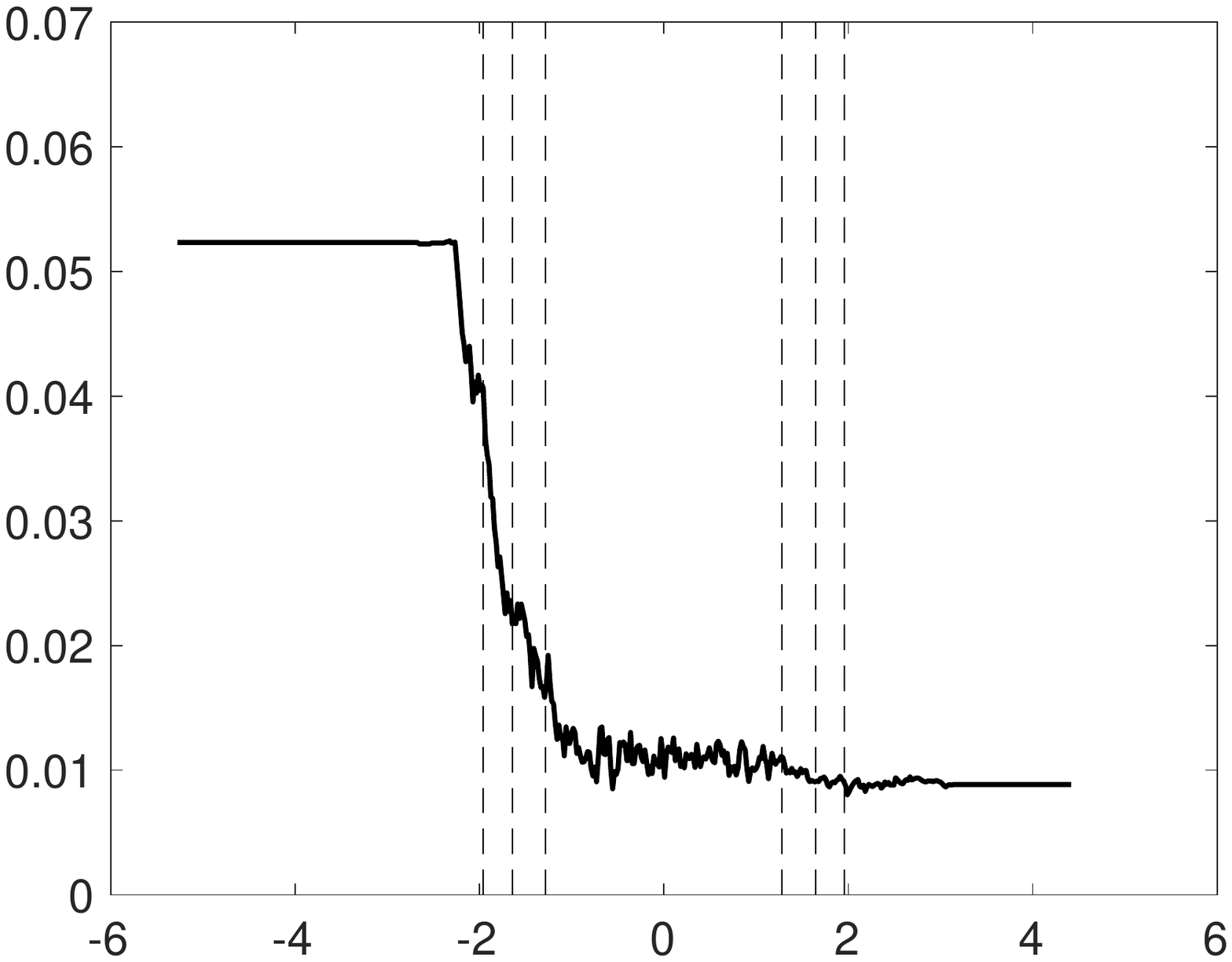}
\end{subfigure}\hfill
\begin{subfigure}[t]{0.5\columnwidth}
\centering	\caption{}
\includegraphics[trim={6cm 7cm 6cm 7cm}, scale=.35]{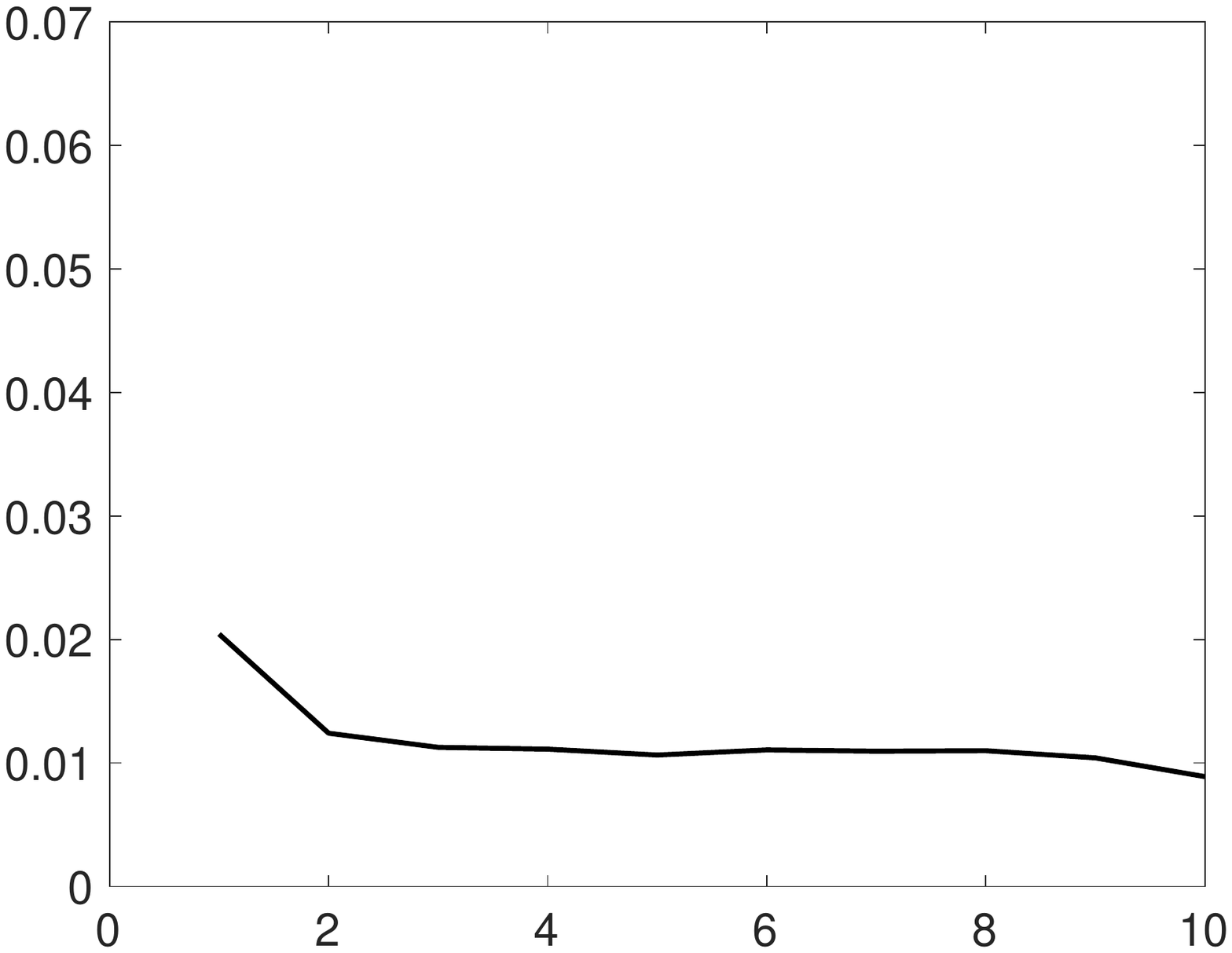}
\end{subfigure}\\[0.25in]
\begin{subfigure}[t]{0.5\columnwidth}
\centering	\caption{\textit{1967--2015}}
\includegraphics[trim={6cm 7cm 6cm 7cm}, scale=.35]{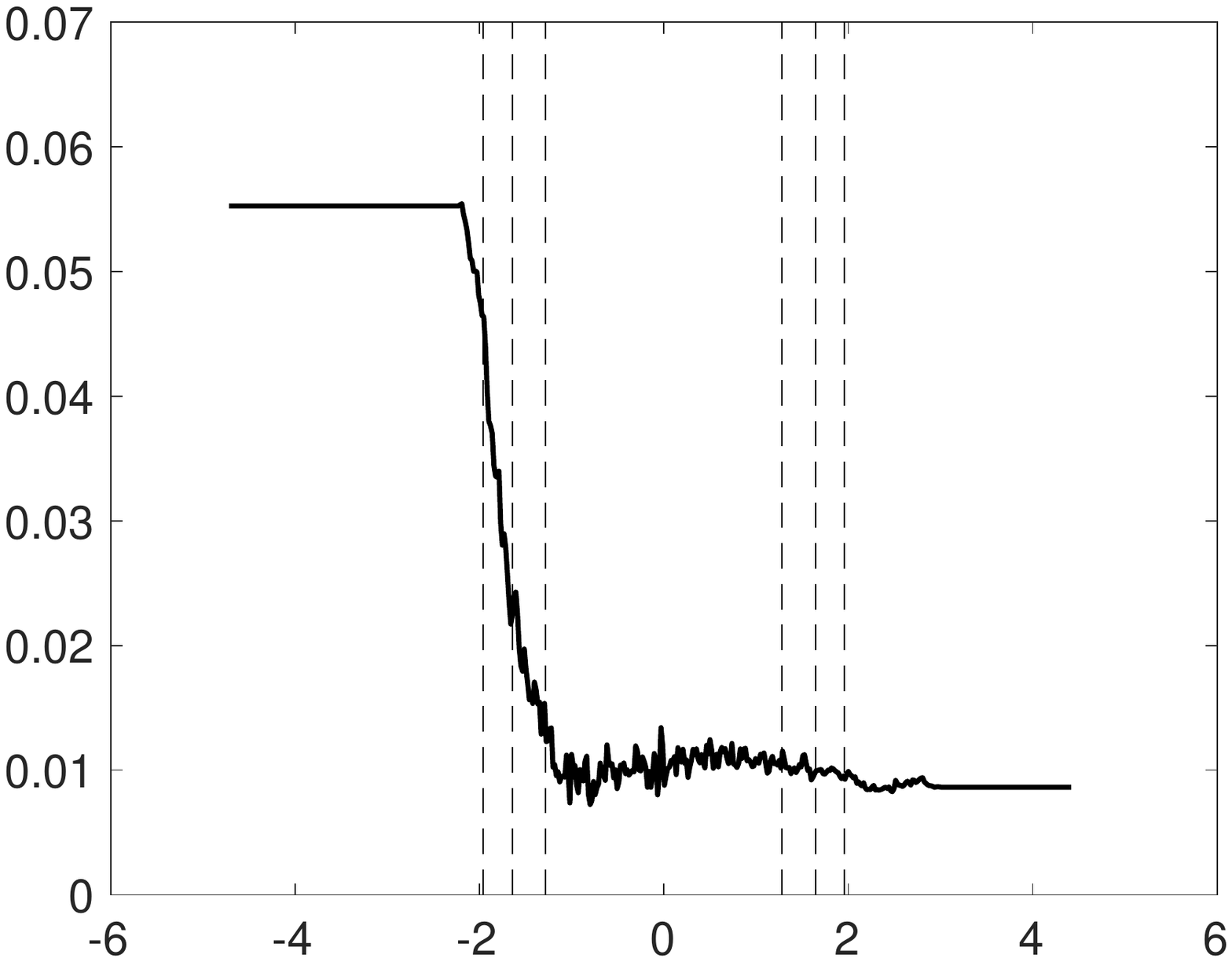}
\end{subfigure}\hfill
\begin{subfigure}[t]{0.5\columnwidth}
\centering	\caption{}
\includegraphics[trim={6cm 7cm 6cm 7cm}, scale=.35]{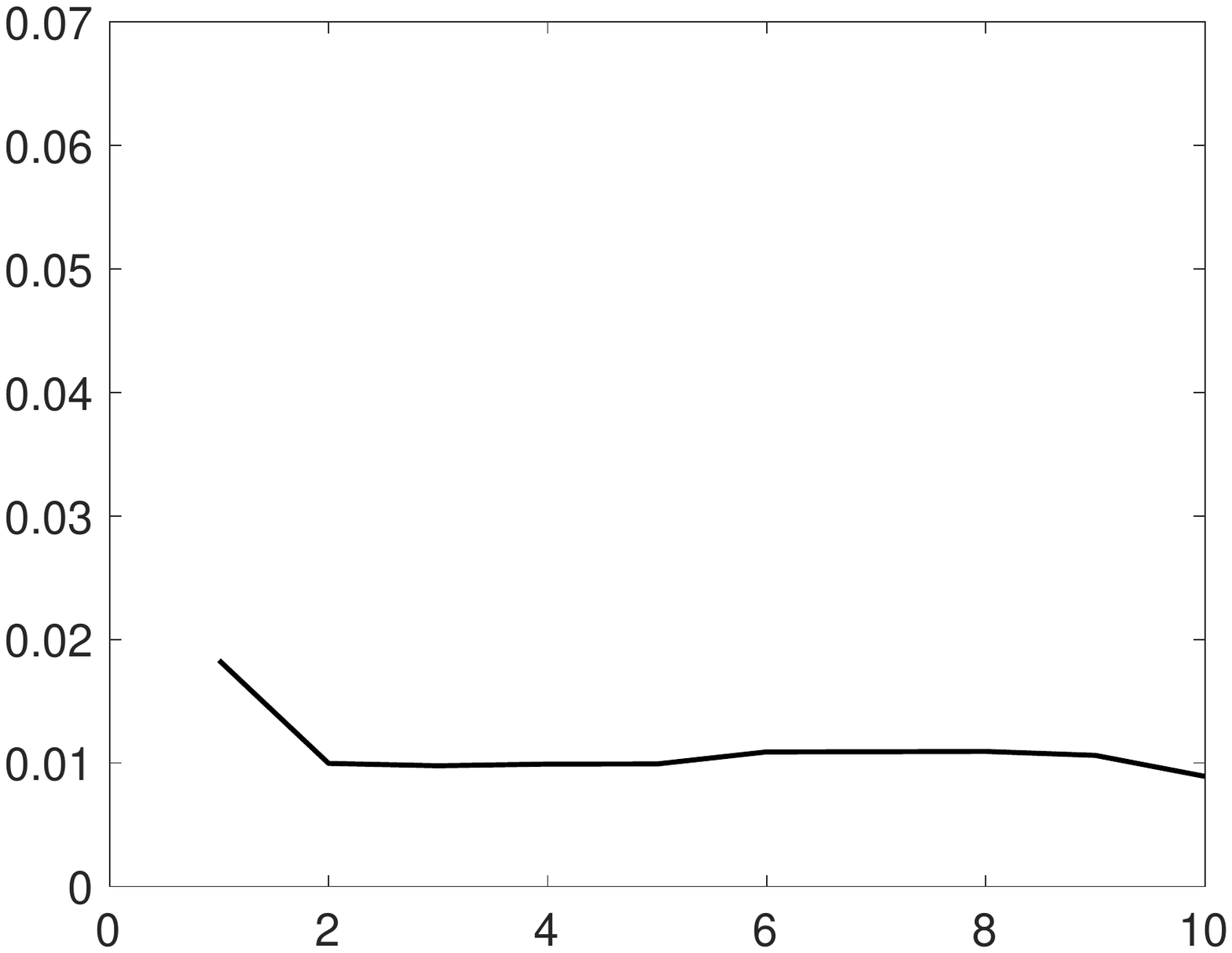}
\end{subfigure}\\[0.25in]
\begin{subfigure}[t]{0.5\columnwidth}
\centering	\caption{\textit{1980--2015}}
\includegraphics[trim={6cm 7cm 6cm 7cm}, scale=.35]{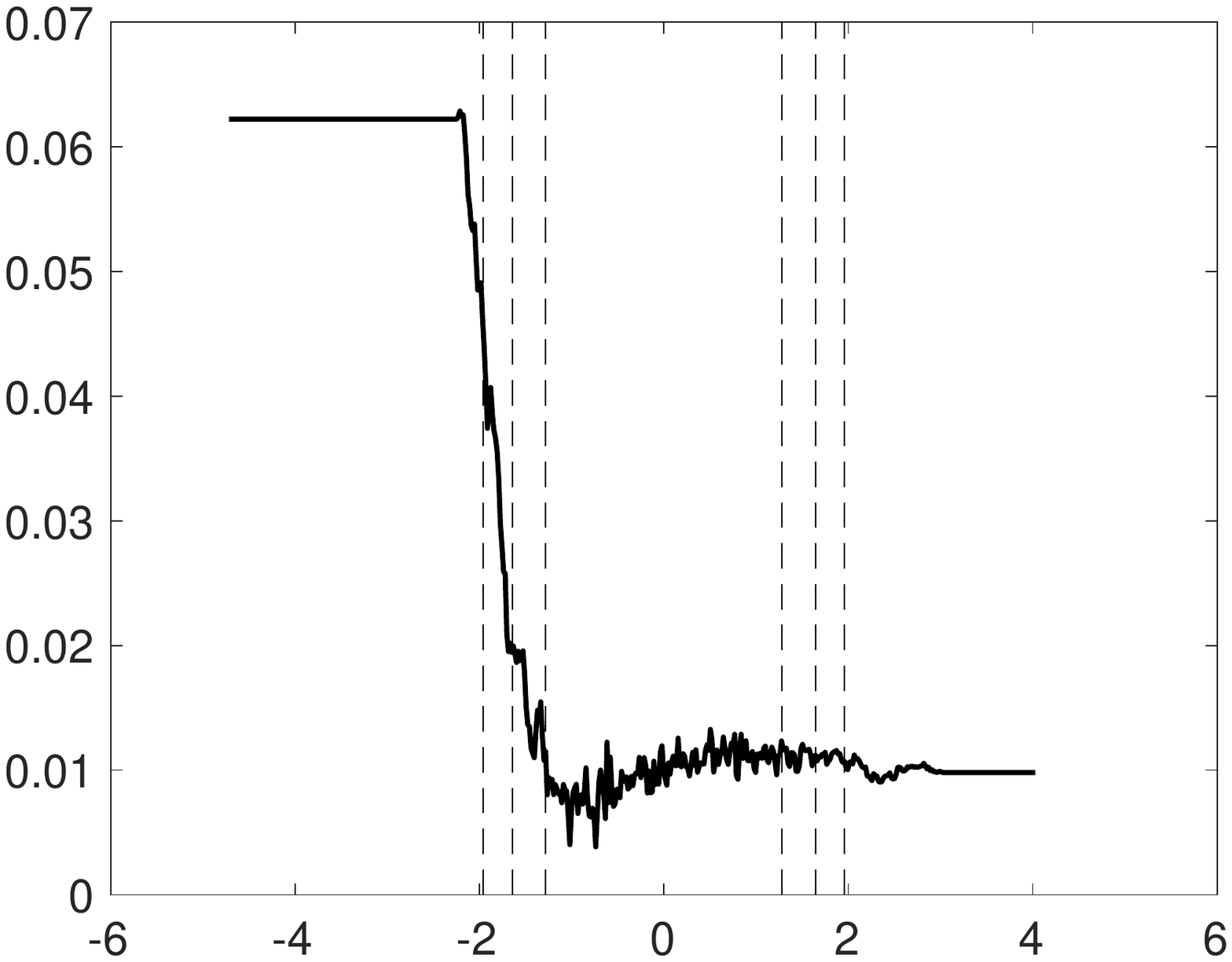}
\end{subfigure}\hfill
\begin{subfigure}[t]{0.5\columnwidth}
\centering	\caption{}
\includegraphics[trim={6cm 7cm 6cm 7cm}, scale=.35]{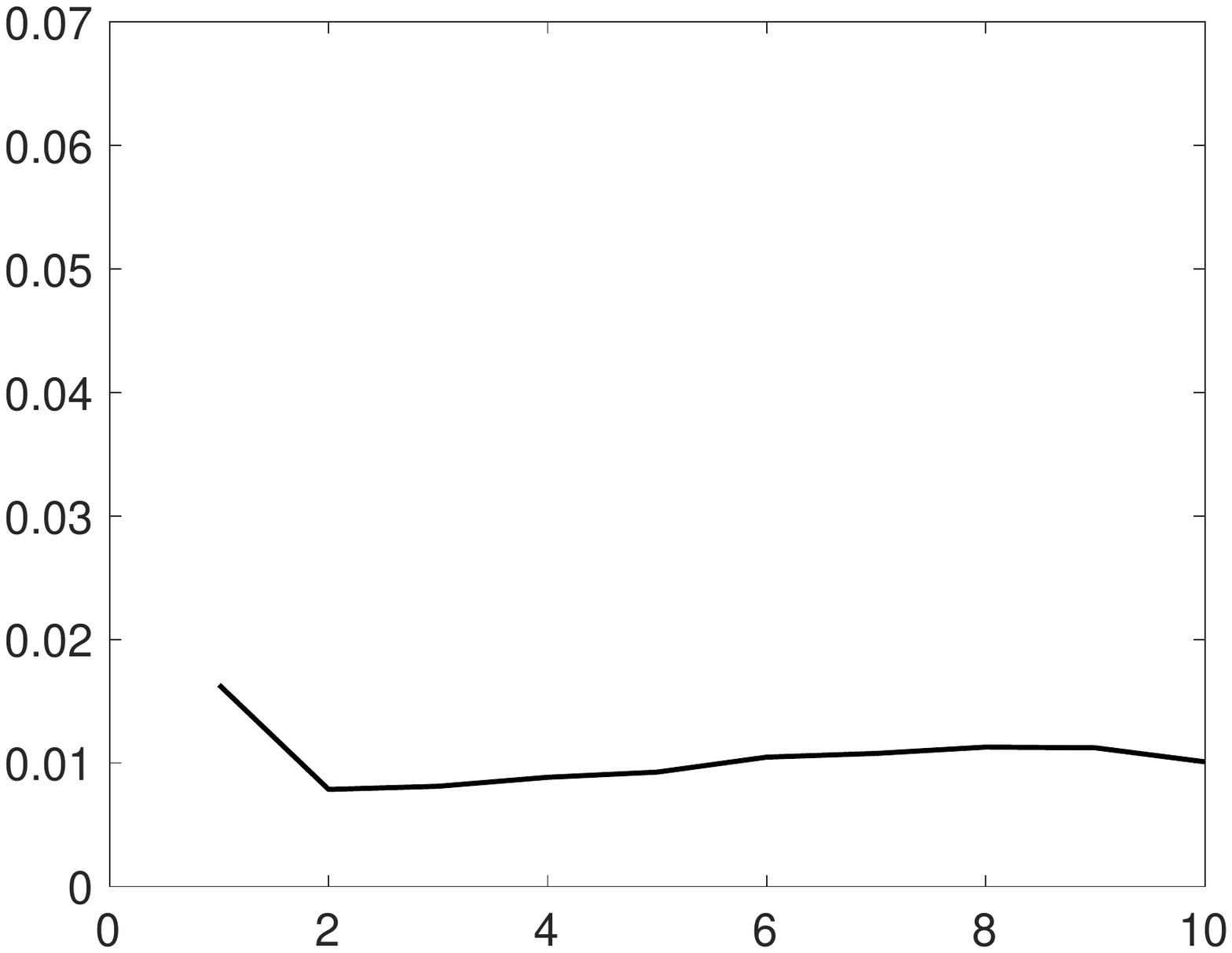}
\end{subfigure}\hfill
\end{figure}

Figure \ref{fig:sizeAll} provides the estimates of the relationship between returns and firm size. The left column shows the estimate, $\{\hat{\mu}(z): z \in \mathcal{Z}\}$, based on equation \eqref{eqn:muhat} whereas the right column plots the average return in each of ten portfolios formed based on the conventional approach currently used in the literature. The portfolio breakpoints for the standard approach are commonly chosen using either deciles of the sub-sample of firms listed on the NYSE or deciles based on the entire sample. Here we choose deciles based on the latter as they ensure better comparability across estimators.  To ensure comparability both estimates have been placed on the same scale. As is clear from the figure, the conventional approach produces an attenuated return differential between average returns and size. One important reason for this is that the standard approach relies on the same number of portfolios regardless of changes in the cross-sectional sample size. As we have shown in Sections \ref{sec:framework} and \ref{sec:first order}, it is imperative that the choice of the number of portfolios is data-driven, respecting the appropriate rate conditions, in order to deliver valid inference.  The standard approach will tend to produce a biased estimate of the return differential and will compromise power to discern a significant differential in the data. This issue will always arise in any unbalanced panel, but is exacerbated by the highly unbalanced nature of these data where the number of firms has been trending strongly over time. 

The estimate, $\{\hat{\mu}(z): z \in \mathcal{Z}\}$, is shown for three different subsamples in Figure \ref{fig:sizeAll}, namely, 1926--2015, 1967--2015, and 1980--2015. The estimated shape between returns and size is generally very similar across the three sub-periods with a relatively flat relationship except for small firms where there is a sharp monotonic rise in average returns as size decreases. The peak average return for the smallest firms appears to have risen over time, at approximately $5$\% over the full sample, $5.5$\% over the sample from 1967--2015 and slightly above $6$\% over the sample 1980--2015.  

Table \ref{tab:empirical} shows the associated point estimates and test statistics corresponding to the graphs in Figure \ref{fig:sizeAll}.  We display results for a number of different choices of the pairs $(z_h$, $z_L)$, namely, $(\Phi^{-1}(.975),\Phi^{-1}(.025))$, $(\Phi^{-1}(.95),\Phi^{-1}(.05))$, and $(\Phi^{-1}(.9),\Phi^{-1}(.1))$, where $\Phi(\cdot)$ is the CDF of a standard normal random variable, shown as vertical lines in Figure \ref{fig:sizeAll}. The table also shows the point estimates and corresponding test statistics from the conventional approach using ten portfolios. Over all three sub-periods, the difference between the function evaluated at the two most extreme evaluation points, $(\Phi^{-1}(.975),\Phi^{-1}(.025))$, is associated with a strongly statistically significant effect of size on returns. Even in the shortest subsample, 1980--2015, the t-statistic is $-5.46$.  This is also the case when the evaluation points are shifted inward to $(\Phi^{-1}(.95),\Phi^{-1}(.05))$.  As shown in Figure \ref{fig:sizeAll}, this result is driven by very small firms. However, the conventional estimator would suggest that the size effect is no longer statistically distinguishable from zero over the last $35$ or so years. Instead, what has happened is that ``larger'' small firms are no longer producing higher returns in the last sub-sample. This pattern can be seen in the innermost set of evaluations points, $(z_H, z_L) = (\Phi^{-1}(.9),\Phi^{-1}(.1))$, where the size effect is estimated to be reversed, albeit statistically indistinguishable from zero.

\begin{figure}[t]
\caption{\doublespacing\textbf{Size Anomaly: NYSE Only}\\[0.05in] \normalsize{This figure shows the estimated relationship between the cross section of equity returns and lagged market equity (equation \eqref{eqn:size}) for NYSE firms. The left column displays $\hat{\mu}(\cdot)$ where $J_t$ is based on equation \eqref{eqn:HOMSE},  $z_H = \Phi^{-1}(.975)$, $z_L = \Phi^{-1}(.025)$. The right column displays the estimated relation using the standard portfolio sorting implementation with $J=10$.  All returns are in monthly  changes and all portfolios are value weighted based on lagged market equity.}}
\label{fig:sizeNYSE}
\begin{subfigure}[t]{0.5\columnwidth}
\centering	\caption{\textit{1926--2015}}
\includegraphics[trim={6cm 7cm 6cm 7cm}, scale=.35]{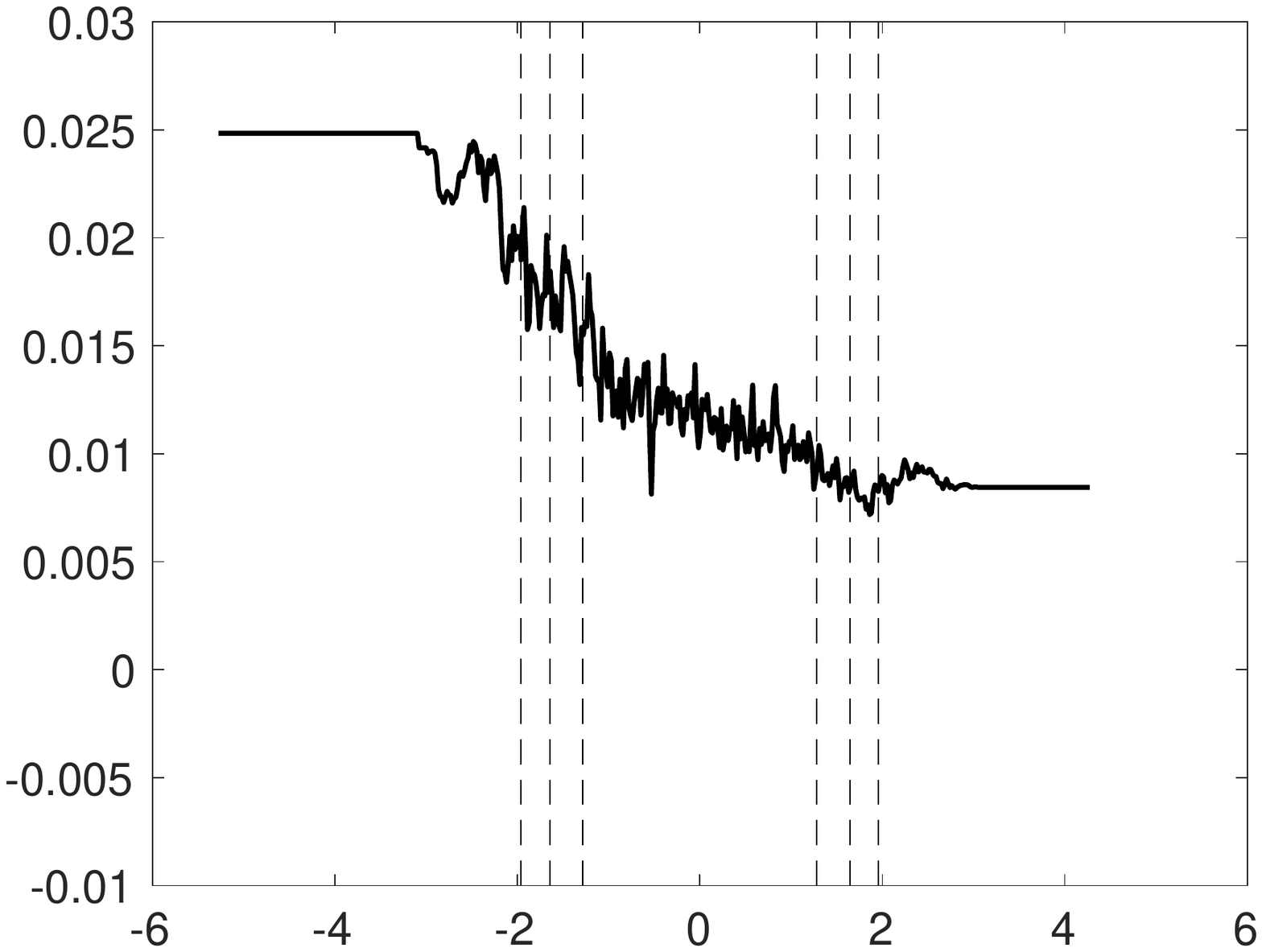}
\end{subfigure}\hfill
\begin{subfigure}[t]{0.5\columnwidth}
\centering	\caption{}
\includegraphics[trim={6cm 7cm 6cm 7cm}, scale=.35]{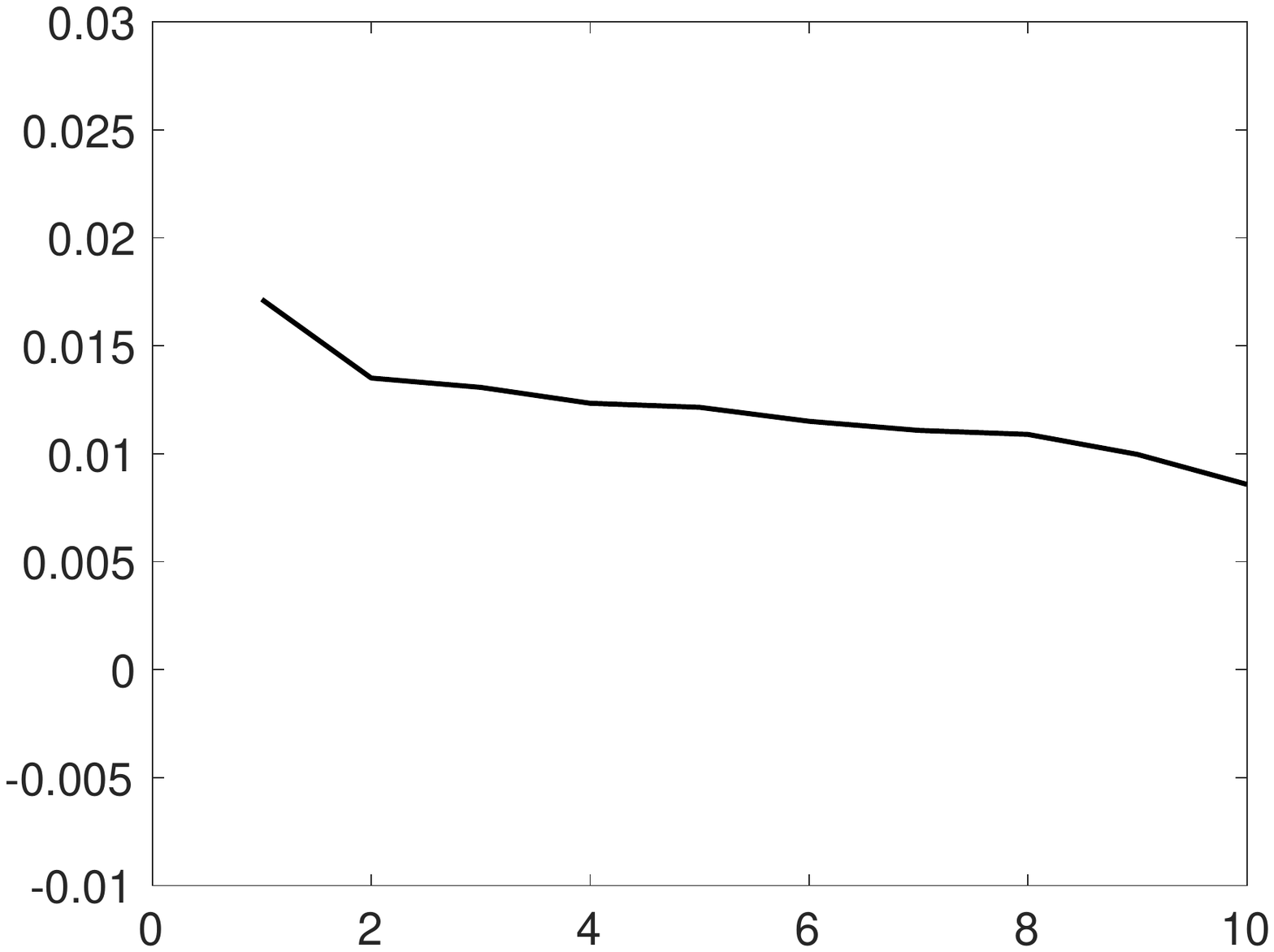}
\end{subfigure}\\[0.25in]
\begin{subfigure}[t]{0.5\columnwidth}
\centering	\caption{\textit{1967--2015}}
\includegraphics[trim={6cm 7cm 6cm 7cm}, scale=.35]{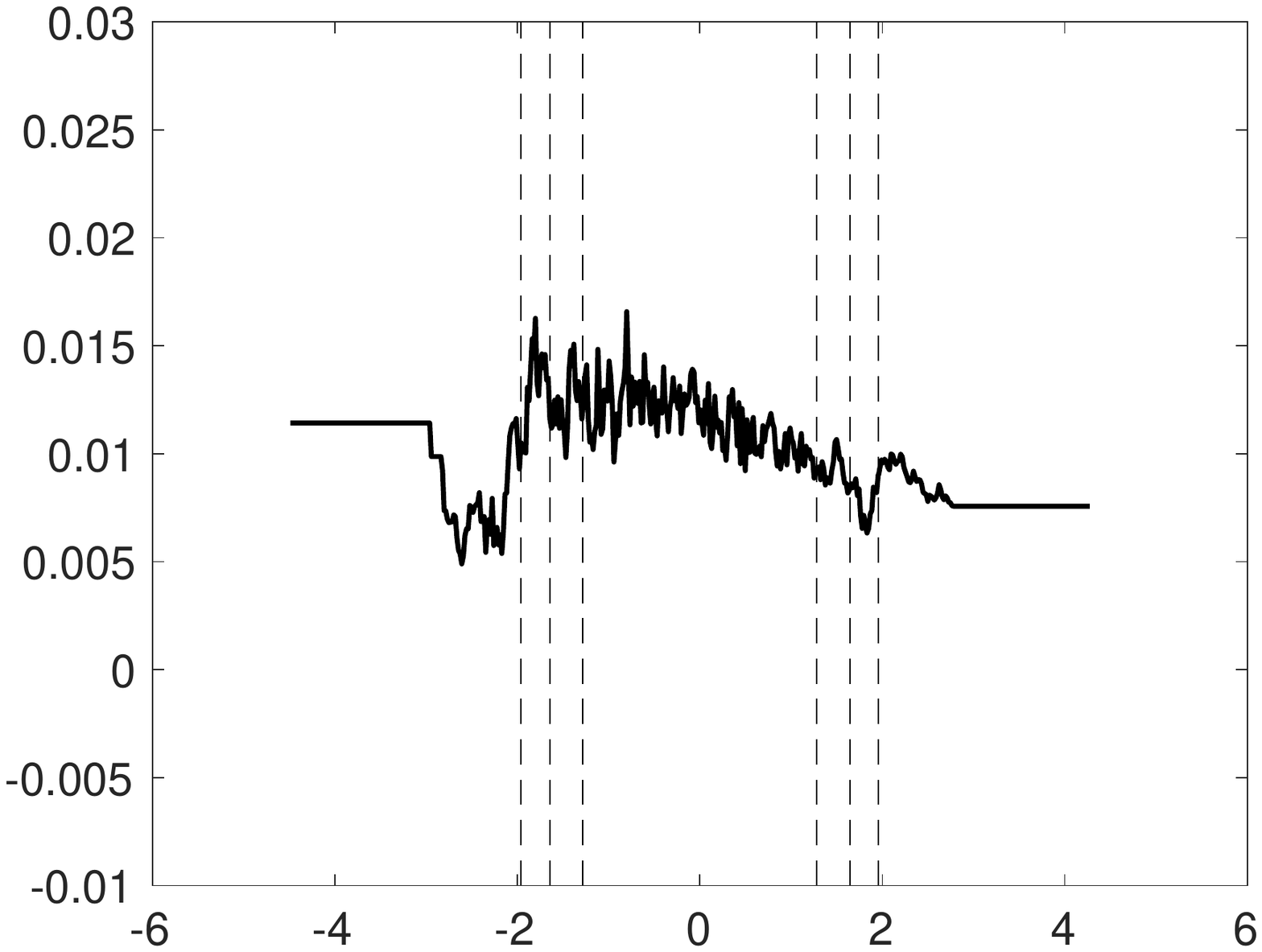}
\end{subfigure}\hfill
\begin{subfigure}[t]{0.5\columnwidth}
\centering	\caption{}
\includegraphics[trim={6cm 7cm 6cm 7cm}, scale=.35]{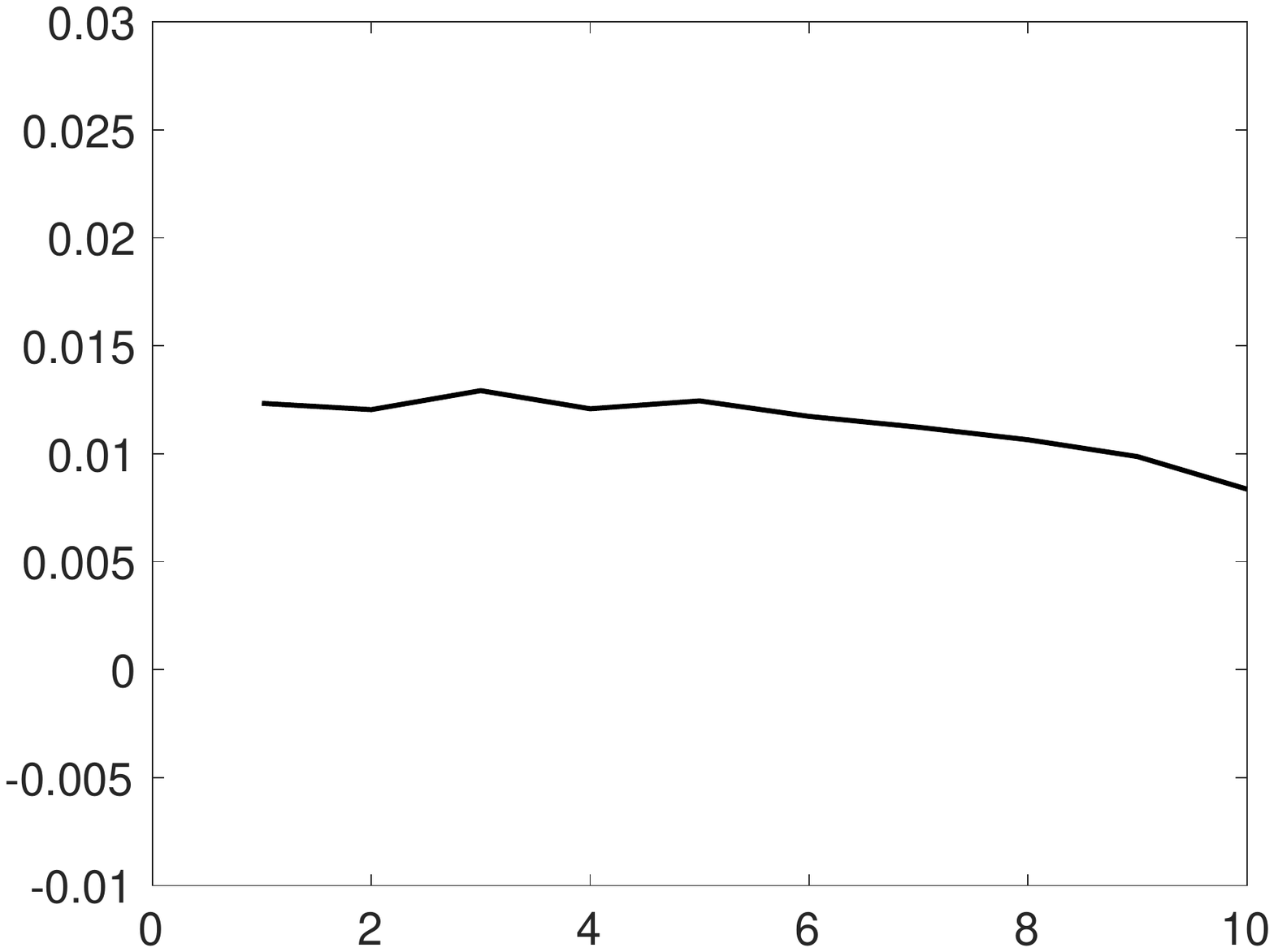}
\end{subfigure}\\[0.25in]
\begin{subfigure}[t]{0.5\columnwidth}
\centering	\caption{\textit{1980--2015}}
\includegraphics[trim={6cm 7cm 6cm 7cm}, scale=.35]{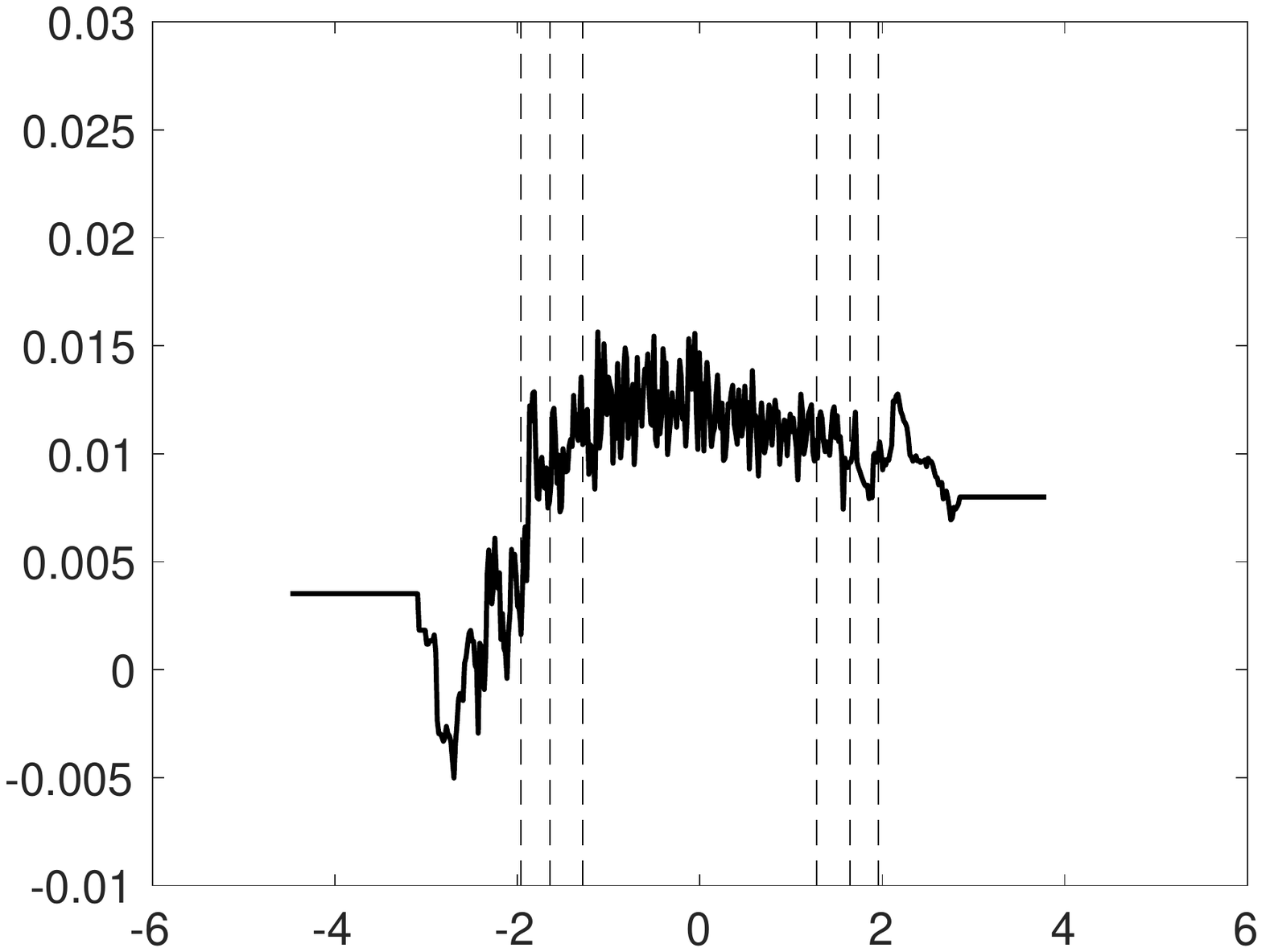}
\end{subfigure}\hfill
\begin{subfigure}[t]{0.5\columnwidth}
\centering	\caption{}
\includegraphics[trim={6cm 7cm 6cm 7cm}, scale=.35]{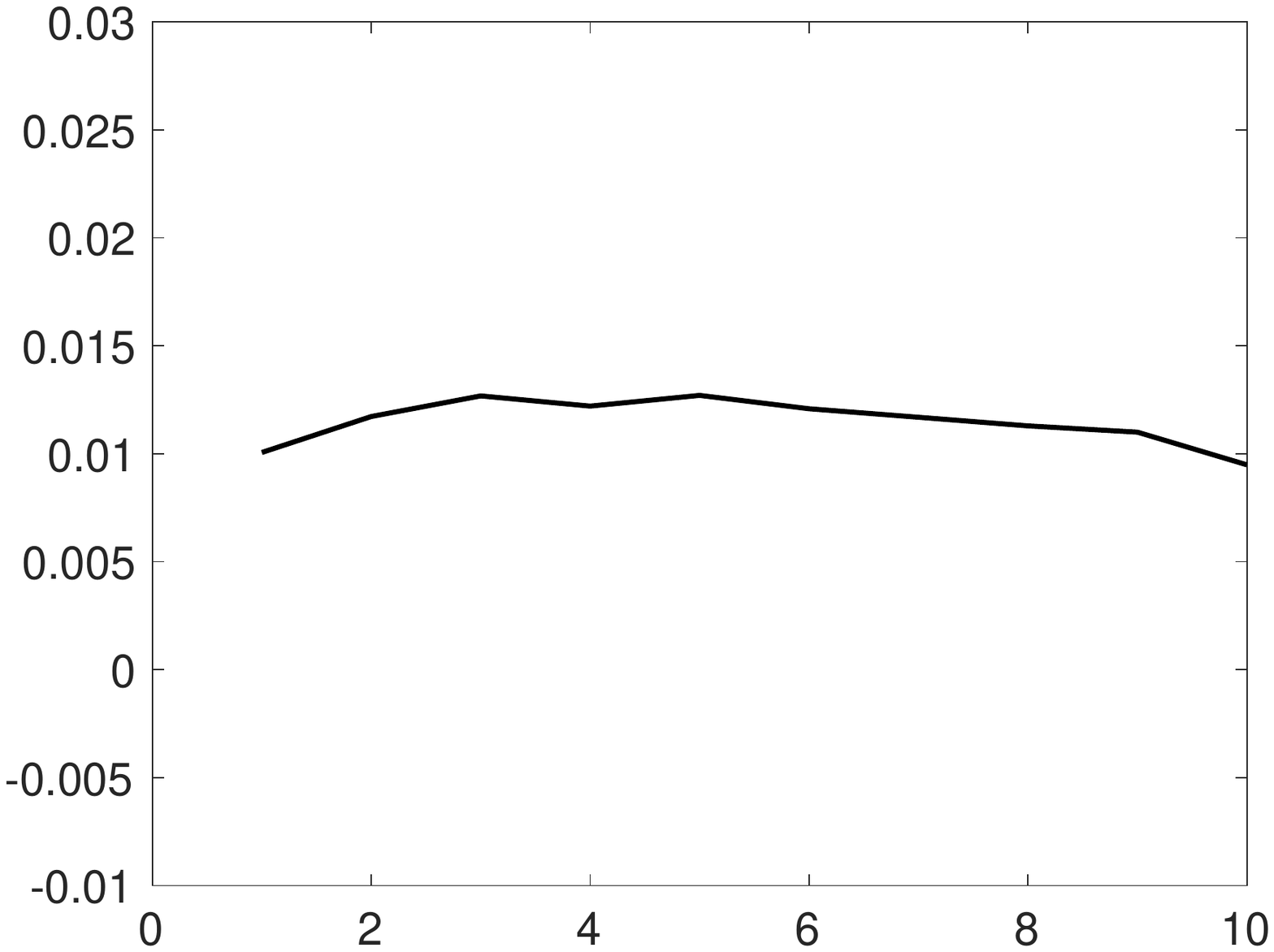}
\end{subfigure}\hfill
\end{figure}

To further investigate the results of Table \ref{tab:empirical} we reconsider the estimates for the relationship between returns and firm size using only firms listed on the NYSE in Figure \ref{fig:sizeNYSE}. In this case, the shape of the estimated relationship changes markedly in the full sample versus the most recent subsamples. In the full sample, the estimated relationship appears very similar to the shape shown in the three charts in Figure \ref{fig:sizeAll}---a sharp downward slope from smaller firms to larger firms. However, over the samples 1967--2015 and 1980--2015, the estimated shape changes demonstrably toward an upside-down ``U'' shape. It is important to emphasize that the standard approach implies a very different shape and pattern of the relationship between returns and size for this sample of firms---especially for the sample from 1967--2015 and 1980--2015.


The left panel of Figure \ref{fig:JtStar} shows time series plots of the optimal number of portfolios in the sample for the size anomaly chosen based on equation \eqref{eqn:HOMSE}, using data for our three sub-periods and based on $z_H = \Phi^{-1}(.975)$, $z_L = \Phi^{-1}(.025)$. Notably, the optimal number of portfolios is substantially larger than the standard choice of ten. Instead, the optimal choice is approximately $250$ in the largest cross section and around $50$ in the smallest cross section.  Furthermore, in all three samples, there is substantial variation in the optimal number of portfolios, again, reflecting the strong variation in cross-sectional sample sizes in these data.  The charts also show the optimal number of portfolios in the NYSE-only sample. In this restricted sample the cross-sectional sample sizes are lower which, all else equal, will reduce the optimal choice of number of portfolios. However, the bias-variance trade-off also changes in the NYSE-only sample and so it is not always the case that the restricted sample has a smaller value for the optimal number of portfolios. In the 1980--2015 sample, the optimal choice of portfolios is slightly larger (at its peak) than the case using all stocks reinforcing the point that the appropriate choice of number of portfolios will be strongly affected by the features of the data being used.

\subsection{Momentum Anomaly}

We next consider the momentum anomaly---where firms which have had better relative returns in the nearby past also have higher relative returns on average. As discussed in Section \ref{sec:framework}, the generality of our sampling assumptions means that our results apply to anomalies such as momentum where lagged returns enter in the unknown function of interest. Specifically
\begin{equation}
R_{it} = \mu ( \textsc{mom}_{it} ) + \varepsilon_{it}, \qquad i = 1,\ldots,n_t, \quad t = 1, \ldots, T.
\label{eqn:momentum}
\end{equation}
Here, $\textsc{mom}_{it}$, represents the 12-2 momentum measure of firm $i$ at time $t$ transformed in the following way: at each cross section $t=1,\ldots,T$, 12-2 momentum is demeaned and normalized by the inverse of the cross-sectional standard deviation (i.e., a zscore is applied). Unlike in the case of the size anomaly, no transformation is necessary to satisfy Assumption \ref{ass:dgp}(c). We chose to normalize each cross-section in this way as it is the natural counterpart in our setting to the standard portfolio sorting approach to the momentum anomaly. Moreover, the results based directly on 12-2 momentum are similar.

\begin{figure}[t]
\caption{\doublespacing\textbf{Momentum Anomaly}\\[0.05in] \normalsize{This figure shows the estimated relation between the cross section of equity returns and 12-2 momentum (equation \eqref{eqn:momentum}). The left column displays $\hat{\mu}(\cdot)$ where $J_t$ has been chosen based on equation \eqref{eqn:HOMSE}, $z_H = \Phi^{-1}(.975)$, $z_L = \Phi^{-1}(.025)$. The right column displays the estimated relation using the standard portfolio sorting implementation with $J=10$. All returns are in monthly  changes and all portfolios are value weighted based on lagged market equity.}}
\label{fig:momentum}
\begin{subfigure}[t]{0.5\columnwidth}
\centering	\caption{\textit{1927--2015}}
\includegraphics[trim={6cm 7cm 6cm 7cm}, scale=.35]{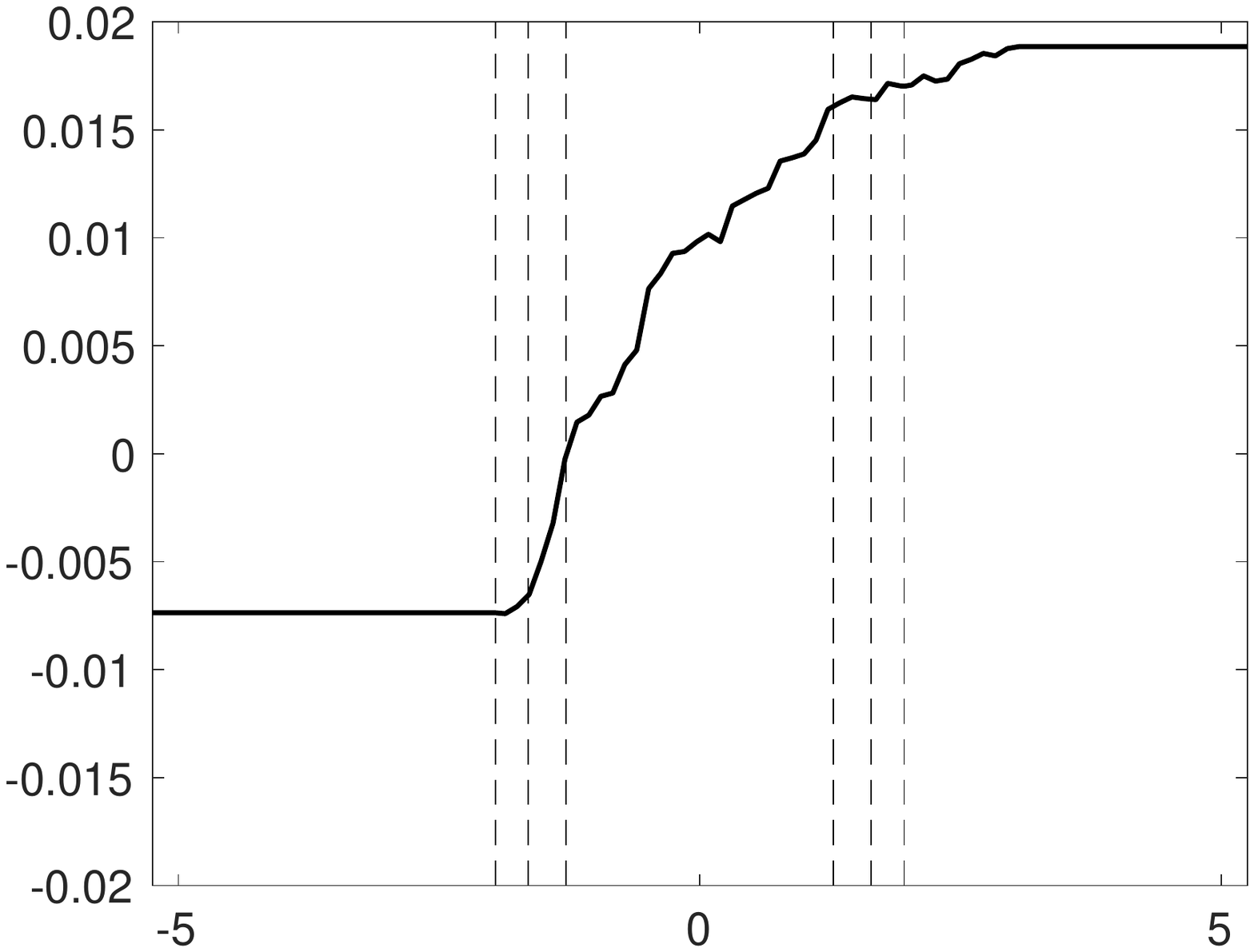}
\end{subfigure}\hfill
\begin{subfigure}[t]{0.5\columnwidth}
\centering	\caption{}
\includegraphics[trim={6cm 7cm 6cm 7cm}, scale=.35]{charts/momentum_muEstStandard_ind0_spec11.pdf}
\end{subfigure}\\[0.25in]
\begin{subfigure}[t]{0.5\columnwidth}
\centering	\caption{\textit{1967--2015}}
\includegraphics[trim={6cm 7cm 6cm 7cm}, scale=.35]{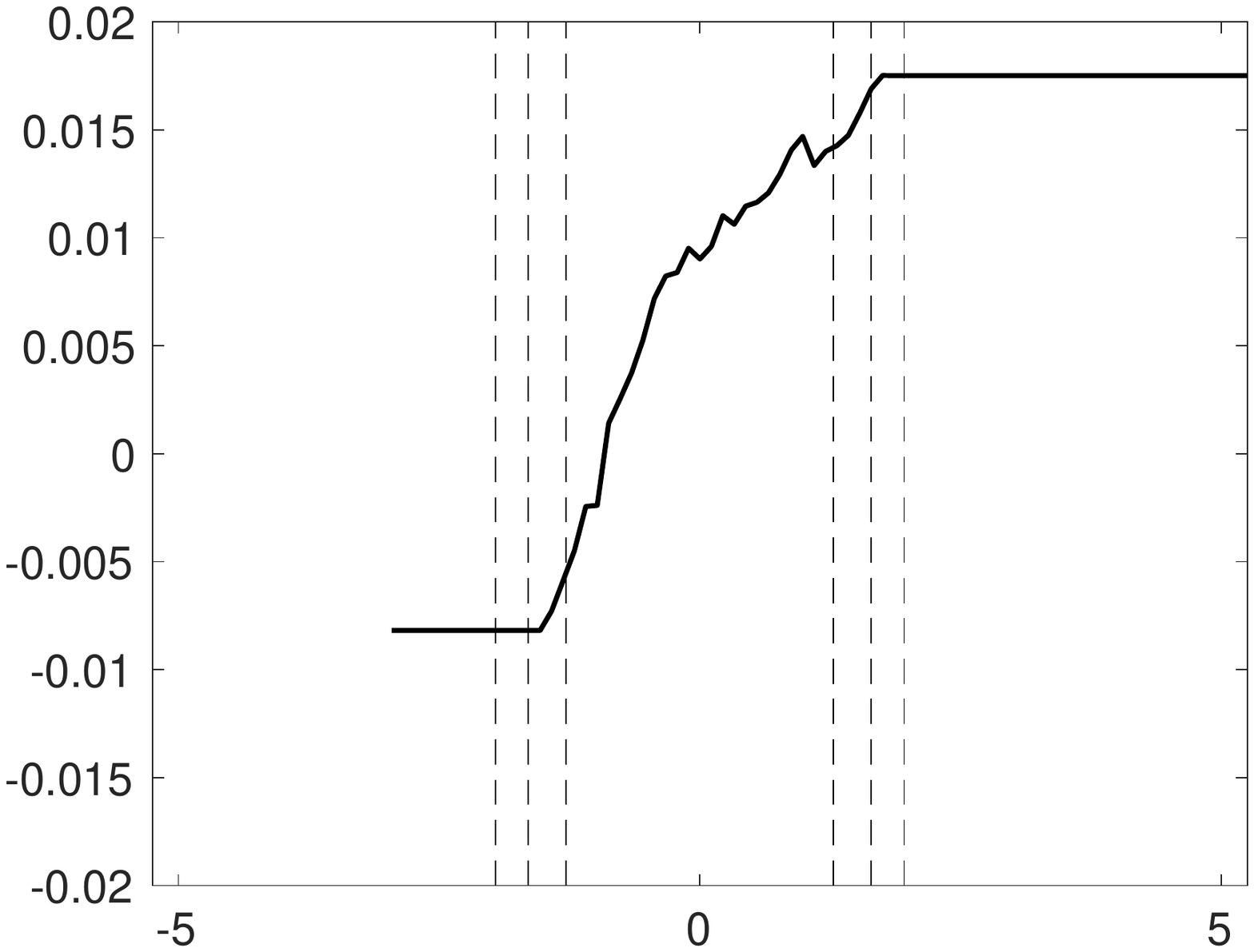}
\end{subfigure}\hfill
\begin{subfigure}[t]{0.5\columnwidth}
\centering	\caption{}
\includegraphics[trim={6cm 7cm 6cm 7cm}, scale=.35]{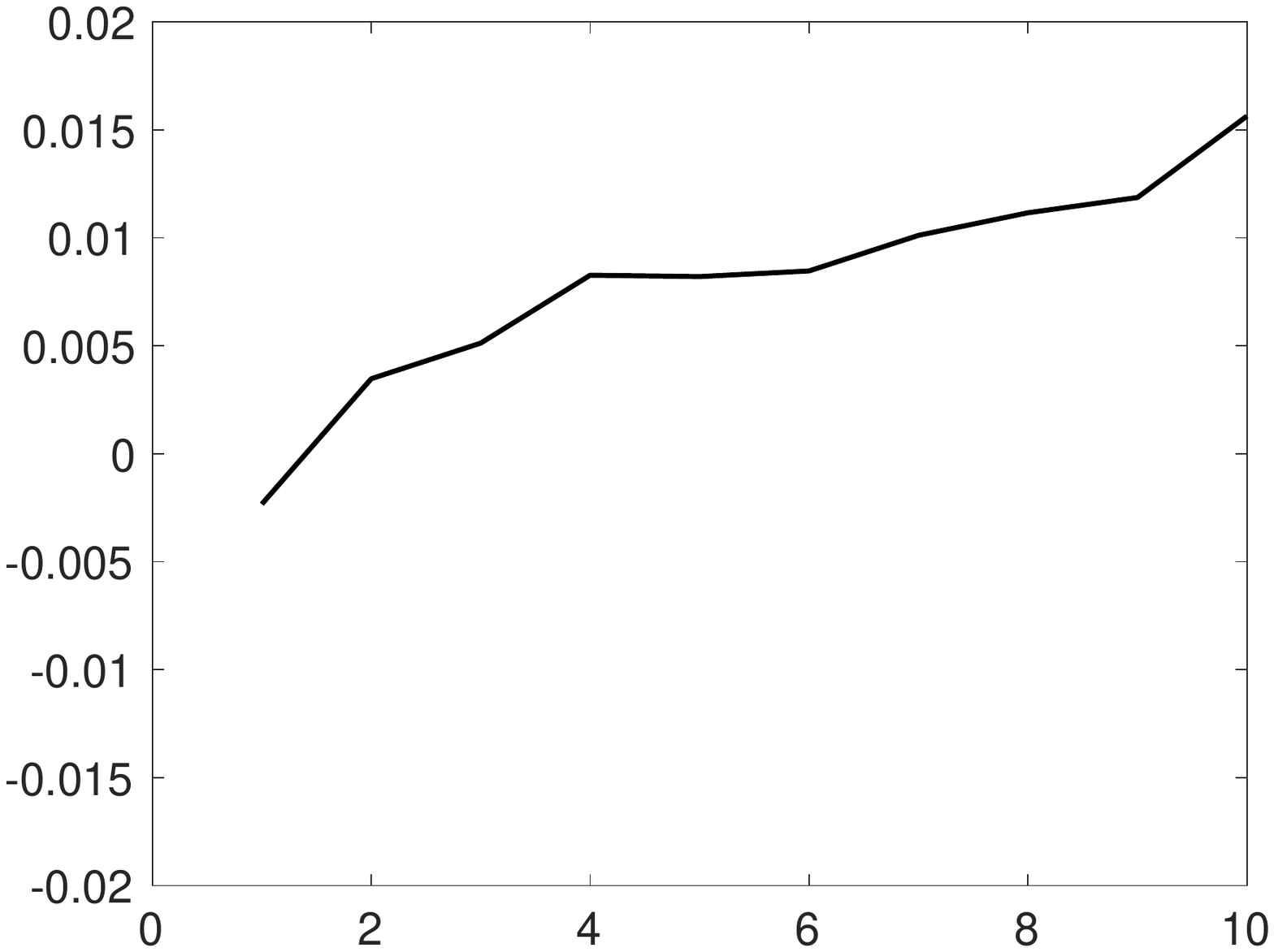}
\end{subfigure}\\[0.25in]
\begin{subfigure}[t]{0.5\columnwidth}
\centering	\caption{\textit{1980--2015}}
\includegraphics[trim={6cm 7cm 6cm 7cm}, scale=.35]{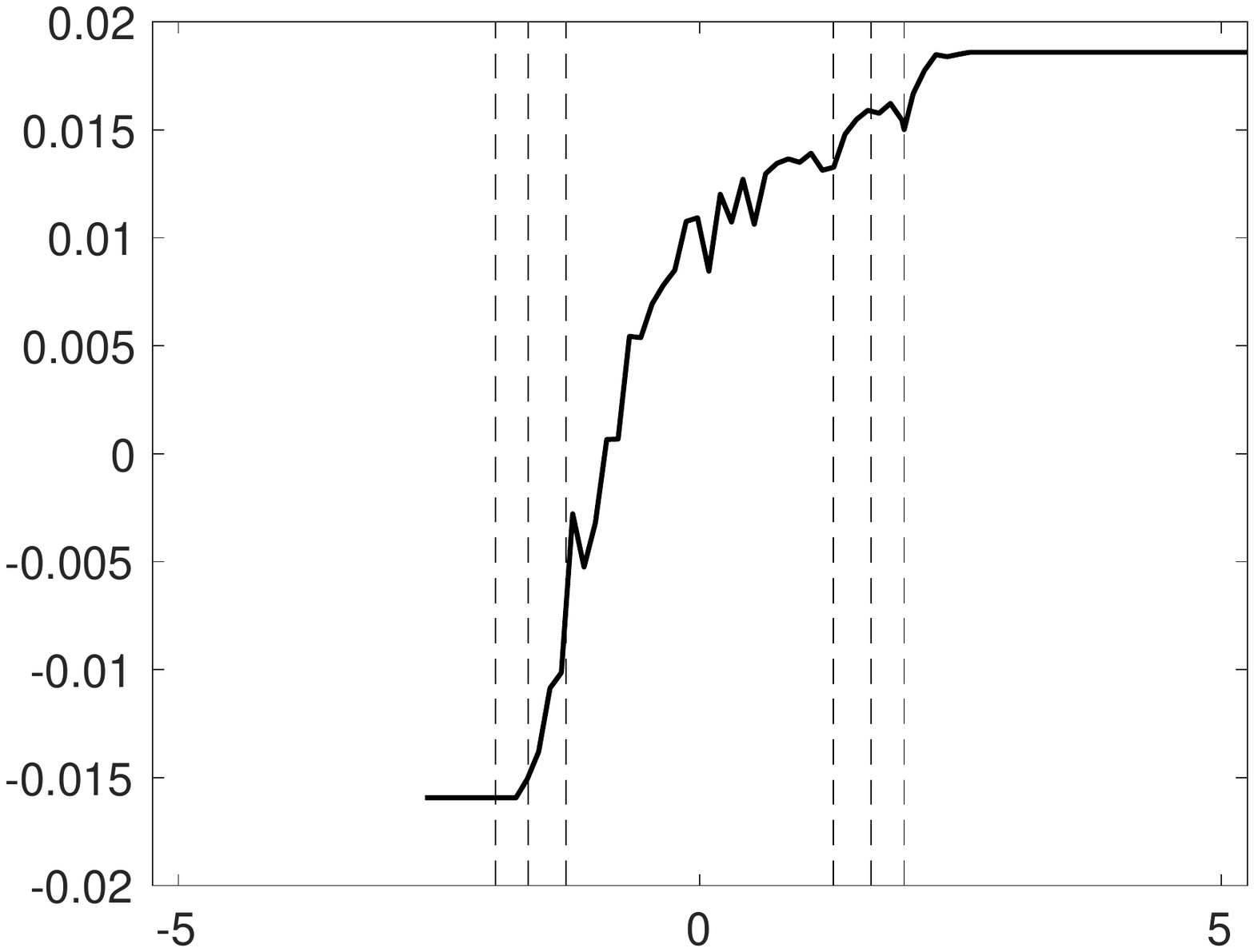}
\end{subfigure}\hfill
\begin{subfigure}[t]{0.5\columnwidth}
\centering	\caption{}
\includegraphics[trim={6cm 7cm 6cm 7cm}, scale=.35]{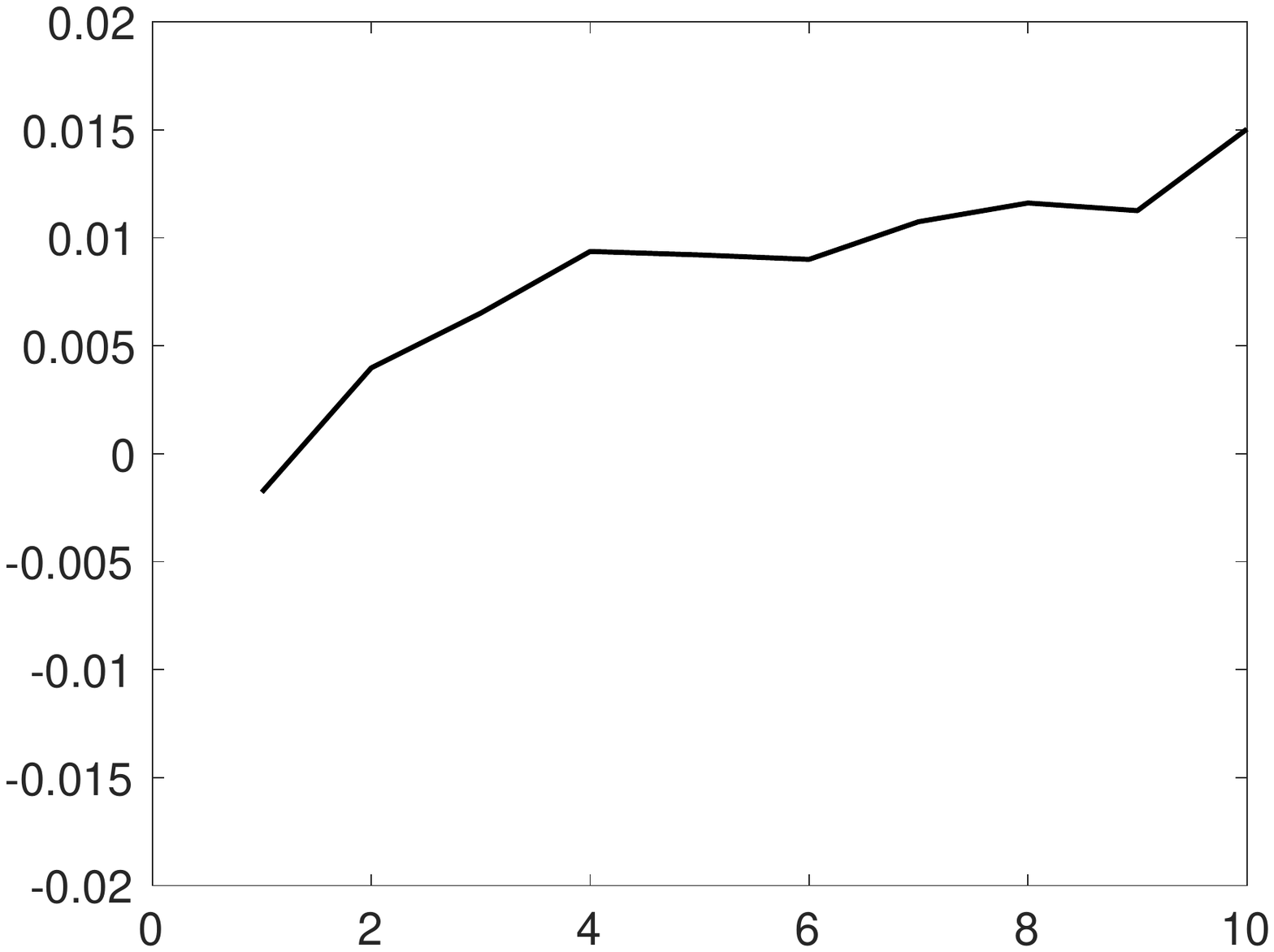}
\end{subfigure}\hfill
\end{figure}

Figure \ref{fig:momentum} shows the estimates of the relationship between returns and momentum. Even more so than in the case of the size anomaly, we observe that $\{\hat{\mu}(z): z\in \mathcal{Z}\}$ is very similar across subsamples. The relationship appears concave with past ``winners'' (i.e., those with high 12-2 momentum values) earning about $2$\% in returns, on average.  The strategy of investing in past ``losers'' (i.e., those with low 12-2 momentum values) on the other hand, has resulted in increasing losses in the later subsamples.  The nadir in the estimated relationship occurs at approximately $-0.8$\% in the full sample, slightly less than that in the 1967--2015 subsample, and $-1.5$\% in the 1980--2015 subsamples. This suggests that the short side of buying the spread portfolio appears to have become more profitable in recent years. This conclusion is robust to excluding the financial crisis and its aftermath. The right column of Figure \ref{fig:momentum} shows that this insight could not be gleaned by using the conventional estimator. Furthermore, the conventional estimator suggests an approximately linear relationship between returns and momentum with a distinctly compressed differential between the average returns of winners versus losers. This underscores how our more general approach leads to richer conclusions about the underlying data generating process.

The bottom panel of Table \ref{tab:empirical} shows the corresponding point estimates and test statistics for the momentum anomaly. The results strongly confirm that momentum is a robust anomaly.  Across all three pairs of evaluation points and the three different samples, the spread is highly statistically significant (last column). Focusing separately on $\mu(z_H)$ and $\mu(z_L)$ we find that the point estimates are positive and negative, respectively, across all our specifications.  In fact, the short end of the spread trade, represented by $\mu(z_L)$, appears to have become stronger in the latter samples (see also Figure \ref{fig:momentum}), producing t-statistics which have the largest magnitude in the 1980--2015 sample when evaluated at either $(\Phi^{-1}(.975),\Phi^{-1}(.025))$ or $(\Phi^{-1}(.95),\Phi^{-1}(.05))$. In contrast, the conventional implementation finds that the short side of the trade is never significant across any of the subsamples and a t-statistic of only $-0.36$ in the 1980--2015 sample.

Cross-sectional regressions are, by far, the most popular empirical alternative to portfolio sorting (see discussion in Remarks \ref{rem:csregressions} and \ref{rem:csregressions2}).  Arguably the most appealing feature of cross-sectional regressions to the empirical researcher is the ability to include a large number of control variables.  Given that we have combined the two approaches in a unified framework it is natural to consider an example.  Here we will consider the nonparametric relationship between returns and momentum while controlling for industry momentum. This empirical exercise is similar in spirit to \citet*{MoskowitzGrinblatt1999}. The model then becomes,
\begin{equation}
\label{eqn:momentumWithControls}
R_{it} = \mu ( \textsc{mom}_{it} ) + \beta_1 \cdot \textsc{Imom}_{it} + \beta_2 \cdot \textsc{Imom}_{it}^2
+ \beta_3 \cdot \textsc{Imom}_{it}^3 + \varepsilon_{it},
\end{equation}
where $\textsc{Imom}_{it}$ is the industry momentum of firm $i$ at time $t$. We also include the square and cube of industry momentum as a flexible way to allow for nonlinearities in this control.

\begin{figure}[t]
\caption{\doublespacing\textbf{Momentum Anomaly: Controlling for Industry Momentum}\\[0.05in] \normalsize{This figure shows the estimated relation between the cross section of equity returns and 12-2 momentum. The left column displays $\hat{\mu}(\cdot)$ controlling for $\textsc{Imom}_{it}$, $\textsc{Imom}_{it}^2$ and  $\textsc{Imom}_{it}^3$ (solid line) as in equation \eqref{eqn:momentumWithControls} where $J_t$ has been chosen based on equation \eqref{eqn:HOMSE}, $z_H = \Phi^{-1}(.975)$, $z_L = \Phi^{-1}(.025)$. The dash-dotted line shows $\hat{\mu}(z)$ without control variables as in equation \eqref{eqn:momentum} for the same $J_t$. The right column displays the estimated relation using the standard portfolio sorting implementation with $J=10$ and no controls. All returns are in monthly  changes and all portfolios are value weighted based on lagged market equity.
}}
\label{fig:momentumWithControls}
\begin{subfigure}[t]{0.5\columnwidth}
\centering	\caption{\textit{1927--2015}}
\includegraphics[trim={6cm 7cm 6cm 7cm}, scale=.35]{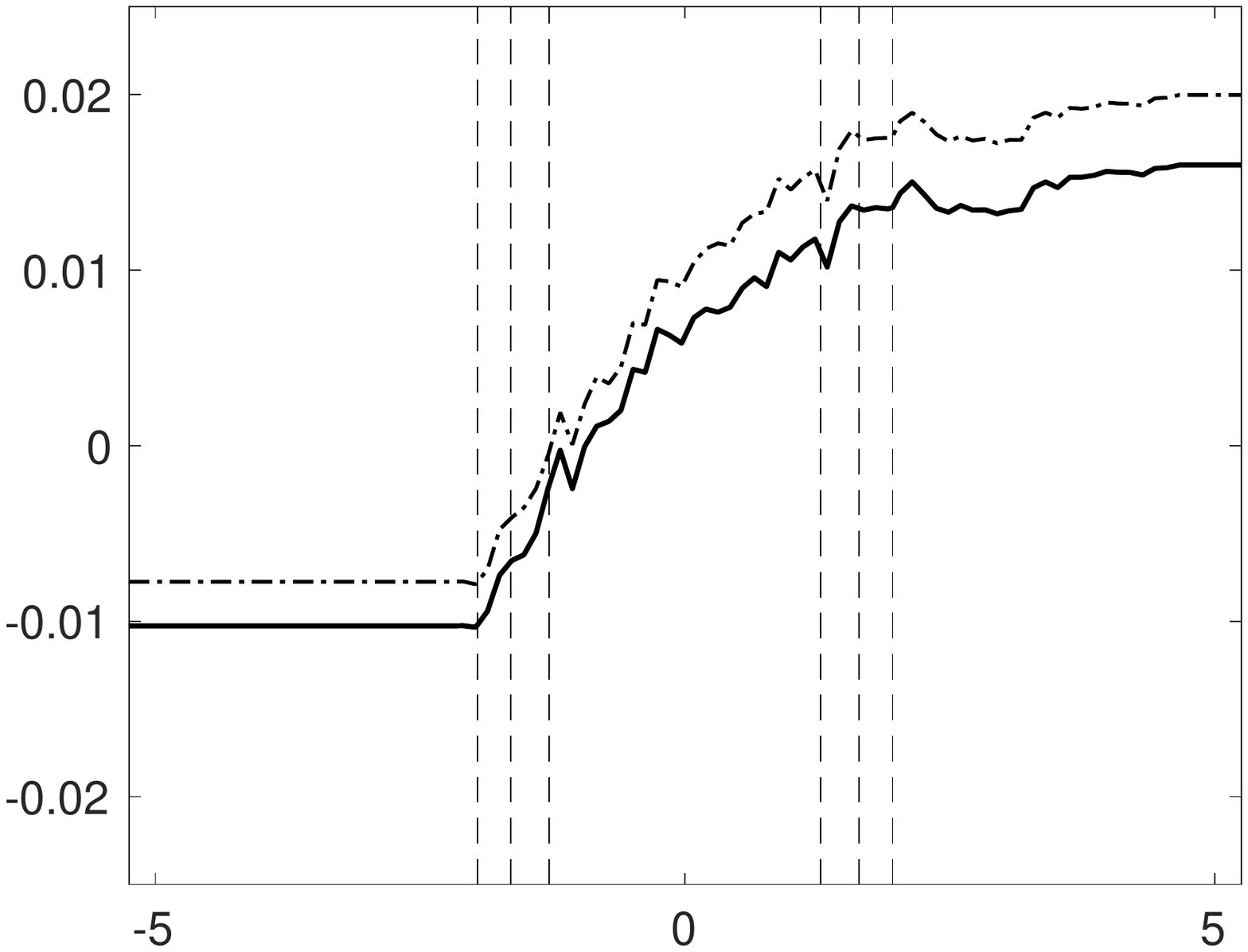}
\end{subfigure}\hfill
\begin{subfigure}[t]{0.5\columnwidth}
\centering	\caption{}
\includegraphics[trim={6cm 7cm 6cm 7cm}, scale=.35]{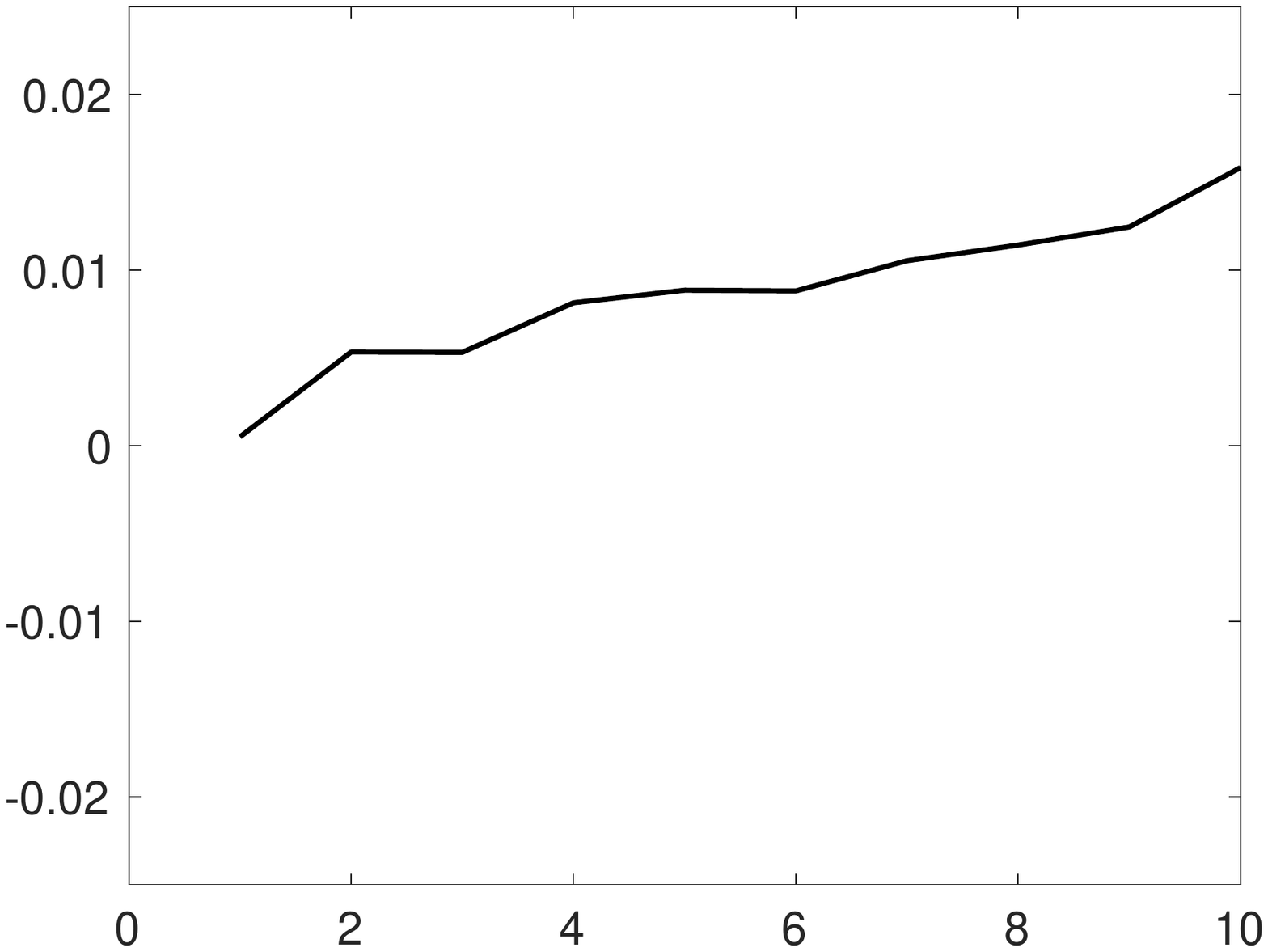}
\end{subfigure}\\[0.25in]
\begin{subfigure}[t]{0.5\columnwidth}
\centering	\caption{\textit{1967--2015}}
\includegraphics[trim={6cm 7cm 6cm 7cm}, scale=.35]{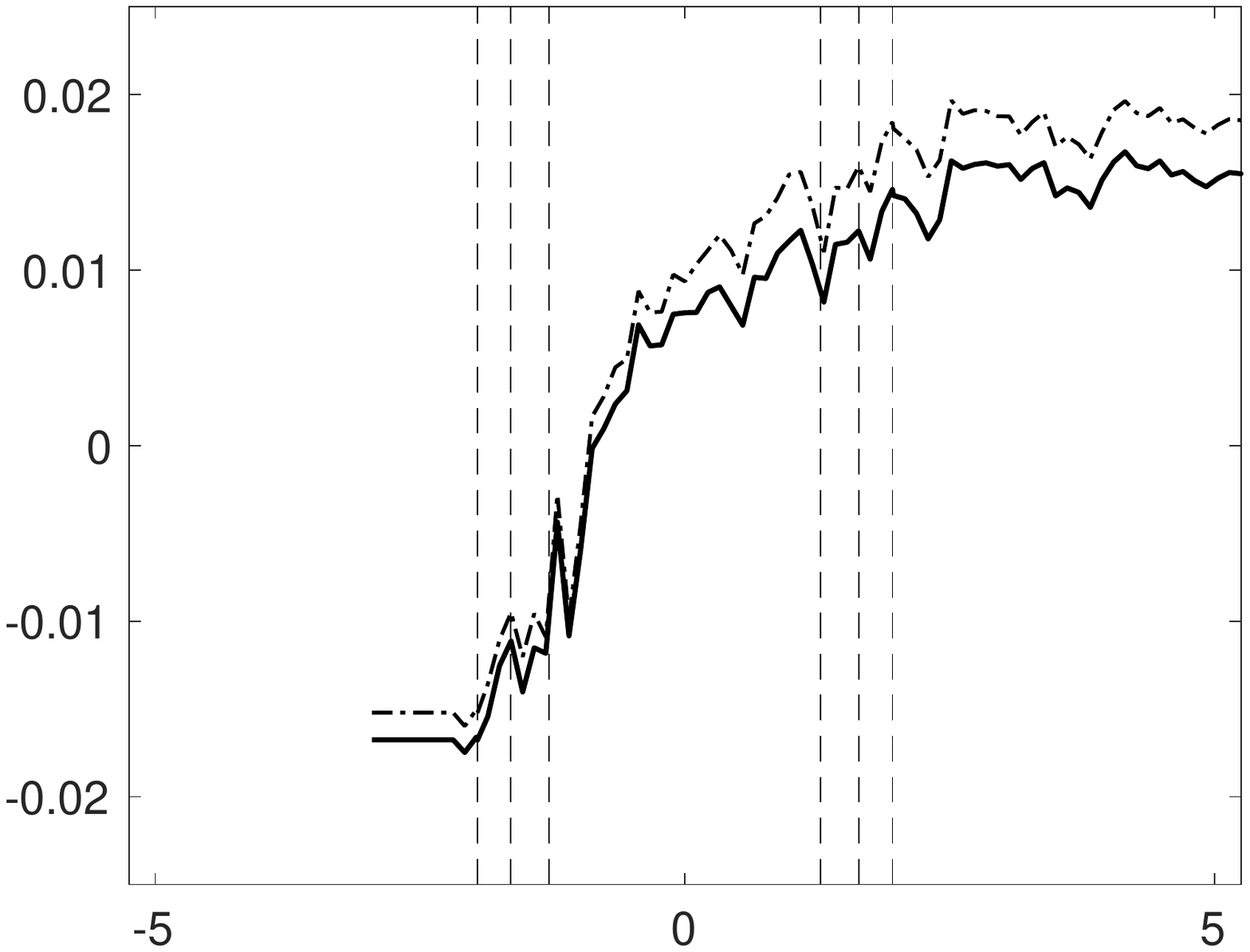}
\end{subfigure}\hfill
\begin{subfigure}[t]{0.5\columnwidth}
\centering	\caption{}
\includegraphics[trim={6cm 7cm 6cm 7cm}, scale=.35]{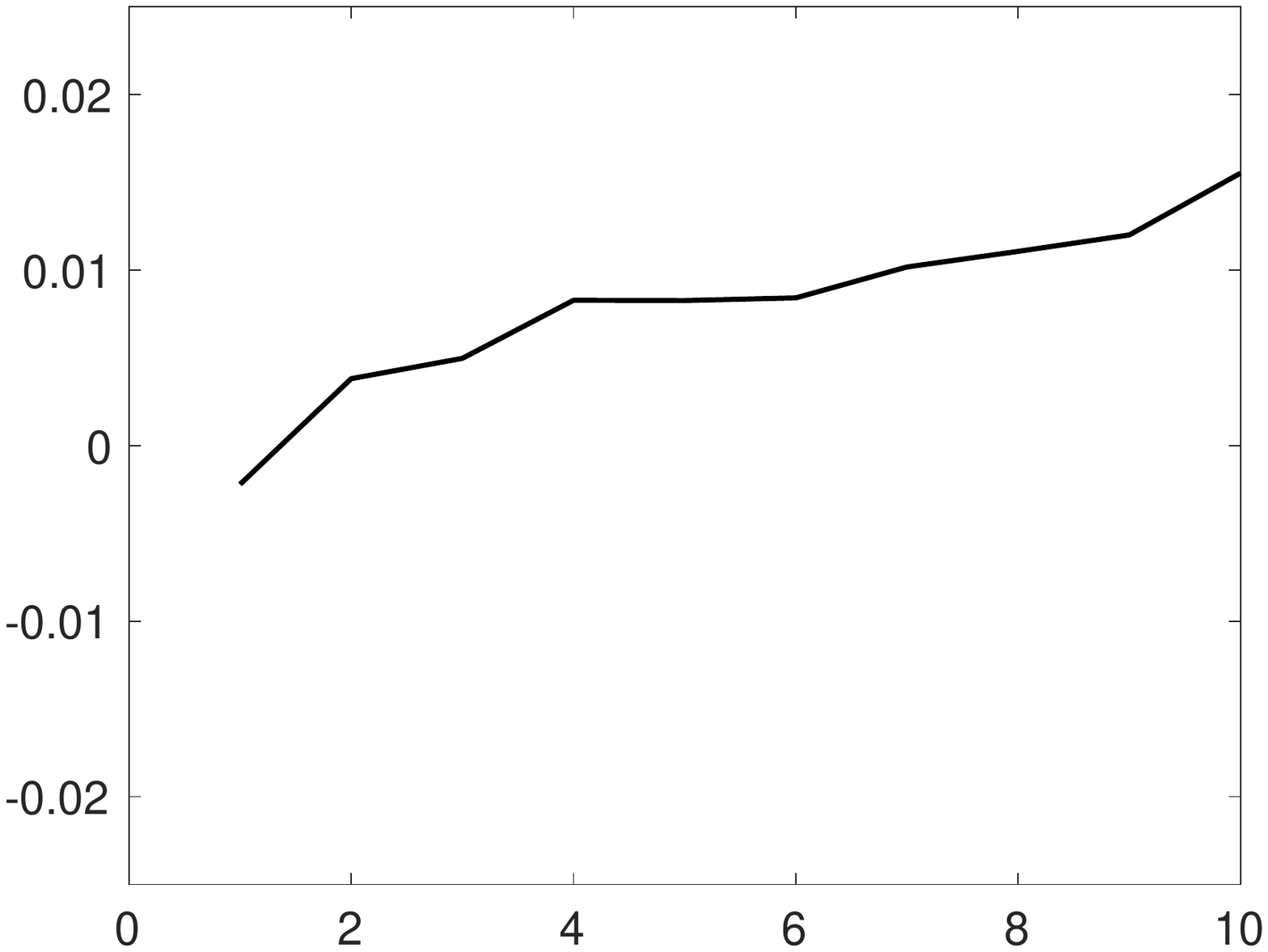}
\end{subfigure}\\[0.25in]
\begin{subfigure}[t]{0.5\columnwidth}
\centering	\caption{\textit{1980--2015}}
\includegraphics[trim={6cm 7cm 6cm 7cm}, scale=.35]{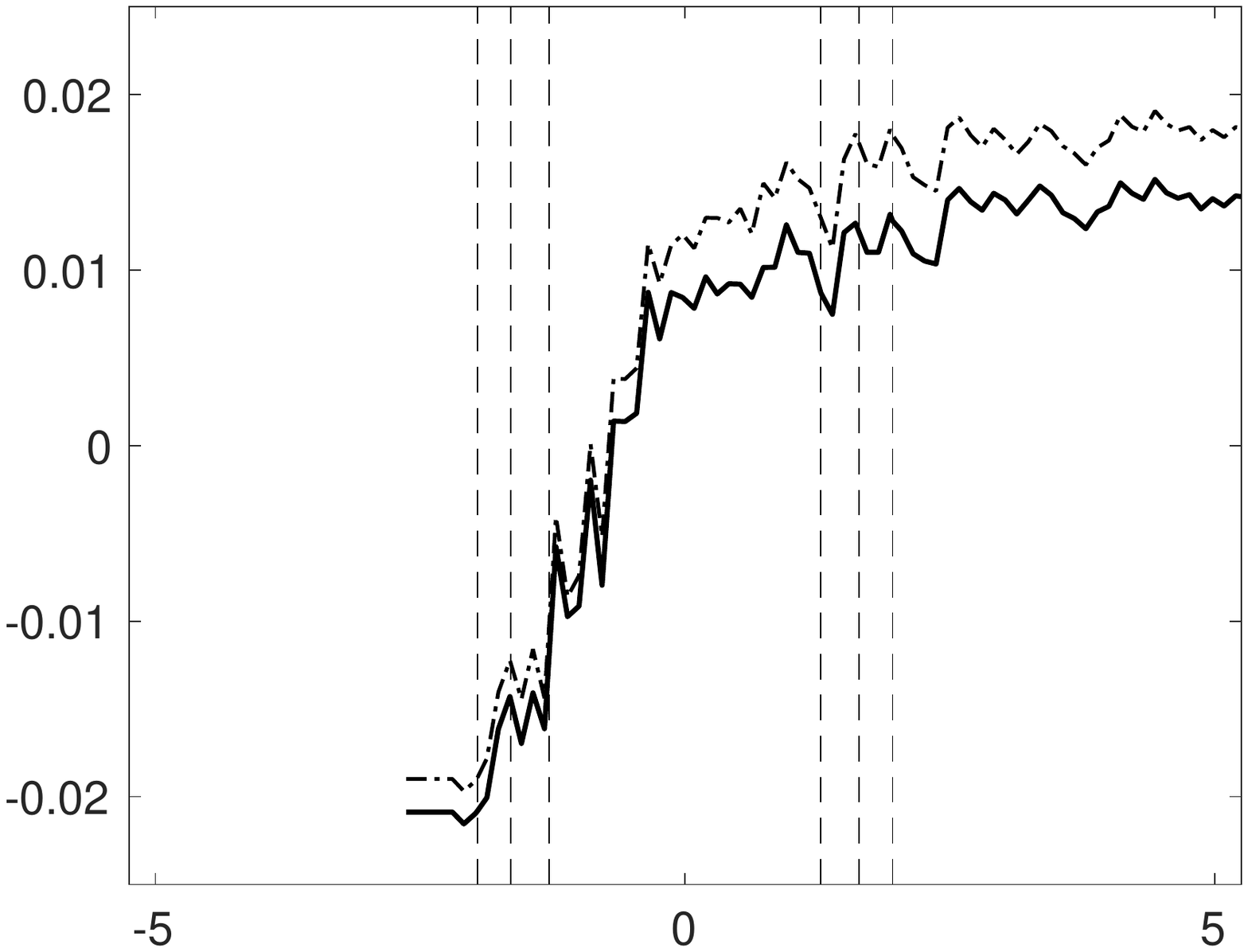}
\end{subfigure}\hfill
\begin{subfigure}[t]{0.5\columnwidth}
\centering	\caption{}
\includegraphics[trim={6cm 7cm 6cm 7cm}, scale=.35]{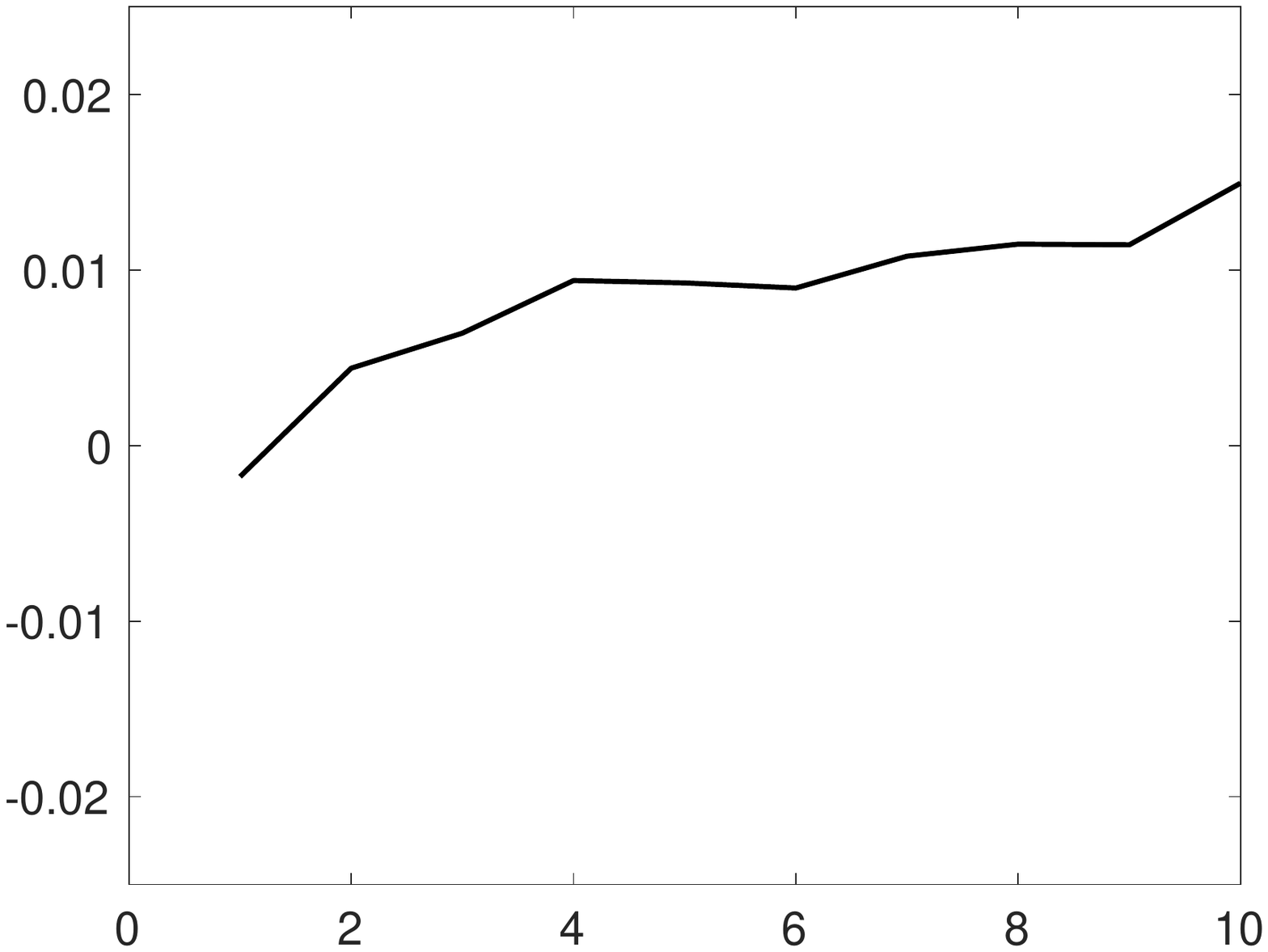}
\end{subfigure}\hfill
\end{figure}

Figure \ref{fig:momentumWithControls} shows the estimates of the relationship between returns and momentum controlling for industry momentum as in equation \eqref{eqn:momentumWithControls} (solid line). For reference the plots in the left column also include $\{ \hat{\mu}(z): z \in \mathcal{Z}\}$ (dash-dotted line) with no control variables (i.e., based on equation \eqref{eqn:momentum}) for the same choice of the number of portfolios at each time $t$. To improve comparability, the estimated function without control variables uses the same sequence of $\{J_t: t = 1\ldots T\}$ as in the case with control variables. Thus, this estimated function differs from that presented in Figure \ref{fig:momentum}. The difference between the two estimated functions tends to be larger for larger values of 12-2 momentum and accounts for, at most, approximately 0.5 percentage point of momentum returns in the full sample. In the two more recent subsamples the differences are smaller but economically meaningful. That said, the broad shape of the relationship between returns and stock momentum is unchanged by controlling for industry momentum. This suggests that, for this choice of specification, momentum of individual firms is generally distinct from momentum within an industry \citep*{MoskowitzGrinblatt1999,GrundyMartin2001}.

The bottom panel of Table \ref{tab:empirical} provides point estimates and associated test statistics based on equation \eqref{eqn:momentumWithControls} in the rows labelled ``w/ controls''. First, it is clear that the inclusion of industry momentum does have a noticeable effect on inference. In general, the magnitude of the t-statistics for the high evaluation point, low evaluation point, and difference are shrunk toward zero. For both the high evaluation point and the difference this is uniformly true and, in all cases, results in t-statistics with substantially larger associated p-values. That said, for all subsamples the difference at the high and low evaluation points results in statistically significant return differential at the 5\% level.  This exercise illustrates the usefulness of our unified framework as it allows for the additional of control variables in a simple and straightforward manner.

Finally, the right panel of Figure \ref{fig:JtStar} shows time series plots of the optimal number of portfolios in the sample for the momentum anomaly. Just as in the case of the size anomaly, the optimal number of portfolios is well above ten. However, a number of specifications result in a maximum number of portfolios of approximately 55. This is much smaller, in general, than for the size anomaly. The charts also show the optimal number of portfolios across time when controlling for industry momentum. These are much larger than the corresponding row in the left column. Intuitively, the inclusion of controls soaks up some of the variation in returns previously explained only by 12-2 momentum. This lower variance results in a higher choice of $J$ (see equation \eqref{eqn:optimal J}).  This example makes clear that the appropriate choice of the number of portfolios reflects a diverse set of characteristics of the data such as cross-sectional sample size, the number of time series observations, the shape of the relationship, and the variability of the innovations. 

\section{Conclusion}
\label{sec:Conclusion}

This paper has developed a framework formalizing portfolio-sorting based estimation and inference. Despite decades of use in empirical finance, portfolio sorting has received little to no formal treatment. By formalizing portfolio sorting as a nonparametric procedure, this paper made a first step in developing the econometric properties of this widely used technique. We have developed first-order asymptotic theory as well as mean square error based optimal choices for the number of portfolios, treating the most common application, testing high vs. low returns based on empirical quantiles. We have shown that the choice of the number of portfolios is crucial to draw accurate conclusions from the data and, in standard empirical finance applications, should vary over time and be guided by other aspects of the data at hand.  We provided practical guidance on how to implement this choice. In addition, we showed that once the number of portfolios is chosen in the appropriate, data-driven way, inference based on the ``Fama-MacBeth'' variance estimator is asymptotically valid.

One of the key challenges in the empirical finance literature is sorting in a multi-characteristic setting where the number of characteristics is quickly limited by the presence of empty portfolios. Instead, researchers often resort to cross-sectional regressions thereby imposing a restrictive parametric assumption. Here, we bridged the gap between the two approaches proposing a novel portfolio sorting estimator which allows for linear conditioning variables.

We have demonstrated the empirical relevance of our theoretical results by revisiting two notable stock-return anomalies identified in the literature---the size anomaly and the momentum anomaly. We found that the estimated relationship between returns and size appears to be monotonically decreasing and convex, with a significant return differential between the function evaluated at extreme values of the size variable. However, the statistical significance is generated by very small firms and the results are no longer robust once the smallest firms have been removed from the sample.  We also found that the estimated relationship between returns and past returns is appears to be monotonically increasing and concave, with a significant and robust return differential. We found that the ``short'' side of the momentum spread trade has become more profitable in later sub-periods. In both empirical applications the optimal number of portfolios varies substantially over time and is much larger than the standard choice of ten routinely used in the empirical finance literature.


\vspace{-.5cm}

\section{References}
\vspace{-.5cm}
\singlespacing
\begingroup
\renewcommand{\section}[2]{}	
\begin{small}
\bibliography{CCFS}{}
\end{small}
\bibliographystyle{jasa}
\endgroup

\clearpage
\captionsetup{justification=raggedright, singlelinecheck=false}

\vspace{-.25cm}


\begin{table}[H]
\vspace{-1cm}
\caption{\doublespacing\textbf{Empirical Results}\\[0.05in] \normalsize{This table reports point estimate and associated test statistics from the models specified in equation \eqref{eqn:size} (top panel) and equations \eqref{eqn:momentum} and \eqref{eqn:momentumWithControls} (bottom panel) using $J^{\star}_t$. The standard estimator refers to the standard implementation with $J=10$. Test statistics are formed using $\hat{V}_\mathtt{FM}$ for the variance estimator.  All returns are in monthly changes and all portfolios are value weighted based on lagged market equity.}}
\label{tab:empirical}
\centering
\scriptsize{}
\begin{tabular}{ll ccc|ccc}
\\
\multicolumn{4}{l}{\textbf{Size Anomaly}} \\
\hline
\\
& \multicolumn{1}{l}{} & \multicolumn{3}{l}{\textit{Point Estimate}} & \multicolumn{3}{l}{\textit{Test Statistic}} \\
& \multicolumn{1}{c}{$(z_H,z_L)$} & High & Low & Difference & High & Low & Difference\\
\hline
\\[-1mm]
1926--2015 & $(\Phi^{-1}(.975),\Phi^{-1}(.025))$ & 0.0089 & 0.0407 & -0.0317 & 5.38 & 8.77 & -6.45 \\[1mm]
 & $(\Phi^{-1}(.95),\Phi^{-1}(.05))$ & 0.0088 & 0.0232 & -0.0144 & 5.03 & 5.82 & -3.31 \\[1mm]
 & $(\Phi^{-1}(.9),\Phi^{-1}(.1))$ & 0.0107 & 0.0147 & -0.0039 & 5.91 & 4.41 & -1.04 \\[1mm]
 & Standard Estimator & 0.0089 & 0.0204 & -0.0115 & 5.81 & 6.00 & -3.09 \\[2mm]

1967--2015 & $(\Phi^{-1}(.975),\Phi^{-1}(.025))$ & 0.0095 & 0.0464 & -0.0369 & 4.70 & 8.68 & -6.46 \\[1mm]
 & $(\Phi^{-1}(.95),\Phi^{-1}(.05))$ & 0.0096 & 0.0227 & -0.0131 & 4.63 & 6.36 & -3.17 \\[1mm]
 & $(\Phi^{-1}(.9),\Phi^{-1}(.1))$ & 0.0103 & 0.0137 & -0.0034 & 4.83 & 4.32 & -0.88 \\[1mm]
  & Standard Estimator & 0.0089 & 0.0183 & -0.0094 & 4.93 & 5.59 & -2.51 \\[2mm]

1980--2015 & $(\Phi^{-1}(.975),\Phi^{-1}(.025))$ & 0.0107 & 0.0453 & -0.0346 & 4.62 & 7.67 & -5.46 \\[1mm]
 & $(\Phi^{-1}(.95),\Phi^{-1}(.05))$ & 0.0111 & 0.0238 & -0.0127 & 4.63 & 5.35 & -2.51 \\[1mm]
 & $(\Phi^{-1}(.9),\Phi^{-1}(.1))$ & 0.0108 & 0.0092 & 0.0016 & 4.45 & 2.58 & 0.36 \\[1mm]
  & Standard Estimator & 0.0101 & 0.0163 & -0.0062 & 4.79 & 4.52 & -1.49 \\[2mm]
\hline

\\[.25cm]
\multicolumn{4}{l}{\textbf{Momentum Anomaly}} \\
\hline
\\
& \multicolumn{1}{l}{} & \multicolumn{3}{l}{\textit{Point Estimate}} & \multicolumn{3}{l}{\textit{Test Statistic}} \\
& \multicolumn{1}{c}{$(z_H,z_L)$} & High & Low & Difference & High & Low & Difference\\
\hline
\\[-1mm]
1926--2015 & $(\Phi^{-1}(.975),\Phi^{-1}(.025))$ & 0.0170 & -0.0074 & 0.0244 & 7.39 & -1.83 & 5.25 \\[.5mm]
 & \textit{\quad w/ controls} & 0.0136 & -0.0102 & 0.0238 & 3.57 & -1.75 & 3.42 \\[1.2mm]
 & $(\Phi^{-1}(.95),\Phi^{-1}(.05))$ & 0.0172 & -0.0062 & 0.0234 & 7.74 & -1.46 & 4.87 \\[0.5mm]
 & \textit{\quad w/ controls} & 0.0138 & -0.0041 & 0.0179 & 3.31 & -0.62 & 2.32 \\[1.2mm]
 & $(\Phi^{-1}(.9),\Phi^{-1}(.1))$ & 0.0143 & -0.0000 & 0.0152 & 6.64 & -0.23 & 3.37 \\[0.5mm]
 & \textit{\quad w/ controls} & 0.0115 & -0.0021 & 0.0136 & 3.03 & -0.42 & 2.13 \\[1.2mm]
 & Standard Estimator & 0.0159 & 0.0000 & 0.0155 & 7.70 & 0.13 & 4.05 \\[2mm]

1967--2015 & $(\Phi^{-1}(.975),\Phi^{-1}(.025))$ & 0.0175 & -0.0082 & 0.0257 & 5.60 & -1.76 & 4.58 \\[0.5mm]
 & \textit{\quad w/ controls} & 0.0146 & -0.0168 & 0.0314 & 3.44 & -2.01 & 3.35 \\[1.2mm]
 & $(\Phi^{-1}(.95),\Phi^{-1}(.05))$ & 0.0163 & -0.0047 & 0.0210 & 5.48 & -1.07 & 3.94 \\[0.5mm]
 & \textit{\quad w/ controls} & 0.0131 & -0.0125 & 0.0255 & 3.28 & -1.77 & 3.16 \\[1.2mm] 
 & $(\Phi^{-1}(.9),\Phi^{-1}(.1))$ & 0.0157 & -0.0063 & 0.0220 & 5.65 & -1.41 & 4.20  \\[0.5mm]
  & \textit{\quad w/ controls} & 0.0083 & -0.0131 & 0.0214 & 2.02 & -2.35 & 3.09 \\[1.2mm]
  & Standard Estimator & 0.0156 & -0.0023 & 0.0180 & 5.62 & -0.58 & 3.66  \\[2mm]

1980--2015 & $(\Phi^{-1}(.975),\Phi^{-1}(.025))$ & 0.0150 & -0.0159 & 0.0309 & 4.13 & -2.45 & 4.15 \\[0.5mm]
 & \textit{\quad w/ controls} & 0.0128 & -0.0208 & 0.0336 & 2.35 & -1.90 & 2.74 \\[1.2mm]
 & $(\Phi^{-1}(.95),\Phi^{-1}(.05))$ & 0.0143 & -0.0127 & 0.0270 & 4.09 & -2.06 & 3.82  \\[0.5mm]
 & \textit{\quad w/ controls} & 0.0117 & -0.0152 & 0.0269 & 2.20 & -1.47 & 2.31 \\[1.2mm]
 & $(\Phi^{-1}(.9),\Phi^{-1}(.1))$ & 0.0144 & -0.0073 & 0.0216 & 4.47 & -1.32 & 3.40 \\[0.5mm]
 & \textit{\quad w/ controls} & 0.0093 & -0.0098 & 0.0191 & 1.67 & -1.35 & 2.08 \\[1.2mm] 
 & Standard Estimator & 0.0150 & -0.0018 & 0.0168 & 4.59 & -0.36 & 2.84 \\[2mm]
\hline

\end{tabular}

\end{table}


\clearpage

\clearpage

\clearpage

\clearpage

\clearpage

\end{document}